\documentclass{emulateapj}
\usepackage{apjfonts}
\newcommand{\scaleup}{\epsscale{1.1}}
\newcommand{\scaleupp}{\epsscale{1.15}}
\newcommand{\plotter}{\plotone}
\newcommand{\plotterr}{\plotone}
\newcommand{\plotterrr}{\plotone}
\newcommand{\breaker}{}

\newcommand{\longtabler}{\LongTables}

\newcommand{\tableset}{deluxetable}

\newcommand{\etal}{et al.}

\newcommand{\mstar}{M_{\ast}}
\newcommand{\msun}{M_{\sun}}
\newcommand{\lstar}{L_{\ast}}

\newcommand{\re}{R_{\rm e}}
\newcommand{\fsb}{f_{\rm sb}}
\newcommand{\paperone}{Paper \textrm{I}}
\newcommand{\papertwo}{Paper \textrm{II}}
\newcommand{\dmu}{\sigma_{\mu}}

\shorttitle{Dissipational Signatures in Core Ellipticals}
\shortauthors{Hopkins \etal}
\slugcomment{Submitted to ApJ, June 5, 2008}
\begin{document}

\title{Dissipation and Extra Light in Galactic Nuclei: \textrm{III}.\ 
``Core'' Ellipticals and ``Missing'' Light}
\author{Philip F. Hopkins\altaffilmark{1}, 
Tod R.\ Lauer\altaffilmark{2}, 
Thomas J. Cox\altaffilmark{1,3}, 
Lars Hernquist\altaffilmark{1}, 
\&\ John Kormendy\altaffilmark{4}
}
\altaffiltext{1}{Harvard-Smithsonian Center for Astrophysics, 
60 Garden Street, Cambridge, MA 02138}
\altaffiltext{2}{National Optical Astronomy Observatory, Tucson, AZ 85726}
\altaffiltext{3}{W.~M.\ Keck Postdoctoral Fellow at the 
Harvard-Smithsonian Center for Astrophysics}
\altaffiltext{4}{Department of Astronomy, University of Texas, 1 
University Station C1400, Austin, Texas 78712-0259}

\begin{abstract}

We investigate how ``extra'' or ``excess'' central light in the
surface brightness profiles of cusp or power-law elliptical galaxies
relates to the profiles of ellipticals with cores.  The envelopes of
cusp ellipticals are established by violent relaxation in mergers
acting on stars present in gas-rich progenitor disks, while their
centers are structured by the relics of dissipational, compact
starbursts.  Ellipticals with cores are formed by the subsequent
merging of the now gas-poor cusp ellipticals, with the fossil
starburst-components combining to preserve a dense, compact component
in these galaxies as well (although mixing of stars smooths the
transition from the outer to inner components in the profiles).  By
comparing extensive hydrodynamical simulations to observed profiles
spanning a broad mass range, we show how to observationally isolate
and characterize the relic starburst component in core ellipticals.
Our method recovers the younger starburst population, demonstrating
that these dense concentrations survive spheroid-spheroid mergers and
reflect the degree of dissipation in the initial mergers that formed
the penultimate galaxy progenitors.  The degree of dissipation in the
mergers that produced the cusp ellipticals is a strong function of
stellar mass, roughly tracing the observed gas fractions of disks of
the same mass over redshifts $z\sim0-2$.  The strength of this
component strongly correlates with effective radius at fixed mass:
systems with more dissipation are more compact, sufficient to explain
the discrepancy in the maximum phase-space densities of ellipticals
and their progenitor spirals. The survival of this component together
with scattering of stars into the envelope in re-mergers naturally
explain the high-Sersic index profile shapes characteristic of very
massive core ellipticals. This is also closely related to the kinematics 
and isophotal shapes: only systems with matched starburst 
components from their profile fits also reproduce the observed 
kinematics of boxy/core ellipticals. The final ``core-scouring'' phase of core
formation occurs when a black-hole binary formed in the merger
scatters stars out of the innermost regions of the extra-light
component. It is therefore critical to adopt a physically-motivated
profile-decomposition that accounts for the fossil starburst component
when attempting to quantify scouring.  We show that estimates of the
scoured mass that employ single-component forms fitted to the entire
galaxy profile can be strongly biased.

\end{abstract}


\section{Introduction}
\label{sec:intro}

Early work to incorporate gas physics and star formation in models of
the formation of bulges and elliptical galaxies through mergers
demonstrated \citep{mihos:cusps,mihos:gradients,mihos:starbursts.94}
that dissipation in disk-disk mergers leads to central starbursts like
those observed in ULIRGs \citep{joseph85}.  \citet{mihos:cusps}
predicted that this process should leave an observable signature by
imprinting a two-component structure into the surface brightness
profiles of merger remnants.  That is, the compact stellar remnant of
the starburst should have a steeper density profile than the inward
extrapolation of the outer \citet{devaucouleurs} $r^{1/4}$-law shape
of the main body of the galaxy.  The ``extra light'' in the starburst
remnant is required \citep{hernquist:phasespace} to explain the high
mass and phase space densities of bulges and elliptical galaxies
\citep{ostriker80,carlberg:phase.space,gunn87,kormendy:dissipation}
and their fundamental plane correlations \citep{dressler87:fp,dd87:fp,
kormendysanders92,rothberg.joseph:kinematics,hopkins:cusps.fp}.

When predicted by \citet{mihos:cusps}, extra light was not known to exist 
in normal ellipticals. \citet{sersic:profile} functions were replacing 
$r^{1/4}$ laws as standard machinery to 
interpret brightness profiles \citep{caon:sersic.fits}.  
The steepest central profiles were 
called ``power~laws'' \citep{ferrarese:type12,kormendy94:review,
lauer:95,byun96:ell.profiles}. 
Both descriptions could be taken to imply, absent cores or nuclear star 
clusters,\footnote{
It is important to distinguish nuclear clusters 
(stellar ``nuclei'') from extra light components -- we demonstrate the 
clear distinction in \citet{hopkins:cusps.ell}.
Nuclei are extremely small (typical radii $\sim1-10\,$pc, 
and mass fractions $\sim10^{-3}\,M_{\rm gal}$) relative to 
starburst/extra light components (radii $\sim0.2-2\,$kpc 
and mass fractions $\sim 0.1\,M_{\rm gal}$). The structural scalings and 
fundamental plane relations of stellar nuclei 
are similar to those of globular clusters, nearly perpendicular to and with 
an opposite physical sense from extra light components. Likewise 
their stellar population ages, metallicities, and kinematics are very 
different from extra light and elliptical galaxies \citep{carollo99:nuclei.scalings,
boker04:nuclei.scalings,walcher06:nuclei.ssp,cote:virgo,hopkins:cusps.ell}.
}
that the brightness profiles of elliptical galaxies are almost power laws with no 
breaks indicative of two-component structure. However, 
\citet{kormendy99} discovered in a small sample of normal ellipticals -- 
and later \citet{jk:profiles}, in all the known low-luminosity ellipticals in the Virgo cluster -- 
that the inner ``power-law'' or ``cusp'' indeed represented a central 
``extra light'' component -- the inner profile of a 
dense, centrally concentrated rise above the inward extrapolation of an outer 
Sersic profile. 
\citet{ferrarese:profiles} and \citet{cote:smooth.transition} identified qualitatively 
similar features -- ``luminosity excess'' in the central regions of ellipticals 
\citep[for a detailed comparison of these features, see the discussion in][]{hopkins:cusps.ell}. 
Extension of this to the profiles of recent gas-rich merger remnants 
observed by \citet{rj:profiles} showed similar signatures. 
The ``power law'' profiles identified by previous authors form a 
part of this extra light; the conclusion that there 
are two components, however, did not become clear until observations from a variety of 
telescopes made it possible to create profiles with sufficiently large dynamic 
range. These authors suggested that the observed extra light is the signature of 
merger-induced starbursts, as predicted by \cite{mihos:cusps}, 
and it is now becoming 
empirically established that essentially all ``power-law'' or ``cusp'' ellipticals 
show this behavior. 

In \citet{hopkins:cusps.mergers} (hereafter \paperone) we showed that observed extra 
light can indeed be identified with the central density excess produced in 
simulations of gas-rich mergers.  We developed a formalism to fit surface 
brightness profiles with two-component models in a manner that accurately recovers
the {\em physical} decomposition between a central dissipational component
(the remnant of the merger-induced starburst) and an outer dissipationless
component (the product of violent relaxation acting on stars formed in the 
pre-merger disks before the coalescence of the merging galaxies).  We used 
this machinery in \citet{hopkins:cusps.ell} (hereafter \papertwo) to analyze
observed ``cusp
ellipticals''\thinspace\footnote{The term ``cusp elliptical'' is used in various ways
in the literature.  Unless otherwise stated, we use this name to refer to 
ellipticals that do not have a central resolved core.  In published papers, 
these are called ``power-law ellipticals'', ``coreless ellipticals'', 
or ``extra light ellipticals.''}.
We showed that this method could be used to recover the dissipational components 
in cusp ellipticals, and that such components were ubiquitous in the local 
cusp elliptical population (in samples of $\sim100$s of such galaxies from 
\citet{lauer:bimodal.profiles} and \citet{jk:profiles}). 
In essentially every case, simulated dissipative merger remnants provided 
good matches to the observed profiles of cusp ellipticals. 
Mutually consistent decompositions of the observed galaxy 
profiles and the simulated merger remnants demonstrated that the structure of cusp 
ellipticals as a function of mass corresponds to that predicted for the remnants 
of gas-rich, spiral-spiral mergers with realistic properties.  That is, the degree
of dissipation needed for mergers to explain the densities and scaling laws of 
ellipticals corresponds well with our empirically recovered starburst components
and agrees with the gas fractions available in observed progenitor disk galaxies
of appropriate masses.

However, it has been argued that the situation may be different
for more massive ellipticals, which appear to exhibit central ``cores,'' 
and are known to display boxy isophotal shapes and slow rotation in 
contrast to the disky isophotal shapes and rapid rotation 
seen in less massive cusp ellipticals \citep{kormendy:bulge.rotation,
davies:faint.ell.kinematics,davis:85,jedrzejewski:87,
bender:88.shapes,bender89,bender:ell.kinematics,peletier:profiles}. 
Cores were first seen in ground-based observations of nearby, high-luminosity 
ellipticals 
as central regions of nearly constant surface brightness 
\citep{king78,young78,lauer85:cores,kormendy85:profiles,kormendy:cores.review}. 
They were cuspier than isothermal cores, but the functional form
of the density profile as $r\rightarrow0$ was unknown.  Later, {\it HST} images
showed that nearly all ellipticals have singular starlight distributions in the sense
that the surface brightness diverges as $\Sigma(r)\sim r^{-\gamma}$
\citep{lauer91,lauer92:m32,lauer92,crane93,kormendy94:review,ferrarese:type12,lauer:95}.
In low-luminosity, early-type galaxies, $\gamma$ typically decreases only
slowly as the center is approached, and a steep $\gamma>0.5$ cusp continues
in to the {\it HST} resolution limit. \citet{lauer:95} classified these as
``power-law'' galaxies.  In more luminous ellipticals, the steep outer density 
profile shows a robust break to a shallow, inner 
power-law cusp with $\gamma<0.3$  \citep{lauer:95}. 
The ``break radius'' $r_b$ corresponds to the 
``core radius'' $r_c$ measured in ground-based observations \citep{kormendy94:review}.  
\citet{lauer:95} continued to call these ``core galaxies,'' even though 
the shallow cusps in projected brightness imply steep and singular cusps in 
luminosity density \citep{lauer:95,gebhardt96}.  

The division of central 
structure into two families was motivated by the observed bimodal distribution 
of cusp slopes $\gamma$ \citep{gebhardt96,lauer:bimodal.profiles}, 
but this aspect of central slopes participates in a larger, 
longer-recognized division of the elliptical population: 
the typical giant, core elliptical has different physical properties from the 
typical normal-luminosity, cusp elliptical 
\citep[e.g.][]{davies:faint.ell.kinematics,davis:85,
jedrzejewski:87,bender:88.shapes,bender89,bender:ell.kinematics,
peletier:profiles,kormendybender96,faber:ell.centers,
simien:kinematics.1,simien:kinematics.2,simien:kinematics.3,
emsellem:sauron.rotation.data,emsellem:sauron.rotation,
mcdermid:sauron.profiles,cappellari:anisotropy}. 
Massive giant ellipticals rotate slowly and have 
``boxy'' (rectangularly-distorted) isophotal shapes, 
characteristic of systems supported by anisotropic velocity dispersions 
\citep{schwarzschild:orbit.structure,dezeeuw:orbits,binneytremaine}; 
\citet{faber:ell.centers} showed that these relate 
to the observed ``core'' population. 
In contrast, less massive ellipticals (and S0 galaxies), where 
the power-law population predominates, rotate 
more rapidly; they have more isotropic 
velocity dispersions and look 
like they have embedded disks (``disky'' isophotal shapes). 

These differences thus led naturally to the idea, developed, in e.g.\ 
\citet[][and references therein]{faber:ell.centers,kormendy99,
quillen:00,rest:01,ravindranath:01,laine:03,lauer:centers,
lauer:bimodal.profiles,ferrarese:profiles,cote:smooth.transition,jk:profiles},
that disky, rapidly rotating cusp ellipticals are direct 
products of gas-rich (``wet'') mergers, whereas boxy, slowly rotating core 
ellipticals have been 
shaped by subsequent dissipationless (``dry'') re-mergers of 
two or more (initially cuspy gas-rich merger remnant) ellipticals. 
Several questions therefore 
arise.  How did core ellipticals form?  What were their progenitors?  
It has been shown 
that mergers of bulgeless disks fail to reproduce the shapes and kinematic properties 
of these galaxies \citep[see e.g.][]{barnes:disk.halo.mergers,
hernquist:bulgeless.mergers,hernquist:bulge.mergers,naab:gas,cox:kinematics}. 
Furthermore, if disks are their progenitors, then
these systems would not be able to avoid dissipation, because spiral galaxies 
contain gas.  If lower-mass cusp ellipticals are the progenitors of core
ellipticals, what happened to the extra light in those cusp ellipticals? 

Nearly all numerical experiments find that light profile shapes are,
to lowest order, preserved in dissipationless mergers
\citep[e.g.,][]{boylankolchin:mergers.fp}.  So, core ellipticals
should be ``extra light'' ellipticals in the same sense as cusp ellipticals -- 
i.e. their profile reflects a combination of an outer, low phase-space 
density violently relaxed remnant of stellar disks, and an inner, compact 
component originally formed via dissipational processes (in whatever 
process formed the progenitors, that would allow 
gas to lose angular momentum and reach these densities in the first place). 
If their last merger was dissipationless, is the
amount of dissipation in the ``original'' spheroid forming merger
(i.e.\ that which formed their progenitors, if these are re-merger
remnants) important or relevant to the properties of the $z=0$ galaxy?
Is any memory of the merger history preserved, and is there any way to 
observationally recover this information?

There have been suggestions of interesting behavior in the observations, 
but they have largely lacked an interpretive context. 
Models of core galaxies generally presume that the projected stellar
brightness decreases monotonically outside the core with a
monotonically increasing logarithmic slope -- the fact that 
many core ellipticals can be reasonably well-fitted by such descriptions 
has made the multi-component structure of such objects more ambiguous. 
A closer look at the
existing data, however, shows evidence in some galaxies 
of characteristic features such as inflection points 
in the slope outside the core, similar to a ``smoothed'' version of the 
features often seen in cusp ellipticals where the profile 
transitions from being dominated by a dissipational (``extra light'') 
component superimposed on a background dissipationless 
outer envelope \citep[see e.g.][]{ratcliff82,barbon84,lauer85:cores,
lauer:bimodal.profiles,jk:profiles}. 
Moreover, even where profiles are smooth and monotonic, 
the inner portions of core galaxies typically rise well above 
an $r^{1/4}$ law fitted to the
envelope. Instead, single profiles fitted to the galaxies are forced to 
higher Sersic indices that rise more steeply at small radii, 
reflecting and including the dense, high surface brightness central extra light 
component, and yielding the high Sersic indices ($n_{s}\sim6-8$) 
characteristic of core ellipticals fitted in this manner 
\citep[e.g.][]{prugniel:fp.non-homology,trujillo:sersic.fits,
ferrarese:profiles,jk:profiles}. 
\citet{cote:smooth.transition} point out that there appears to be a smooth 
transition from low-$n_{s}$ outer profiles with a steep central 
rise above the corresponding 
inward extrapolation, to high-$n_{s}$ outer profiles with 
a corresponding rise implicit in the Sersic profile, and hence not 
explicitly appearing above this threshold. 
Without a physical motivation 
for decomposing this inner rise and outer envelope, interpretation of this 
phenomenon has been restricted to the empirical notation of 
the best-fit Sersic indices. 
In the context of the present work,
however, it highlights the potentially composite nature of core
galaxies.

It is generally believed that the connection between the merger history 
of galaxies and their nuclear profile slope (``cusp'' or ``core,'' 
on scales much less than the effective radius or the scales of the extra light) 
arises because of ``scouring'' by a binary black hole 
\citep[for a review, see][]{gualandrismerritt:scouring.review}. 
\citet{begelman:scouring} first pointed out that binary black holes 
coalescing in a dissipationless
galaxy merger stall (i.e.\ are no longer efficiently 
transported to the center via dynamic friction) at radii $\sim$\,pc, 
larger than the radii at which gravitational radiation can efficiently 
dissipate energy and merge the binary -- the so-called 
``last parsec problem.'' They noted that significant gas content 
can provide a continuous source of drag and friction and 
solve this problem in gas-rich mergers, but that in ``dry'' mergers, the binary 
will remain stalled for some time, and will harden by scattering stars in the 
nucleus in three-body interactions. This will continue, flattening the 
nuclear slope, until sufficient mass in stars 
($\sim M_{\rm BH}$, by simple scaling arguments) is ejected to merge 
the binary. It is therefore of particular interest to estimate the 
stellar mass which must be scattered to explain the slopes of cores, 
as a test of scouring models and (in such models) a probe of the 
galaxy merger history. However, such estimates have been ambiguous 
\citep[and often controversial; see e.g.][]{ferrarese:profiles,lauer:bimodal.profiles,
lauer:massive.bhs,cote:smooth.transition,jk:profiles}, 
in large part because of the lack of an 
{\em a priori} physical model for the profile shape. Understanding the global 
profile shapes and extra light in core ellipticals is therefore important to 
reliably estimating how their nuclear profiles have, in detail, been 
modified by scouring.

It is also important to recognize that there can be both continuity and 
bimodality in the cusp/core populations. Because the expected number of major 
mergers in the formation history of 
a typical elliptical is not large ($\sim$ a couple), it should be a relatively 
Poisson process: a significant number of ellipticals (especially at low masses) 
will have experienced only the original, single major gas-rich merger 
that transformed them into ellipticals since $z\sim2-3$ 
\citep[see e.g.][]{maller:sph.merger.rates,
hopkins:groups.qso,hopkins:groups.ell,somerville:new.sam,lin:mergers.by.type}; 
others (especially at high masses) will have experienced 
$\sim$one or two subsequent major mergers, which will tend to be 
``dry'' spheroid-spheroid mergers. Although there might be some intermediate 
cases, to the extent that properties (such as kinematics 
and isophotal shapes) are affected by the last major 
merger, there should be significant differences 
between those whose last merger was dissipational (with a disk that contains 
some mass in gas) or dissipationless (spheroid-spheroid).

However, although the last merger may be dissipational or dissipationless, 
the total 
amount of dissipation in the formation history -- the mass fraction formed 
dissipationally, from gas losing angular momentum in mergers and participating 
in nuclear starbursts -- should be continuous across either population 
(for e.g.\ a given mass and original formation time). Although some objects 
may have had subsequent dry mergers, they still formed stars in a dense 
central concentration in the original, gas-rich merger that formed the 
progenitor spheroid; this process will be the same regardless of whether or not 
the system is destined for a future dry merger. As a function of mass, this 
should broadly reflect the gas fractions of ultimate progenitor disks at the 
spheroid formation times, and as such is a continuous function of 
mass, star formation history, and formation time. This continuity in 
dissipational content, to the extent that it effects the structure, fundamental 
plane correlations, and stellar populations of spheroids, is reflected 
in the continuity of e.g.\ the fundamental plane \citep[recently, see][]{
cappellari:fp,bolton:fp.update,hyde:stellar.mass.fp}, 
the stellar age and metallicity versus mass relations \citep{trager:ages,
nelan05:ages,thomas05:ages,gallazzi:ssps,gallazzi06:ages}, 
and the color-magnitude relation \citep{strateva:color.bimodality,
baldry:bimodality}. As there has been 
considerable observational debate regarding the degree of 
continuity or bimodality between cusp and core populations 
\citep[see e.g.][]{ferrarese:profiles,lauer:bimodal.profiles,
cote:smooth.transition,jk:profiles}, 
it is clearly of interest to identify properties 
that are or are not expected to be continuous across detailed 
merger and re-merger histories within the spheroid population. 

In order to understand the structure and formation history of the 
core elliptical population, we therefore extend our study of 
merger remnants and cusp ellipticals in \paperone\ and \papertwo\ 
to the core population in this paper. We wish to test the hypotheses 
that these systems were, in fact, originally formed (i.e.\ their progenitors 
were formed) 
in gas-rich mergers (albeit potentially modified in gas-poor re-mergers), 
and that the original degree of dissipation can be empirically 
recovered, and is the critical parameter 
that can explain their densities, scaling relations, 
profile shapes, and sizes. 

In \S~\ref{sec:sims} and
\S~\ref{sec:data} we describe our set of merger simulations
and the observational data sets we consider, respectively.  In
\S~\ref{sec:profile.evol} we study how light profiles of 
gas-rich mergers, which we studied in detail in \paperone\ and 
\papertwo, evolve in subsequent re-mergers of such ellipticals, 
and demonstrate that our fitting procedures designed to recover 
the original dissipational/starburst component in gas-rich merger 
remnants can be applied to re-merger remnants. 
In \S~\ref{sec:gradients} we investigate how gradients in stellar 
populations, imprinted by the extra light and original 
gas-rich merger, are affected by re-mergers. 
Readers interested primarily in our comparison of the properties 
and scalings of dissipational components in observed core 
ellipticals may wish to skip to \S~\ref{sec:fitting}, where we 
compare our simulations with and apply our fitted galaxy
decomposition to a wide range of observed systems. In
\S~\ref{sec:scaling} we use these comparisons to study how
structural parameters of the outer stellar light and inner extra light
component scale with galaxy properties, 
and compare them with the extra light components in 
gas-rich merger remnants and cusp ellipticals, and examine 
how the existence and strength of the extra light component
is related to galaxy structure 
and drives galaxies along the fundamental plane. 
In \S~\ref{sec:kinematics} we consider the global isophotal 
shapes and kinematic properties of re-merger remnants 
and how they depend on extra light content.
In \S~\ref{sec:outer.sersic} we examine how 
re-mergers and issues of profile fitting 
relate to the outer Sersic profiles of core ellipticals and re-merger 
remnants, and compare results obtain with different 
choices of empirical fitting functions. In \S~\ref{sec:missing.light} 
we demonstrate that this can affect estimates of 
``missing light'' on small scales, and explain how this 
effect arises and 
what it means for a proper physical understanding of 
nuclear light profiles. Finally,
in \S~\ref{sec:discuss} we discuss our results and outline future
explorations of these correlations.

Throughout, we adopt a $\Omega_{\rm M}=0.3$, $\Omega_{\Lambda}=0.7$,
$H_{0}=70\,{\rm km\,s^{-1}\,Mpc^{-1}}$ cosmology, and appropriately
normalize all observations and models shown, but note that this has
little effect on our conclusions.  We also adopt a
\citet{chabrier:imf} initial mass function (IMF), and convert all
stellar masses and mass-to-light ratios to this choice. The exact IMF
systematically shifts the normalization of stellar masses herein, but
does not substantially change our comparisons. All magnitudes are in
the Vega system, unless otherwise specified.

\breaker
\section{The Simulations}
\label{sec:sims}

Our merger simulations are described in \paperone\ and \papertwo\ 
\citep[for details, see][]{springel:gadget,springel:entropy,
springel:multiphase,springel:models}. They are fully hydrodynamic 
simulations of galaxy-galaxy mergers, including physical and empirically 
motivated models for black hole accretion, star formation, a multi-phase 
interstellar medium, and feedback from supernovae, stellar winds, and 
black hole growth (we show in \papertwo\ that 
our conclusions are robust to variations in these prescriptions). The typical gravitational 
softening is $\sim20-50\,$pc ($\sim 0.01\,R_{e}$), and 
hydrodynamic gas smoothing lengths in the starbursts are smaller, 
which we show in \paperone\ and \papertwo\ 
is sufficient to properly resolve not only the mass 
fractions but also the spatial extent of the extra light components of 
interest here. 

We consider a series of several hundred simulations of colliding
galaxies, described in \citet{robertson:fp,robertson:msigma.evolution} and
\citet{cox:xray.gas,cox:kinematics}.  We vary the numerical resolution, the orbit of the
encounter (disk inclinations, pericenter separation), the masses and
structural properties of the merging galaxies, initial gas fractions,
halo concentrations, the parameters describing star formation and
feedback from supernovae and black hole growth, the 
equation of state of the interstellar medium gas, initial merger 
redshifts, halo virial velocities, merger mass ratios\footnote{The results 
here are primarily from equal-mass mergers; however the behavior 
does not change dramatically for mass ratios to about 3:1 or 4:1, 
appropriate mass ratios for comparison with the observations of ellipticals 
used in this paper. At higher mass ratios, 
the result is a small bulge in a still disk-dominated galaxy 
\citep[see e.g.][]{younger:minor.mergers,hopkins:disk.survival,
hopkins:disk.heating},
which we do not study here.}, and initial black hole masses.

To this, we add a subset of spheroid-spheroid 
``re-mergers,'' representative of gas-poor or ``dry'' mergers of 
elliptical galaxies. In these cases, we collide two remnants of previous 
disk-disk mergers, in order to explore how their properties are 
modified through re-merging. We typically merge two identical remnants (i.e.\ two 
identical copies of the remnant of a given disk-disk merger), but 
have also explored re-mergers of various mass ratios (from 
1:1 to 4:1), and mixed morphology re-mergers (i.e.\ merging an 
elliptical remnant with an un-merged gas-rich disk). In the former case, 
we generally find a similar division in mass ratio at which a ``major'' 
merger is significant. In the latter, we find the properties are more akin to 
those of other gas-rich (disk-disk) mergers, and the remnant should 
for most purposes still be 
considered the direct product of a gas-rich merger. 
In our re-merger series, we vary the orbital parameters, both of the 
initial gas-rich merger and re-merger, and consider systems with 
a range of initial gas fractions in the (pre gas-rich merger) progenitor disks. 
Our re-mergers 
span a similar range in virial velocities and final stellar masses to 
our gas-rich mergers. 

Each simulation is evolved until the merger is complete and the remnants are 
fully relaxed, then analyzed 
following \citet{cox:kinematics} in a manner designed to mirror 
the methods typically used by observers (for details see \papertwo). 
The projected surface brightness profile is fitted with standard elliptical isophotal 
fitting algorithms \citep[e.g.][]{bender:87.a4,bender:88.shapes} and 
radial deviations 
of the iso-density contours from the fitted ellipses determine 
the boxyness or diskyness of each contour (the $a_{4}$ parameter). 
Throughout, we show profiles and quote our results in 
terms of the major axis radius. 
The effective radius $\re$ is the projected half-mass stellar 
effective radius\footnote{This differs from what is sometimes adopted 
in the literature, where $\re$ is determined from the best-fitting
Sersic profile, but because 
we are fitting Sersic profiles to the observed systems we usually quote both.}, 
and the stellar mass $M_{\ast}$ refers to the total stellar mass of the galaxy.
When we plot quantities such as $\re$, we 
typically show just 
the median value for each simulation across $\sim100$ sightlines. The sightline-to-sightline 
variation in these quantities is typically smaller than the 
simulation-to-simulation scatter, but we explicitly note where it is large.

\breaker
\section{The Data}
\label{sec:data}

We compare our simulations to and test our predictions on an ensemble
of observed surface brightness profiles of ellipticals 
from \citet{jk:profiles} and \citet{lauer:bimodal.profiles}. The 
important aspects of the observations for our purposes are also 
summarized in \papertwo. Briefly, first
is the $V$-band Virgo elliptical survey of \citet{jk:profiles}, based
on the complete sample of Virgo galaxies down to extremely faint
systems in \citet{binggeli:vcc} 
\citep[the same sample studied in][]{cote:virgo,ferrarese:profiles}. 
The HST images alone,
while providing information on the central regions, typically extend
to only $\sim1$\,kpc outer radii, which is insufficient to fit the
outer profile. 
\citet{jk:profiles} therefore combine observations from a
large number of sources
\citep[including][]{bender:data,bender:06,caon90,caon:profiles,davis:85,kormendy:05,
lauer:85,lauer:95,lauer:centers,liu:05,peletier:profiles} 
and new photometry from McDonald Observatory, the HST archive, and 
the SDSS 
for each of their objects which enables accurate surface brightness measurements
over a wide dynamic range: 
profiles spanning $\sim12-15$ magnitudes in surface brightness,
corresponding to a range of nearly four orders of magnitude in
physical radii from $\sim10\,$pc to $\sim100\,$kpc. 

We also add surface brightness profiles from \citet{lauer:bimodal.profiles}, 
further supplemented by \citet{bender:data}. 
\citet{lauer:bimodal.profiles} compile $V$-band measurements of a
large number of nearby systems for which HST imaging of the galactic
nuclei is available.  These include the 
\citet{lauer:centers} WFPC2 data-set, the \citet{laine:03} WFPC2 BCG
sample (in which the objects are specifically selected as brightest
cluster galaxies from \citet{postmanlauer:95}), and the \citet{lauer:95}
and \citet{faber:ell.centers} WFPC1 compilations (note that issues of 
completeness and environment are not important for any of our conclusions).
This extends our sampling of the
high-mass end of the mass function, but at the cost of some dynamic
range in the data: \citet{lauer:bimodal.profiles} 
combine these data with ground-based
measurements to construct profiles that typically span
physical radii from $\sim10\,$pc to $\sim10-20$\,kpc. 
We have compared these data with additional 
profiles used in \citet{bender:data,bender:ell.kinematics,bender:ell.kinematics.a4,
bender:velocity.structure}, and in some cases subsequently updated. 
These are more limited, extending from $\sim30-50\,$pc out to $\sim$ a few
kpc, but they allow us to construct multicolor ($V$, $R$, $I$) 
surface brightness, ellipticity, and $a_{4}/a$
profiles, which we use to estimate our sensitivity to 
the observed waveband and photometric quality/dynamic range.

When comparing to cusp populations and gas-rich mergers 
from \paperone\ and \papertwo, we occasionally refer to the 
sample of local remnants of recent gas-rich
mergers from \citet{rj:profiles}. Details of these observations 
are summarized in \paperone; they are $K$-band profiles of 
remnants spanning a range in merger
stage, from ULIRGs and (a few) unrelaxed systems to shell
ellipticals. As demonstrated therein, these systems will almost all
become (or already are, depending on the classification scheme used)
typical $\sim \lstar$ cusp ellipticals. 

Here, we are specifically interested in testing the 
hypothesis that core ellipticals retain some memory of extra light, even 
if they have subsequently been modified by 
dry re-mergers. We therefore divide our sample into those 
systems which are confirmed via HST observations as
being either cusp or core ellipticals. We will specifically focus on 
the core ellipticals, but will compare their properties 
to those of cusp ellipticals and gas-rich merger remnants which 
we study in detail in \paperone\ and \papertwo. 
We exclude spheroidals, as they are not
believed to form in major mergers as are ellipticals \citep[e.g.][]{kormendy:spheroidal1,
kormendy:spheroidal2,jk:profiles} (they also predominate as satellites at 
extremely low masses, and not as core galaxies). 
We also exclude S0 galaxies: these likely form a continuous family with 
low-luminosity cusp ellipticals (and their properties 
are consistent with our gas-rich merger remnants in \papertwo), but 
robustly separating the dissipational extra 
light and violently relaxed components in such objects would require ideal 
subtraction of the large-scale disk (and fitting three-component models 
involves large degeneracies). Furthermore, S0 galaxies are almost uniformly 
classified as cusp ellipticals in the literature. 

This yields a final sample of $\approx 110$ unique core elliptical galaxies 
spanning masses $\lesssim0.1\,\mstar$ to $\gtrsim10\,\mstar$, 
with fitted parameters presented in Table~\ref{tbl:core.fits} 
(with a comparison sample of $\approx 80$ cusp ellipticals 
and $\approx50$ gas-rich merger remnants from 
\paperone\ and \papertwo).
There is overlap in the samples used; we have
$\sim300$ surface brightness profiles for our collection of unique
ellipticals, including (for many objects) repeated measurements in 
multiple bands and with various instruments. 
This provides a useful means to quantify error estimates in fits to these
profiles and check for systematic sources of error, as the variations between fit parameters 
derived from observations 
in different bands or made using different
instruments are usually much larger than the formal statistical errors in the
fits to a single profile. For sources with 
multiple independent observations, we define error bars (Table~\ref{tbl:core.fits}) for each 
fit parameter representing 
the $\sim1\,\sigma$ range in 
parameters derived from various observations, typically from three
different surface brightness profiles but in some cases from as many
as $\approx 5-6$ sources (where there are just $2$ sources, the ``error'' 
is simply the range between the two fits)\footnote{In many cases the 
different observations are comparable; 
in some there are clearly measurements 
with larger dynamic range and better resolution: the errors derived 
in this manner should in such cases be thought of as the 
typical uncertainties introduced by lower dynamic range or less 
accurate photometry.}.

We convert the observations to physical units given our adopted cosmology 
and compile global parameters and calculate stellar masses 
as described in \papertwo\ (our conclusions are not sensitive to 
either of these procedures). The data often cover a dynamic range and have resolution
comparable to our simulations, provided we do not heavily weight the
very central ($\lesssim30\,$pc) regions of HST nuclear profiles. 
Experimenting with different smoothings and imposed dynamic range
limits, we find it is unlikely that resolution or seeing differences
will substantially bias our comparisons. They can introduce
larger scatter, however: the robustness of our results increases considerably 
as the dynamic range of the observed profiles is increased.

Throughout, we will usually refer interchangeably to the observed surface
brightness profiles in the given bands and the surface stellar mass
density profile. In \paperone\ and \papertwo\ we test this assumption 
in simulations, and show that once the system is relaxed,
the optical bands become good proxies for the stellar mass
distribution, with $\lesssim20\%$ variation in our $M/L$ over the entire 
fitted range of radii, in good agreement with what is inferred using 
the observed color gradients or full resolved stellar population data 
to determine $M/L$ versus radius (systems have been 
corrected or excluded for the effects of dust, the observed color gradients are weak, 
age and metallicity gradients tend to offset in $M/L$, and the ages of the 
observed systems are uniformly old, all leading to relatively little bias). 
Furthermore, comparison of systems
observed in different bands demonstrates that our conclusions are
unchanged regardless of the observed
bands in which we analyze these systems.

\breaker
\section{Light Profile Evolution in Re-Mergers}
\label{sec:profile.evol}

\subsection{Physical Consequences of the Re-Merger}
\label{sec:profile.evol:physics}

We first examine how the surface brightness profiles of typical merger remnants 
are modified by re-mergers. 
To lowest order, since the gas is mostly exhausted, the re-merger will 
be dissipationless, and merging 
two dissipationless, similar systems $M_{1}$ and $M_{2}$ on a parabolic orbit should
roughly 
preserve their profiles. This leads to the energy conservation equation
\citep{hausman:mergers,hernquist:phasespace}
\begin{equation}
E_{f} = k\,{(M_{1}+M_{2})}\,\sigma_{f}^{2} = E_{i} = k\,M_{1}\,\sigma_{1}^{2} + k\,M_{2}\,\sigma_{2}^{2}
\label{eqn:egy.cons}
\end{equation}
where $\sigma_{f}$ is the velocity dispersion of the final remnant, and $k$ is a constant 
that depends on the shape of the profile. 
For a 1:1 merger of initially similar systems, then, 
the mass doubles and the velocity dispersion is unchanged,
$\sigma_{f}=\sigma_{i}$. 
This implies that the effective radius must double as well. Modulo a simple overall 
rescaling of the profile, we therefore 
have reason to believe that our two component approach, applied successfully to 
cusp ellipticals in \papertwo, should 
be able to identify the original dissipational and dissipationless components 
of the pre-re-merger galaxies. 

\begin{figure}
    \centering
    \scaleup
    \plotterr{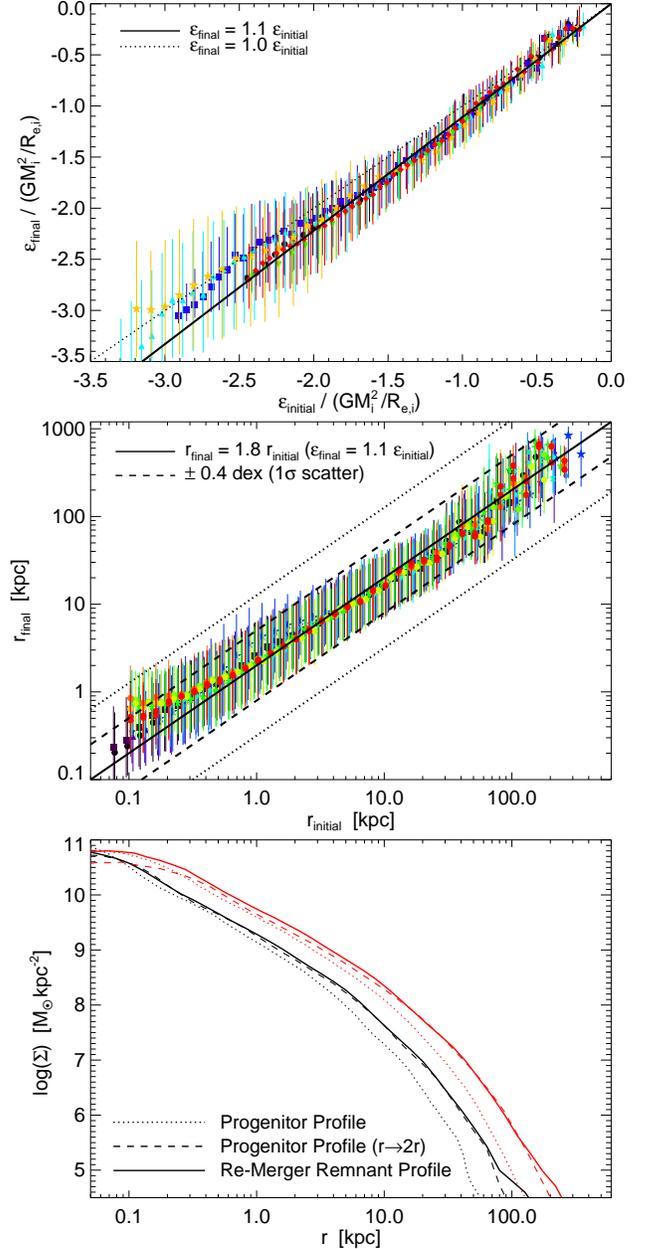}
    \caption{The impact of re-mergers on the stellar light profile. {\em Top:} 
    Final specific binding energy ($\epsilon_{\rm final}$) 
    of stellar particles after a spheroid-spheroid re-merger 
    is plotted relative to their specific binding energy after the first, 
    spheroid forming gas-rich merger ($\epsilon_{\rm initial}$; both $\epsilon$ 
    are in units of the characteristic energy $G\,M_{i}^{2}/R_{e,i}$, where 
    $M_{i}$ and $R_{e,i}$ are the galaxy mass and effective radius before any mergers).
    Each plotted point is the median value for $>100$ particles within $\pm0.025$\,dex 
    in $\epsilon_{\rm initial}$, with error bars showing the $\pm1\,\sigma$ range of $\epsilon_{\rm final}$. 
    Different colors and symbols denote different simulations. 
    Dotted line represents exact conservation of specific binding energy; solid line 
    is the best-fit constant rescaling $\epsilon_{\rm final}\approx1.1\,\epsilon_{\rm initial}$ 
    (representing the fact that energy is preferentially redistributed to the less bound parts of the 
    halo). 
    {\em Middle}: Same, but for the radii of each stellar particle 
    before ($r_{\rm initial}$) and after ($r_{\rm final}$) a re-merger (reflecting their 
    pre and post-merger orbital apocentric radii). 
    Solid line represents the expectation from the $\epsilon_{\rm final}-\epsilon_{\rm initial}$ 
    relation and the virial theorem: $r_{\rm final}\approx1.8\,r_{\rm initial}$. Dashed and 
    dotted lines indicate the $\pm1\,\sigma$ ($0.4$\,dex) and $\pm2\,\sigma$ range.
    {\em Bottom:} Comparison of initial progenitor projected surface mass density profile, 
    and that profile with all radii increased by a uniform factor of $2$ (the total mass 
    also doubled) 
    with that of the re-merger remnant (two representative examples shown). 
    \label{fig:rf.of.ri}}
\end{figure}

Figure~\ref{fig:rf.of.ri} compares this simple expectation to a subset of our re-merger 
simulations. We plot the ``final'' specific binding energy (binding energy per unit mass 
$\epsilon_{\rm final}$ 
of a given stellar particle after the system relaxes following a 1:1 spheroid-spheroid 
re-merger) as a function of ``initial'' specific binding energy (specific binding 
energy $\epsilon_{\rm initial}$ 
before the re-merger, but after the initial spheroid-forming merger). 
Equation~(\ref{eqn:egy.cons}) generalized to the individual stars implies that this 
should be conserved in a re-merger, and indeed, we see a tight correlation 
of the form $\epsilon_{\rm final}\approx\epsilon_{\rm initial}$. It has been 
well-established for some time that 
particles tend to preserve their rank order in binding energy in 
dissipationless mergers 
\citep{barnes:disk.halo.mergers}, and we confirm this here (with a 
scatter in $\epsilon_{\rm final}/\epsilon_{\rm initial}$ of $\sim0.15$\,dex). 
In detail, fitting the correlations in our simulations gives 
$\epsilon_{\rm final}\approx1.1\,\epsilon_{\rm initial}$. 
The highly bound particles are slightly more bound after the re-merger, 
because we are considering 
the stellar mass here, not the total mass -- relatively more of the energy is transferred to the 
less bound outer halo compared to the tightly bound inner spheroid 
(something like this must happen, since some material is scattered to positive energies in 
the re-merger, removing energy from the more bound particles). 

If the particle orbital distributions are conserved, then, we expect from our 
estimate above that their radii (or, for non-circular orbits, their apocentric radii) 
should approximately double after a re-merger. We therefore compare 
the ``final'' and ``initial'' radii of stellar particles in the same manner. 
Note that although technically an instantaneous three-dimensional stellar particle 
radius has no inherent physical meaning for non-circular 
orbits, even an instantaneous snapshot (e.g.\ viewing a galaxy at some moment) will 
capture most particles near their apocenter (in the mean over any ensemble of 
particles, effectively one will recover the apocentric radii, because this is where 
any given orbit spends most of its time). 
We therefore show in Figure~\ref{fig:rf.of.ri} the 
radii at a given instant before and after the re-merger 
but find that measuring a time-averaged radius or 
the apocentric radii averaged over some time interval yields nearly identical results. 
We consider a narrow range of initial radii and 
examine where the stars at these radii end up after a re-merger, plotting the 
median and $\pm1\,\sigma$ range of their final 
radii (each plotted bin contains at least $>100$ 
particles, and most contain $>1000$). To lowest order, the median final effective 
radii are roughly twice the initial radii. 

In detail, fitting the points shown yields a mean 
increase in radius of a factor of $1.8$ (slightly less than $2$, 
because of the fact noted above that energy is preferentially transferred to less bound 
material in the outer halo; in fact this factor $1.8$ corresponds exactly to the 
$\epsilon_{\rm final}=1.1\,\epsilon_{\rm initial}$ change in specific energies, given 
the virial theorem). 
At small radii, there appears to be a deviation from this, but this can be 
explained entirely 
by a combination of resolution effects 
and the fact that the distribution is one-sided (there 
are only positive radii). To lowest order, then, the mass profile is preserved, with all 
radii ``inflated'' or ``puffed up'' by a factor $\sim2$. We show this explicitly in the 
figure; comparing the final profiles with the progenitor profiles expanded by this 
factor. Correspondingly (given mass conservation), 
the effective surface brightness $I_{e}$ decreases by a factor $\sim2$ 
and $\sigma$ is conserved; however, typical elliptical profiles are sufficiently steep 
that this ``puffing up'' leads to a final surface brightness profile that is everywhere 
(at each fixed radius) brighter than the progenitors (with the effect larger at large radii). 

In short, we have recovered the conventional wisdom that 
that particles tend to preserve their rank order in binding energy in 
dissipationless mergers 
\citep{barnes:disk.halo.mergers}, 
which also leads to their mass profiles and central mass concentrations 
being preserved, at least in mergers of systems with certain
distributions,
such as the NFW profile \citep{nfw:profile}, the \citet{hernquist:profile} profile, 
or 
the \citet{devaucouleurs} $r^{1/4}$ law \citep[e.g.][]{boylankolchin:mergers.fp}. 

However, for a given initial orbital apocentric radius or specific binding energy, 
there is in fact a wide range of final orbital apocentric radii, which we will show 
in \S~\ref{sec:missing.light} has important consequences 
for the details of the observed profile. In detail, the distribution of final radii for a given 
initial radius is close to log-normal (out to $\sim3\,\sigma$ in the wings), 
with a median a factor of $\sim1.8$ larger 
than the initial radius and a $1\,\sigma$ dispersion of $0.38$\,dex (factor $\approx2.5$). 
Most of this scatter owes to mixing in specific binding energy 
(the scatter in $\epsilon_{\rm final}/\epsilon_{\rm initial}$ of $\sim0.15$\,dex translates 
to a scatter in $r_{\rm final}/r_{\rm initial}$ of $0.3$ dex), and would be present 
even if individual orbital structures were perfectly conserved; the rest owes to 
mixing altering the nature of individual orbits (i.e.\ changing the isotropy of the 
system and moving stars between tube and box orbits); only a small component 
of the scatter is attributable to the inherent uncertainty in defining an orbital 
``radius'' in a non-trivial potential. 
Figure~\ref{fig:rf.of.ri} shows the $1\,\sigma$ range for each bin in initial radius and 
this fit to the entire distribution.

\begin{figure*}
    \centering
    \scaleup
    \plotone{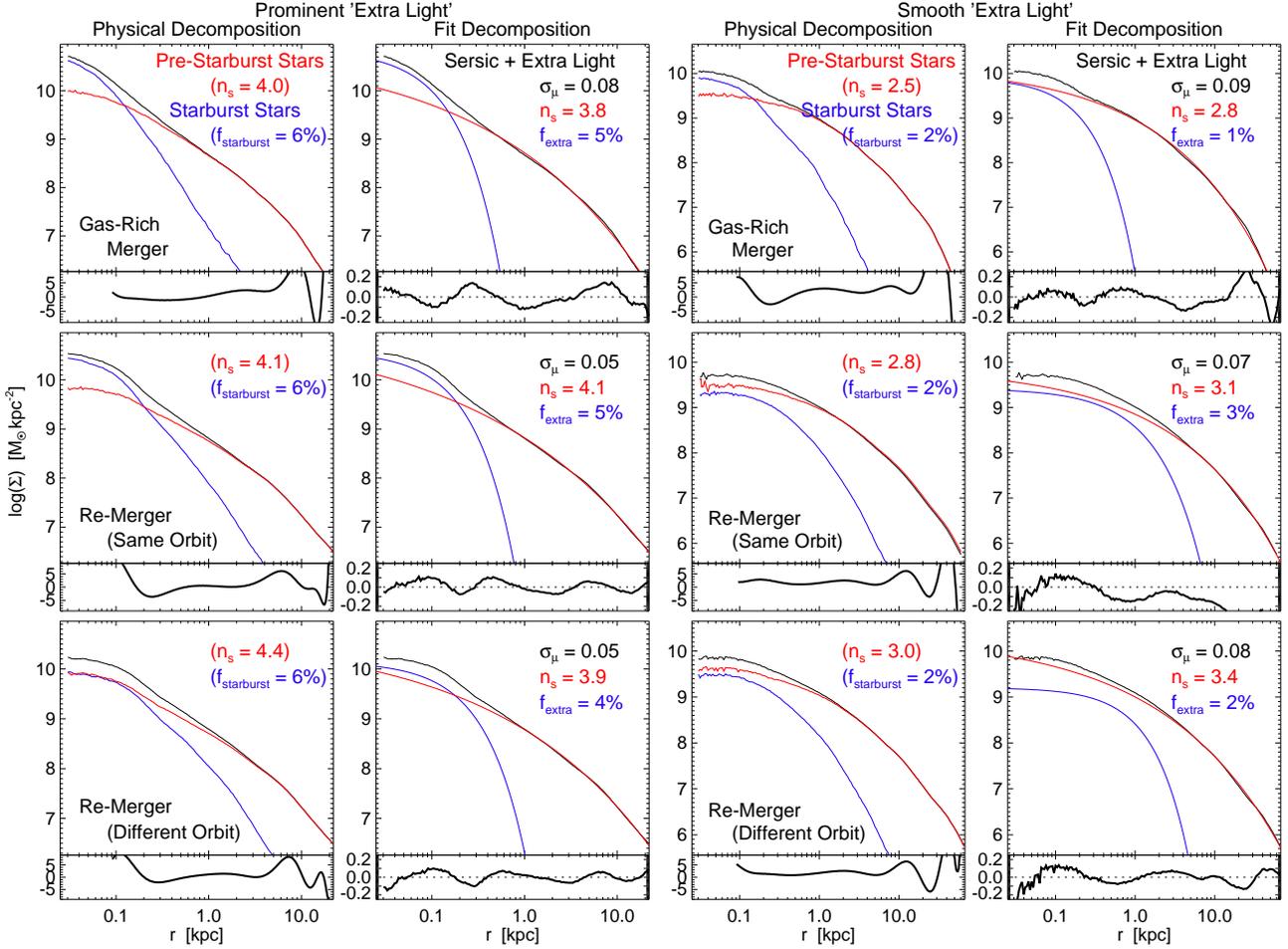}
    \caption{{\em Top:} Surface mass density of a gas-rich merger 
    remnant (black), physically decomposed into stars formed prior to the final merger (which 
    are then violently relaxed; red) and stars formed in the 
    dissipational starburst (blue). 
    The pre-starburst component follows a Sersic law closely, with 
    best-fit index shown, and the starburst mass fraction is labeled. 
    The second derivative of the light profile (${\rm d}^{2}\mu/({\rm d}\log{r})^{2}$) is shown below.
    We compare this with our two-component (Sersic plus cusp or extra light) 
    fit (inner exponential and outer Sersic) to the total light profile, 
    with the Sersic index of the outer component 
    and mass fraction of the inner component, and rms scatter ($\dmu$) about the 
    fit. Residuals from the fit are shown below. 
    {\em Middle:} The same, after a spheroid-spheroid 
    re-merger of the remnant above. The (post re-merger) profile of the 
    stars originally formed in the starburst is now shown, and the re-merger 
    remnant is fitted with the same procedure. 
    {\em Bottom:} Same, but for a re-merger with different 
    orbital parameters. 
    We show a system with a prominent transition to the extra light 
    ({\em left panels}), and a more typical system with a smoother 
    transition between outer and extra light ({\em right panels}).
    Our two-component, 
    cusp plus Sersic function fit accurately 
    recovers the profile of the violently relaxed component and mass fraction 
    of the starburst component, even after a re-merger. The remnant of the 
    original starburst (extra light) is less obvious, however, after re-mergers. 
    \label{fig:demo.rem.appearance}}
\end{figure*}

\begin{figure}
    \centering
    \scaleup
    \plotone{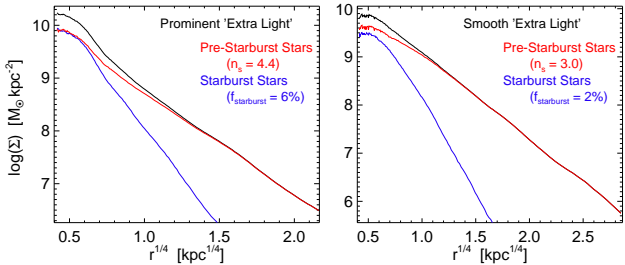}
    \caption{As Figure~\ref{fig:demo.rem.appearance} (bottom panels); 
    showing the re-merger remnant profiles and physical breakdown, in 
    $\mu-r^{1/4}$ instead of $\mu-\log{r}$. The smooth case ({\em right}) 
    retains its extra light in a physical sense (and it is successfully 
    recovered by our decomposition procedure), but the profile shows 
    no obvious upward ``break'' relative to an outer Sersic index. 
    \label{fig:demo.rem.appearance.rquarter}}
\end{figure}

Figures~\ref{fig:demo.rem.appearance} \&\ \ref{fig:demo.rem.appearance.rquarter} 
show how the surface density profile 
of the final galaxy is modified by a re-merger. We show the surface brightness profiles 
for two gas-rich merger remnants, one in which the ``extra light'' 
signature in the center of 
the galaxy is plainly obvious (a good match to e.g.\ 
NGC 4459 in Virgo), and one in which 
it is more subtle (i.e.\ the extra light blends more smoothly into the outer profile; 
analogous to e.g.\ NGC 4886a). 
For each, we also show the profile of those stars specifically 
formed in the central starburst and those not. 
We then show the surface density profiles after a re-merger (for each, we show 
the profile after two different re-mergers with different orbital parameters -- the 
choice of orbital parameters has only a weak impact on the profile). We can follow 
the stars which originally formed in the gas-rich merger-induced starburst, and 
plot their profile as well (even though there is no new starburst in the re-merger, since 
the systems are gas-depleted). 
In both cases, the profiles are roughly preserved, as expected based on the arguments 
above. The starburst light profile, while stretched by a similar factor $\sim2$ by the re-merger, 
remains relatively centrally concentrated and continues to dominate the profile 
at radii $\lesssim1\,$kpc, while the non-starburst stars that were originally relaxed into an 
outer Sersic-like profile continue to follow a similar distribution. 

The light profile is, however, smoothed out by this re-merger. In the case where the 
original remnant shows a sharp departure from the outer profile, the departure remains 
relatively prominent and 
detectable even after a re-merger. But, when the original starburst 
light is more smoothly distributed, the re-merged light profile 
can be smooth, without 
any obvious feature. 
This case demonstrates that because the inner parts of the outer/non-starburst  
component shape (already being ``hot'') are not affected much in 
the re-merger, while the starburst component is heated, we 
often find the starburst component is less dominant in the center 
after the re-merger (they have mixed). It 
is still important to the central 
profile and much more concentrated than the outer light, but the 
distinction is less sharp.
It is therefore interesting to test whether our empirical fitting routines
reliably recover the mass profile of what was, originally, the starburst light, even 
after a re-merger.

\subsection{Decomposing the Remnant Profile}
\label{sec:profile.evol:fitting}

We therefore adopt a two-component approach, described in 
\paperone\ and \papertwo, to fit the 
final (post re-merger) surface brightness profile of each system. 
The methodology arrived at and tested in these papers amounts to a simple 
prescription: we fit the total observed light profile 
(temporarily ignoring the known true physical decomposition of the light profile) 
to the sum of two Sersic components. Based on the arguments 
and tests in \paperone\ and \papertwo, we 
fit to an outer Sersic plus cusp or extra light model, with an outer
component for the original pre-starburst/dissipationless stars with a free Sersic index, and an
inner component reflecting the remnant of the 
dissipational/starburst population. Where data are limited at small radii, 
or in the case of our simulations where we do not resolve the structure 
on very small nuclear scales, we find the best results (in an average sense) 
adopting a fixed Sersic
index $n_{s}=1$ for the inner component, but we note (as demonstrated 
in \papertwo), that the same mean results are recovered for a free 
inner Sersic index, and that where the data are of sufficient quality 
(i.e.\ for systems with HST nuclear profiles extending to very small radii), 
it is possible to free this parameter. 

Our fitting procedure then takes the following form:
\begin{eqnarray} 
\nonumber I_{\rm tot} &=&
I^{\prime}\,\exp{{\Bigl\{}-b_{n}^{\prime}\,{\Bigl(}\frac{r}{R_{\rm extra}}{\Bigr)}^{1/n_{s}^{\prime}}{\Bigr\}}}\\
& &+I_{\rm o}\,\exp{{\Bigl\{}-b_{n}\,{\Bigl(}\frac{r}{R_{\rm outer}}{\Bigr)}^{1/n_{s}}{\Bigr\}}},
\label{eqn:fitfun}
\end{eqnarray} 
where $R_{\rm extra}$ and $R_{\rm outer}$ are the effective radii of the 
inner ($n_{s}^{\prime}\sim1$) and outer (free $n_{s}$) components 
(which we identify with the starburst and
old bulge or pre-starburst components, respectively), $I^{\prime}$ and
$I_{\rm o}$ are the corresponding normalizations, 
$n_{s}^{\prime}$ is the Sersic index of the inner (extra light) component 
(fixed $n_{s}^{\prime}=1$ where resolution limits apply)
and $n_{s}$ is the Sersic
index of the outer bulge or pre-starburst component.  The constant
$b_{n}$ is the appropriate function of $n_{s}$ such that $R_{\rm extra}$ and
$R_{\rm outer}$ correspond to to the projected half-mass radii.
We stress that these fits are intended to 
recover the appropriate profile and fit where these components 
are physically relevant -- i.e.\ at radii $\gtrsim30-50\,$pc where 
our simulations are well-resolved (note that the typical effective 
radii of even the inner component are much larger than this, 
$\sim0.5-1\,$kpc). For most of the galaxy observations 
considered here, this corresponds to radii $\gtrsim0.5''$, a factor 
$\sim10$ larger than the HST diffraction limit. 

The results of applying these fits are shown in Figure~\ref{fig:demo.rem.appearance}. 
Despite the smearing of the transition to the extra light that owes to the re-merger, 
this approach still reliably recovers both the true outer Sersic index of the 
violently relaxed non-starburst stellar material and the mass fraction in the 
dissipational central starburst. We have tested this for our entire library of 
re-merger simulations, and find that while the smoothing of the profile makes the 
object-to-object errors somewhat larger, it does not alter the ability of this 
method to recover, on average, the appropriate breakdown 
between physical components of the galaxy light profile. This is in part because 
the fitting does not inherently rely on a ``break'' in the profile -- if there is a smooth 
but significant change in e.g.\ the effective concentration of the system going 
from inner to outer regions (or equivalently the rate of change of the 
logarithmic slope), it will be reflected in the fits by a dense inner component 
and more extended outer component.

\begin{figure*}
    \centering
    \scaleup
    \plotone{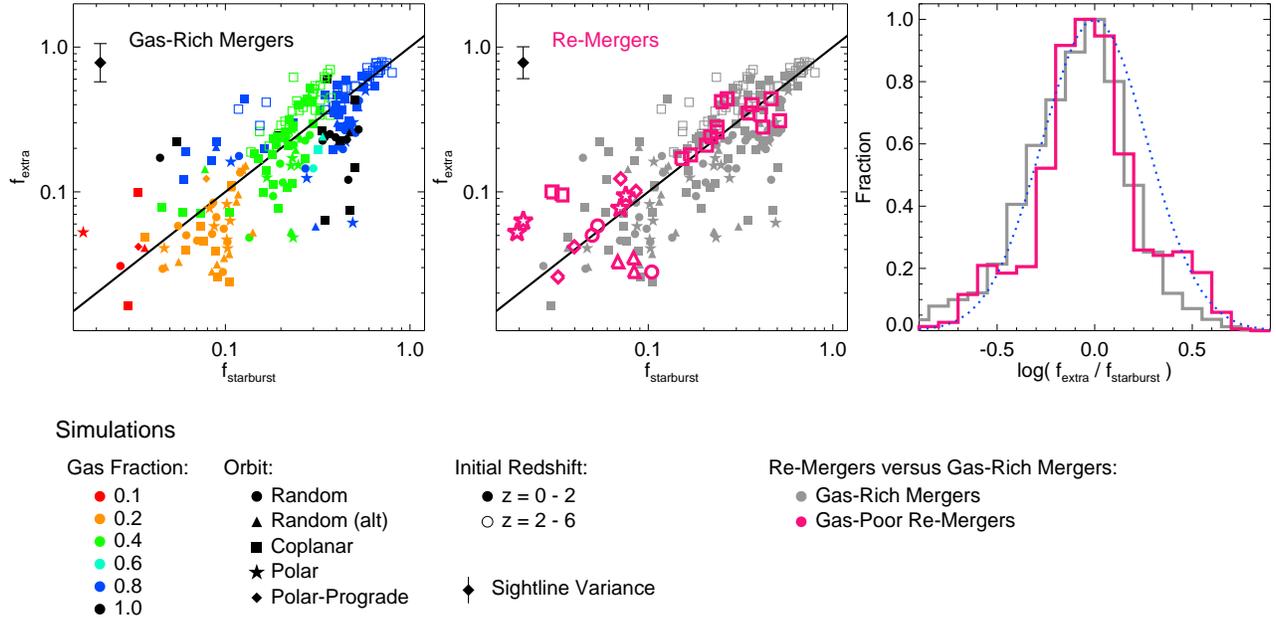}
    \caption{Success of our proposed two-component empirical decomposition 
    at recovering the known physical parameters of the galaxy starburst and pre-starburst 
    (violently relaxed) components.
    {\em Left:} Mass fraction in the fitted ``extra light'' component $f_{\rm extra}$ 
    versus the known mass fraction of the physical starburst $f_{\rm sb}$, 
    in our gas-rich (disk-disk) merger simulations. 
    Each point is the average across $\sim100$ sightlines to a given simulation; 
    median sightline-to-sightline variance is shown ($\approx0.15\,$dex; plotted error bar). 
    Various colors and symbols denote different initial disk gas fractions, orbital 
    parameters, and merger redshifts (key). 
    {\em Center:} Same, but comparing our gas-rich simulations (grey, for clarity) 
    with our gas-poor (dry) re-merger simulations (magenta). 
    The starburst fraction $f_{\rm sb}$ represents the sum of the 
    starburst masses in the two ellipticals re-merged. 
    Symbol shapes denote original orbits of the progenitors in the gas-rich mergers (although 
    the re-mergers follow a mix of orbits). Each of these points is close it its 
    progenitor gas-rich merger (progenitor initial gas fractions can be inferred 
    from the location relative to points in left panel; this is less significant than $f_{\rm sb}$). 
    {\em Right:} Distribution of $f_{\rm extra}/f_{\rm sb}$, including sightline-to-sightline 
    and simulation-to-simulation 
    variation. Histograms are for our gas-rich (grey) and re-merger (magenta) 
    simulations. Dotted blue line shows a Gaussian with $\sigma=0.27\,$dex. 
    The fitted $f_{\rm extra}$ recovers the physical $f_{\rm sb}$ on average, with a 
    factor $\sim2$ scatter, and without any 
    significant bias from any varied simulation parameters. 
    Re-mergers yield a similar result -- the error bars only significantly 
    increase for systems with either very little starburst components (where the smoothing is 
    significant; see Figure~\ref{fig:demo.rem.appearance}) or a large number of 
    re-mergers. 
    \label{fig:rem.recovery}}
\end{figure*}

Figure~\ref{fig:rem.recovery} shows the result of these fits to our gas-rich 
disk-disk merger and gas-poor re-merger simulations. 
We directly compare 
the fitted extra light mass fraction to the mass fraction and size of the known
physical starburst component, and 
find that the fitted components recover the appropriate values in the mean with a 
factor $\sim2$ scatter. This result is robust with respect to e.g.\ the mass, 
orbital parameters, mass ratios, initial gas content, treatment of feedback 
and model for the ISM equation of state, and redshift of our simulations. 
We refer to \paperone\ for a more detailed comparison where we similarly 
demonstrate the ability to recover the starburst effective radius and outer 
component Sersic index. There is no significant bias evident in our re-merger 
sample: although it is considerably more limited, the representative 
systems we have re-merged appear to lie in a similar location in this space to 
their progenitors, with similar scatter. At the very lowest starburst fractions, 
the scatter does increase in re-mergers, owing to the smoothing of the profile 
and mixing of the (small) extra light mass with the outer dissipationless 
mass (shown in Figure~\ref{fig:demo.rem.appearance}).

We note that these simulations do not directly model the process of ``scouring,'' or the 
gravitational scattering of 
individual stars by a coalescing black hole binary. Unfortunately, it 
is prohibitively expensive to model these processes self-consistently in a galaxy-scale simulation, 
as they require resolving the masses of individual stars and sub-parsec spatial scales. 
We instead adopt a standard gravitational softening prescription, and for 
this reason the predicted profiles should not be taken literally within the 
simulation smoothing lengths ($\sim50\,$pc). 
In fact, scouring is expected to alter the galaxy profile significantly only near the 
radius of influence of the central black hole, i.e.\ $\ll 30-50$\,pc 
for most of the systems of interest 
\citep[where the enclosed mass of the galaxy is $\sim M_{\rm BH}$; 
as generally expected from theoretical models and simple scaling 
arguments;][]{milosavljevic:core.mass,
merritt:mass.deficit,sesana:binary.bh.mergers}, 
which is below our simulation resolution limits and 
well below the relevant size scale at which the extra or starburst light we model 
is important ($\sim0.5-1\,$kpc). While a very small fraction of stars at these radii 
may pass near the black hole on centrophilic orbits, there is no scenario, in 
any plausible scouring calculation, in which the profile at radii 
$\sim0.5-1\,$kpc, enclosing a stellar mass $\gtrsim100\,M_{\rm BH}$, 
could be sufficiently altered by this particular process in order to change our 
conclusions that the original extra light is recovered after re-mergers. 
In other words, scouring flattens (``cores out'') the central peak in the 
dissipational or ``extra light'' profile -- it changes the exact slope/shape of the 
central light profile as $r\rightarrow 0$, but not its total mass or $\sim$kpc extent 
(and we have tested in numerical experiments that our adopted fitting 
procedures are robust to changes in the nuclear profile shape). In 
short, the effects of ``scouring'' on this specific calculation are no different than 
the effects of our finite spatial and mass resolution.

\begin{figure*}
    \centering
    \scaleup
    \plotone{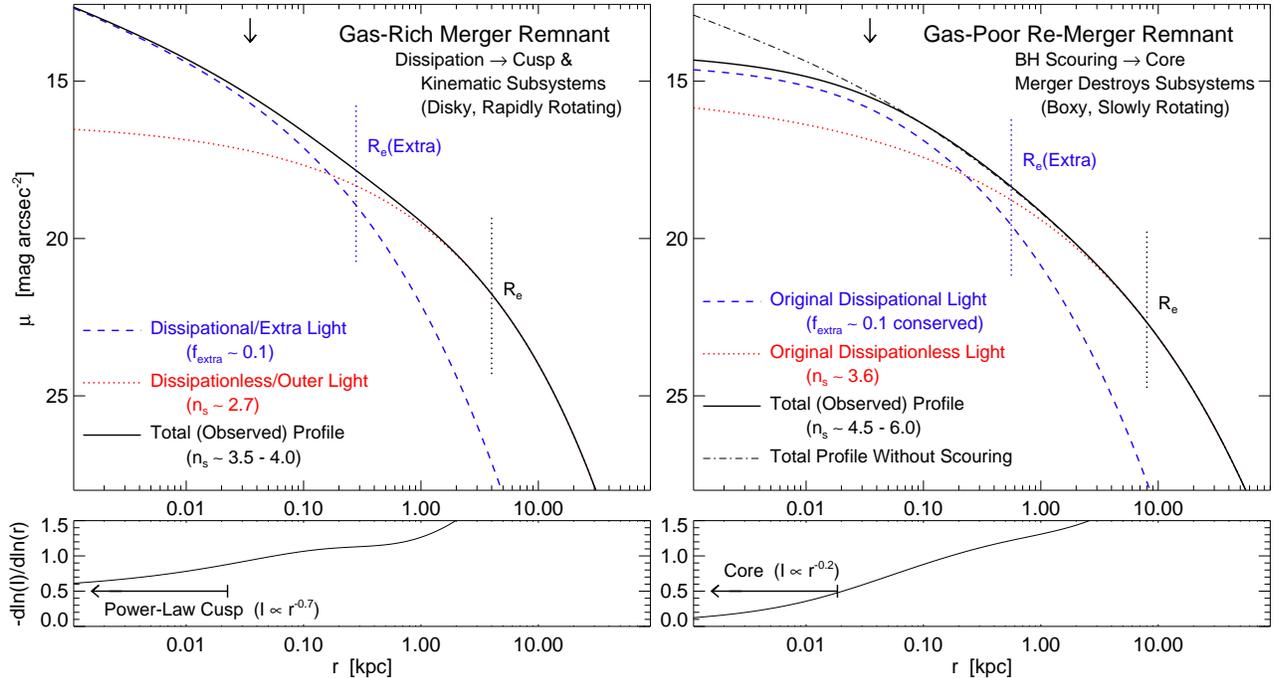}
    \caption{Summary of the physical properties of spheroid profile shapes (here 
    for a typical $\sim\lstar$ elliptical), as 
    formed in a major, gas-rich merger ({\em left}) and then modified by a 
    major gas-poor, spheroid-spheroid re-merger ({\em right}). Lower panels 
    show the logarithmic derivative of the total (observed) profile. 
    In the original merger, stars from the pre-merger disks are violently relaxed 
    into an extended Sersic-like envelope ($n_{s}\sim2.7$). Conservation of 
    phase-space density prevents this component alone from reaching the high densities of 
    observed ellipticals. 
    Gas dissipation, however, 
    yields a nuclear starburst, leaving a dense ``extra light'' component 
    on top of the outer component, dominating the profile at $r\lesssim0.5-1\,$kpc. 
    The nuclear ``cusp'' is the inwards continuation of this dissipational component. 
    Together this yields a global profile with typical $n_{s}\sim4$ and densities 
    of observed ellipticals (much higher than their progenitor spirals). 
    In a re-merger, both components are ``puffed up'' by a factor $\sim2$, and 
    scattering of stars broadens the original dissipationless/envelope component 
    (to $n_{s}\sim3.6$). There is no new dissipation, but the remnant of 
    the original dissipational 
    component continues to dominate the profile within $\sim1\,$kpc and 
    is (in principle) separable. Black hole ``scouring'' scatters stars from the nuclear 
    cusp/extra light -- evacuating $\sim M_{\rm BH}$ worth of stars and flattening the 
    central profile to form a nuclear ``core.'' Although the apparent effects on the 
    extra light profile can be large, the total mass in the ``scoured'' region here 
    (the ``missing mass'') is only $\sim2\%$ of the {\it extra light} mass 
    (most of which is near $\sim0.5-1\,$kpc). 
    Arrows denote the typical resolution limits of our simulations (sufficient 
    to model the structure of the extra light and outer components, but 
    insufficient to simultaneously resolve the transformation of cusps into cores). 
    \label{fig:demo}}
\end{figure*}

Figure~\ref{fig:demo} summarizes the physical properties of the 
surface brightness profiles of gas-rich merger remnants and 
gas-poor (``dry'') re-merger remnants. The original gas-rich merger 
remnant is fundamentally two-component, with a Sersic-like violently relaxed envelope 
(with characteristic Sersic index $n_{s}\sim2.7\pm0.7$, discussed in 
\papertwo) and an inner dissipational/extra light component 
that dominates the light profile inside of $\sim0.5-1$\,kpc. The inwards extrapolation  
of this extra light forms the central cusp, a (roughly) power-law like 
continuation (with characteristic power-law slopes 
${\rm d}\ln{I}/{\rm d}\ln{r}\sim -0.7\pm0.3$). The extra light, and by extension the cusp 
(insofar as it is simply the inwards continuation of the extra light profile), 
as well as embedded kinematic subsystems \citep[see][]{cox:kinematics,naab:gas} 
which make the remnant isophotal shapes more disky and 
yield rapid rotation, are formed by the dissipational star formation event(s). Rapid 
rotation and disky structure are also frequently observed in what we 
now recognize as the extra light regions of cuspy ellipticals 
\citep[e.g.,][]{emsellem:sauron.rotation.data,emsellem:sauron.rotation,jk:profiles}.

In a dissipationless re-merger, both the original dissipational and dissipationless 
components are ``puffed up'' by a uniform factor $\sim2$. The scattering of 
stars (the scatter in Figure~\ref{fig:rf.of.ri}) makes the envelope of the original 
dissipationless component broader, raising its Sersic index 
(discussed in \S~\ref{sec:outer.sersic}). The mass of the original 
dissipational component is conserved, as is its dominance within the central 
$\sim$kpc -- 
this is a consequence of the overall preservation of profile 
shape. However, heating of the inner component and mixing of stars 
according to the scatter in Figure~\ref{fig:rf.of.ri} makes the relative 
prominence (the rise above the outer profile) somewhat more smooth. 
With no scouring mechanism, the central slope would still eventually rise 
about as steeply as the progenitor cusp
(again, reflecting the conservation of profile shape). 
Scouring, or scattering of stars by a nuclear binary black hole, 
will however evacuate the central regions and 
flatten this slope. 

Here we have applied a toy scouring model following 
\citet{gualandrismerritt:scouring.review}, 
which kicks out a total of $\sim2\,M_{\rm BH}$ worth of stars. 
The scoured region (i.e.\ the scoured or missing mass) is therefore 
represented by the area between the scoured and un-scoured profiles. 
Because the extra light dominates the stellar density in the central regions 
(and formed the original cusp), scouring acts primarily on the stars in 
the nuclear region of the extra light. 
This leaves a central core with a shallow slope (${\rm d}\ln{I}/{\rm d}\ln{r}\sim -0.2\pm0.2$). 
The re-merger also tends to destroy the embedded kinematic subsystems 
\citep[see][]{cox:remerger.kinematics} and no new source of 
dissipation is available to re-form them, leaving a remnant with 
boxy isophotal shapes and less rotation.

\subsection{Caveats and Limitations}
\label{sec:profile.evol:caveats}

We emphasize here the clear hierarchy of scales: the large scale profile and 
envelope is 
made primarily from the dissipationless violently relaxed component. 
The extra light dominates the profile within $\sim$kpc scales, a factor of a few 
to an order of magnitude smaller than the effective radius. 
Processes such as scouring and the observational distinctions between 
cusps and cores occur at scales yet another order of magnitude (or more) 
smaller, $<100\,$pc. Because of this hierarchy, although scouring 
can appear to alter the shape of the ``extra light'' profile significantly, 
it has a negligible effect on the estimate of the total extra light mass fraction (here, 
the entire mass content inside the ``scoured'' radius is just $\sim2\%$ of the extra light mass)
or effective radius (this $R_{e}\sim1$\,kpc). In general, we can continue 
to treat galaxy profiles as multi-component (in a physical sense) after a moderate 
number of dry mergers, because the mixing of stars over 
factors $\sim2$ in radii in major re-mergers is insufficient to 
eliminate the differences between such different 
scales. 

This will no longer be true after a sufficiently large number of 
re-mergers (if there is an initial order-of-magnitude difference in the 
scales of the extra and outer light, then simple scalings suggest 
$\sim3$ 1:1 mass ratio re-mergers, or $\sim5-10$ more likely 1:2 $-$ 1:3 
mass ratio re-mergers, could completely ``blend'' the dissipational and 
dissipationless components). However, cosmological simulations, 
clustering measurements, 
and halo occupation models \citep{maller:sph.merger.rates,zheng:hod.evolution,
masjedi:cross.correlations,hopkins:groups.ell} 
suggest this will only be important in the most extreme $M_{\ast}\gtrsim 10^{12}\,M_{\sun}$ 
BCG populations (which constitute only $\sim10-20\%$ of the 
mass density {\em within} the core elliptical population, and only $\sim3-5\%$ of 
the mass density in ellipticals). In this regime, there are other reasons 
to treat our models with caution: such systems almost all experienced their 
first mergers at very high redshifts where the progenitors are less well-understood, 
they live in unusual environments, and at these masses it may be the
case that
growth by a large series of minor mergers becomes more 
important than growth by major mergers. We do not intend, therefore, for 
our models here to be considered as analogs for these systems; our intention is 
to understand the bulk of the core elliptical population at masses 
$\lesssim$ a few $L_{\ast}$ (constituting $\sim80-90\%$ of the mass density in 
core ellipticals), the vast majority of which still live in less
extreme environments \citep[e.g.][]{blanton:env,wang:sdss.hod,
masjedi:merger.rates} and are expected to 
have experienced a small number ($\sim$ a couple) of major re-mergers 
since redshifts $z\sim2-4$. 

Ellipticals with cores, then, if they are the products of 
the re-mergers of cuspy progenitors, should have just 
as much ``extra light'' (dissipational content) as cusp ellipticals, and 
we will show in \S~\ref{sec:fitting} and \S~\ref{sec:scaling} 
that this surviving dissipational component 
obeys the same correlations and has similar properties to the 
extra light in observed cusp systems. However, the 
term ``extra light'' has not generally been applied to core ellipticals, because 
observations typically show less (compared to 
what is seen in cusp ellipticals) of an obvious upward break 
in the profile at small radii relative to the inwards extrapolation of 
their outer Sersic profiles. 
In fact, core ellipticals are sometimes referred to as ``missing light'' 
ellipticals, because they show a deficit in 
their central light profiles, relative to the extrapolation of an 
outer Sersic profile. These differences arise naturally from the 
smoothing and mixing processes in re-mergers, and in fact 
are completely consistent with our advocated formation scenario 
(see \S~\ref{sec:outer.sersic}). 

As discussed above, 
``extra light'' should be physically identified with stellar populations 
originally formed in dissipational starbursts (on top of more 
extended violently relaxed populations from 
dissipationlessly merged stellar disks), whereas ``missing light'' 
is associated with the deficit of stars in the nucleus owing to scattering of 
some small mass by a merging binary black hole. 
In a {\em physical} sense then, {\em ellipticals can be both ``extra'' and 
``missing'' light ellipticals} -- the terms as we mean them 
are not mutually exclusive. Indeed, if 
``extra light'' means stars from dissipation (on top of dissipationlessly 
assembled stars) -- as we use the term -- 
then all ellipticals {\em must} have ``extra light'' at some 
level. And if a black hole binary does not merge sufficiently rapidly (if there 
is little gas in a recent major merger), scouring {\em will} happen. In 
this physical scenario, 
cored or ``missing light'' ellipticals are re-merged extra light ellipticals 
where the inwards extrapolation of the original extra light (which constituted the 
original ``cusp'') has been flattened by scouring, forming a core in the 
center of the dissipational extra light profile. 

For these reasons, we will continue to 
refer to our fitted central components as 
``extra light'' and interpret them as the 
surviving dissipational components from the gas-rich mergers that 
formed the core elliptical progenitors. 
Given our interpretive context and calibration from our re-merger 
simulations, we will show that such components appear ubiquitous in 
core ellipticals and obey well-defined scaling laws closely related to those 
of the dissipational components in cusp ellipticals. 

However, 
that is not to say that 
core ellipticals would necessarily be identified as ``extra light'' galaxies 
in the sense of certain traditional observational metrics. Specifically, 
the smoothing of the transition between inner and outer components 
in re-mergers and raising the outer Sersic profile $n_{s}$ by scattering 
stars to large radii means that the extra light may appear 
less pronounced.
Re-mergers also tend to destroy kinematic subsystems and mix stellar orbits 
\citep{cox:remerger.kinematics}, potentially wiping out obvious changes in isophotal 
shapes or kinematics near the radius where 
the estimated extra light begins to dominate the profile.
Absent such features or a physical model (as we have in the form of our 
simulations) to lend an interpretive physical context and suggest 
a multi-component nature in the first place, there will always be an infinite 
space of functional forms that fit the observed profiles 
with arbitrarily high accuracy and do not include an {\em explicit} 
``extra'' or secondary component. This does not mean, of course, 
that the remnant dissipational populations are not manifest in the 
fits (after all, both our fit and others equally represent the data to a 
meaningful physical accuracy $\sigma_{\mu}\lesssim0.1\,{\rm mag\,arcsec^{-2}}$, 
comparable to the inherent point-to-point variance in simulations and 
observed systems). The parameters of interest here will simply be reflected 
in more indirect fashion (in some combination(s)) in whatever 
fitting function is adopted. We discuss some of these comparisons with 
alternative fitting functions commonly applied to core ellipticals in the 
literature in \S~\ref{sec:outer.sersic}. Given our aims and desire to 
compare simulations and observations on a uniform footing, however, 
we proceed with this specific machinery as a means to 
interpret the observed profiles of core ellipticals.

\breaker
\section{Impact on Stellar Population Gradients}
\label{sec:gradients}

In \papertwo, we demonstrated that dissipation can give rise to 
strong stellar population gradients in ellipticals, and that this relates 
e.g.\ the gradient strengths, sizes, and extra light masses. We therefore 
briefly examine how these gradients might be modified in re-mergers; and 
show that the gradients tend to survive, preserving our predictions 
from \papertwo. 

\begin{figure}
    \centering
    \scaleup
    \plotone{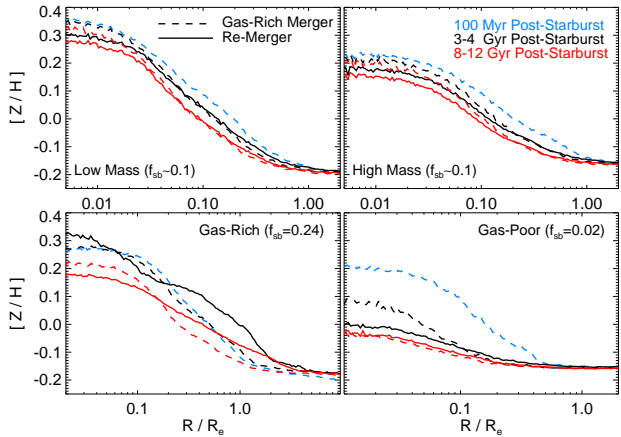}
    \caption{Illustration of the typical effects of a re-merger on the stellar population 
    gradients of spheroids. For four distinct original gas-rich mergers (a $< L_{\ast}$ 
    and $> L_{\ast}$ case with typical starburst mass fractions $f_{\rm sb}\sim10\%$, 
    and a highly gas-rich and very gas-poor $\sim L_{\ast}$ case, as labeled), 
    we show the metallicity as a function of radius (plotted line is the median, variation 
    across $\sim100$ sightlines is $\approx0.05-0.1$ in ${\rm [Z/H]}$ at each $r$). 
    We show the $B$-band light-weighted metallicity, at each of three approximate 
    times after the major starburst corresponding to the original gas-rich merger 
    (the gas-rich merger remnant is evolved passively, the re-merger case experiences a
    re-merger between $t=0-3$\,Gyr). At the same mean stellar population age, 
    the re-merger has had little effect on the profile, washing it out only slightly. 
    The starburst mass fraction and overall stellar age are much more important to 
    the gradient strength. We find similar results for age, $[\alpha/{\rm Fe}]$, 
    and color gradients. 
    \label{fig:grad.fx.demo}}
\end{figure}

Figure~\ref{fig:grad.fx.demo} shows the metallicity as a function of radius 
from several representative simulations, comparing the evolution with time of 
gradients in simulated ellipticals that evolve passively after the original 
gas-rich 
merger to (originally identical) systems that have 
had another (dry) re-merger 
in the intervening time. The details of our methodology 
are described in \papertwo, but, briefly, we model every stellar particle 
as a single burst population with an age and metallicity given self-consistently 
by the star formation model in the simulation. Stars formed before the original 
gas-rich merger (i.e.\ those in our initial conditions) are placed 
on the mass-metallicity relation for disks of the appropriate mass at the 
corresponding redshift for our simulation. We 
project the remnant (along $\sim100$ sightlines, but the 
results depend only weakly on viewing angle) and 
calculate the light-weighted metallicity, stellar population age, and other properties 
in small radial annuli. 

Naively comparing the profile in e.g.\ metallicity or age 
after a re-merger to that immediately after a gas-rich merger, it appears 
that there is a significant difference.  However, this owes largely to age effects 
(the stars in the re-merger have aged significantly during the merger). Clearly 
the correct procedure is to compare at fixed age or post-original gas-rich merger 
time. It is also important to rescale the systems by their effective radii, as 
we expect a roughly uniform puffing up of the remnant in a re-merger. 

Figure~\ref{fig:grad.fx.demo} shows that when we do this, 
there is relatively little smearing out of the gradients by re-mergers. 
There is some effect, but it is generally of the order of the scatter 
in the gradients across sightlines,
and is generally weaker than the effects of age or variation in 
initial gradient strength across our simulations owing to different degrees of 
dissipation. This is expected -- Figure~\ref{fig:rf.of.ri} demonstrates that the 
rank order in radii is preserved in a mean sense. Although there is some
scatter in the particle radii, it is not dramatic as far as gradients are concerned.
Note in Figure~\ref{fig:grad.fx.demo}, 
the change in metallicity with radius is quite smooth, typically 
decreasing from the central metallicity over $\sim2$ orders of magnitude 
in radius. Compared to this dynamic range, 
the typical dispersion in final radii for stellar particles 
at a given initial radius ($\sim0.4\,$dex) is significant, but not large. 
Thus even strong gradients are relatively robust to the 
smoothing effect in re-mergers.

\begin{figure}
    \centering
    \scaleup
    \plotone{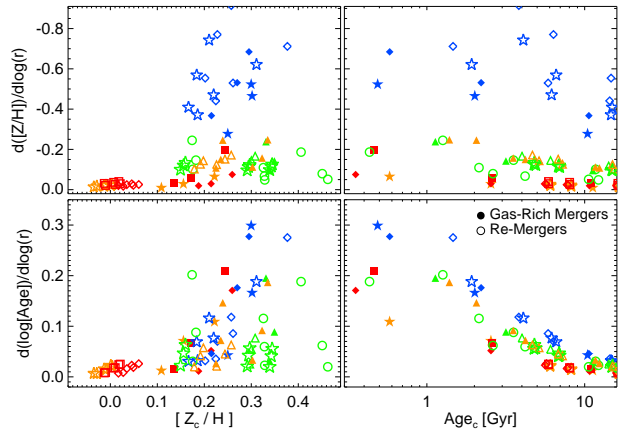}
    \caption{Stellar population gradients in 
    metallicity or age as a function of central metallicity or age, 
    in a subset of gas-rich merger 
    remnants (filled symbols) and re-mergers of these same systems (open 
    symbols). 
    Color denotes the initial gas fraction of the original gas-rich mergers, 
    and symbol type denotes orbital parameters, as in Figure~\ref{fig:rem.recovery}. 
    The dilution of gradients in re-mergers is weak -- they continue 
    to trace the same trends as gas-rich mergers of the same 
    stellar age. The effect of increased dissipation in the original merger 
    is dominant, and the re-merger is less important in general than 
    the simple subsequent aging of the stellar populations by $\sim$ a few 
    Gyr after the gas-rich merger. 
    \label{fig:grad.correlations}}
\end{figure}

Figure~\ref{fig:grad.correlations} summarizes these results for our 
re-merger simulations, plotting the gradient strength (quantified as 
the logarithmic slope in metallicity or age versus radius at 
$\sim0.1-1\,R_{e}$; for details see \papertwo) as a function of 
central (averaged within an aperture of $R_{e}/8$, 
comparable to typical observations) metallicity or 
stellar population age. 
We show both gas-rich merger remnants and the products of 
dry re-mergers of those same systems. 
Quantified in this manner, there is no significant offset between 
the re-mergers and the passively evolved gas-rich merger 
remnants. 

In \papertwo\ we showed that these distributions and 
the gradient strength as a function of mass and velocity dispersion, 
as well as color gradients as a function of elliptical age, 
all agree well with those observed in elliptical populations. 
Because these are not entirely washed out in re-mergers, 
this agreement continues to hold for re-mergers and 
core ellipticals. It has been specifically noted 
observationally \citep[e.g.][]{lauer:centers,ferrarese:profiles,
kuntschner:line.strength.maps,mcdermid:sauron.profiles} 
that core ellipticals and even BCGs exhibit typical color and 
stellar population gradients compared to other galaxies 
of similar mass and age. To the extent that they 
may represent re-mergers of (originally) cusp ellipticals, our 
simulations imply that this is the natural expectation.

\breaker
\section{Dissipational Content in Elliptical Galaxies with Cores}
\label{sec:fitting}

\subsection{Fitting Surface Brightness Profiles}
\label{sec:fitting:fits}

If the preceding discussion is correct, 
extra or starburst light 
survives a galaxy merger in some sense. 
We expect, then, that 
even if core ellipticals have been shaped by spheroid-spheroid 
re-mergers, and their profiles 
smoothed, they should show a central stellar density similar to our simulations 
corresponding to what was (originally) their starburst light content, 
which should be recovered (on average) by our applied decompositions. 
Are observed systems consistent with this scenario?

\begin{figure*}
    \centering
    \scaleup
    \plotone{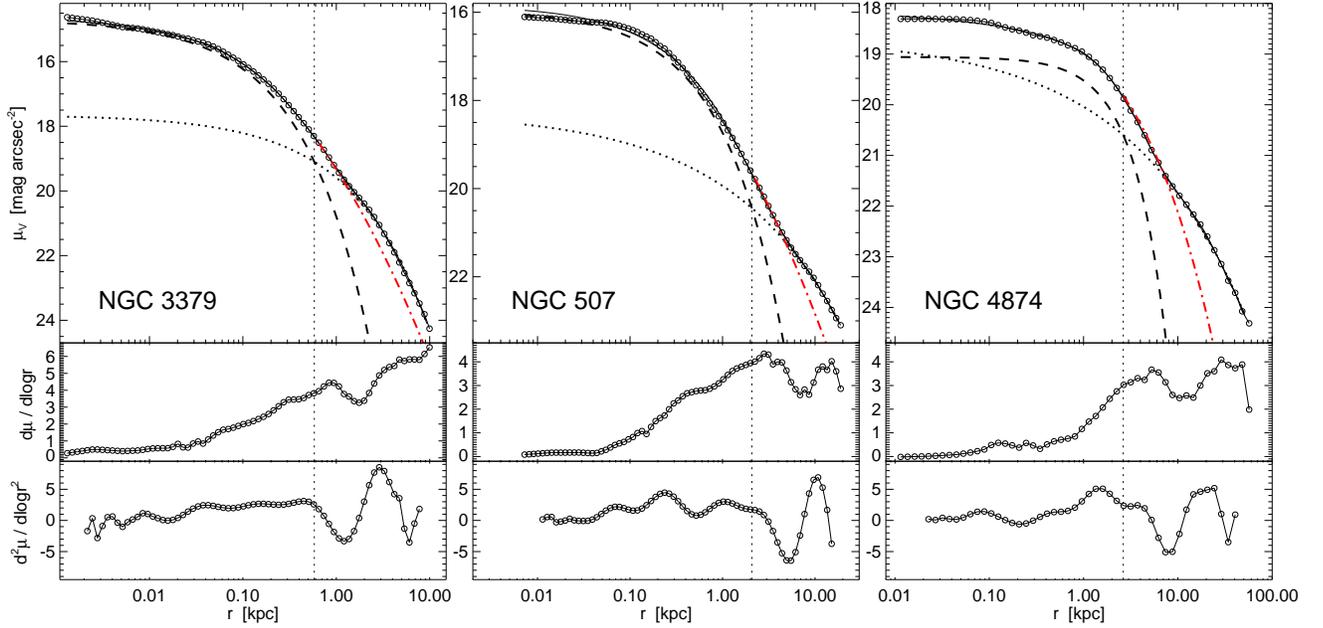}
    \caption{Examples of core ellipticals with possible surviving signatures of 
    starburst or ``extra'' light from cusp progenitors 
    in their nuclei (preserved, albeit smoothed out, from the original, gas-rich spheroid 
    forming merger). {\em Top:} Observed surface brightness profile (open circles) 
    and best-fit two-component model (solid line). The inner (starburst remnant; 
    black dashed line) and outer (violently relaxed dissipationless remnant; black dotted line) 
    components of the best-fit model are shown, with the radius (vertical dotted line) 
    at which the two are equal. Nuker fits (red dot-dashed) 
    to the profile between the central $\sim100$\,pc (inside which the core dominates the fit) 
    and radius of the shoulder are shown -- these fall short where our fitted 
    outer component dominates the light, reflecting a 
    significant change in the curvature of the profile. 
    {\em Middle:} First derivative of the surface brightness 
    with respect to $\log{r}$. {\em Bottom:} Second derivative of the profile. 
    Characteristic features or shoulders in the profiles can survive re-mergers, 
    manifest plainly in a non-parametric fashion in both first and second derivatives, 
    corresponding to the radius at which our method estimates the starburst 
    component 
    begins to dominate. (The specific objects are discussed further in the text.) 
    \label{fig:shoulders}}
\end{figure*}

\begin{figure*}
    \centering
    \scaleup
    \plotone{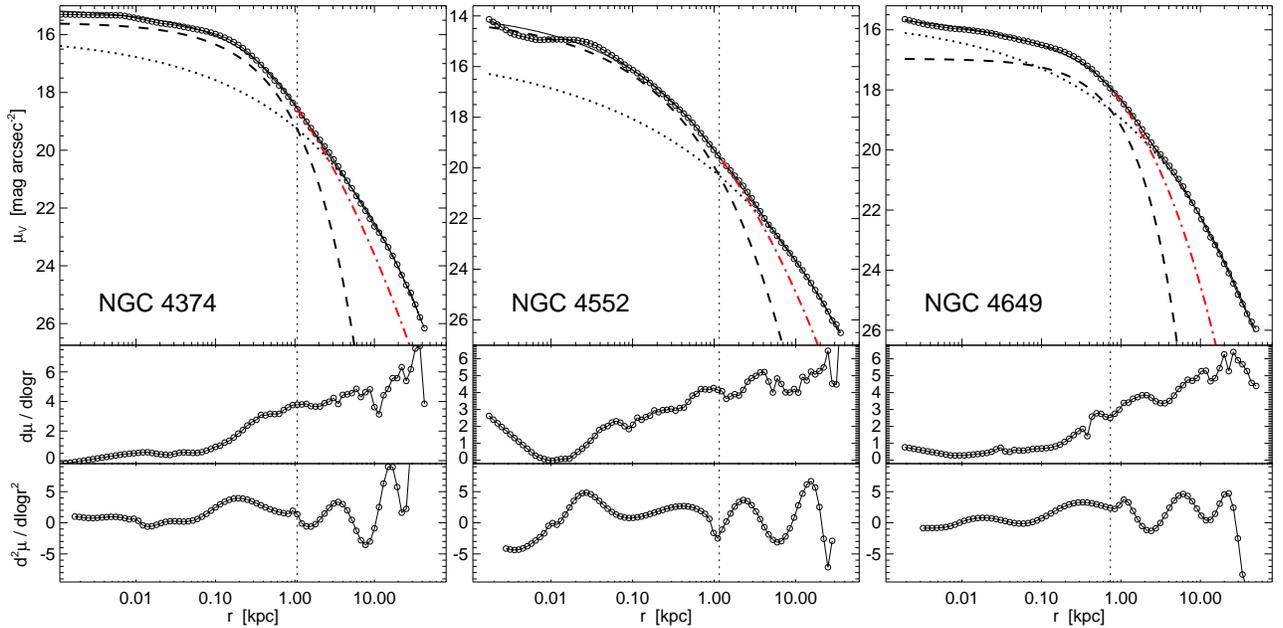}
    \caption{As Figure~\ref{fig:shoulders}, but for systems with less 
    prominent/obvious features. Despite being weaker, a similar 
    transition is seen, and the amplitude in the first and second profile 
    derivatives is comparable. The Nuker fits to the central 
    profile similarly underpredict the outer component, as expected if 
    a second physical component becomes important.
    \label{fig:shoulders.lesser}}
\end{figure*}

Figure~\ref{fig:shoulders} shows the surface brightness profiles of three 
core ellipticals from moderate mass (NGC 3379, a 
$\sim\mstar$ core elliptical) to very massive (NGC 4874, a $\gtrsim10\,\mstar$ 
elliptical). In these systems, there is a distinct shoulder in the light profile, 
which is evident in a completely non-parametric sense in the first and second 
logarithmic derivatives of the profiles -- the slope of the profile is clearly non-monotonic 
with radius over a brief range.  Moreover, the shape of this feature is 
similar in all three cases. There are perhaps 
alternative explanations for these features 
\citep[NGC 3379 may in fact be a face-on S0, see][and 
NGC 4874 is a very massive galaxy with a large envelope, albeit 
not technically a BCG]{lorenzi:3379}
but these are not unique systems -- 
there are a significant number of other such cases, 
as we describe below.

The features in these observed light profiles are similar,
both in their shape and characteristic radii,
to the physical transition in our simulations 
where the extra or starburst light begins to dominate with respect to the 
dissipationless, non-starburst light. 
Applying our two-component fit decomposition 
to these profiles, we find a good fit that (as expected) is able to 
explain these 
shoulder-like features as the transition between extra/starburst light and 
the outer light distribution. 
In a parametric sense, if we fit a Sersic 
or Nuker-law profile to the central regions (specifically, we consider radii between 
$100\,$pc -- within which the core structure 
is important -- and $\sim0.5-2$\,times the radius at which our two component fit determines that the two 
components are equal; the exact range is not important), we find that the 
substantial curvature of the inner profile leads to a fit which falls short of the shoulder, 
and fails to reproduce the outer profile dominated (in our two-component model) by an 
outer, violently relaxed population. 

Figure~\ref{fig:shoulders.lesser} compares three 
galaxies with relatively weak features, in Virgo. 
These are systems where presumably the scattering of stars in 
re-mergers, discussed in \S~\ref{sec:profile.evol} has smoothed any 
strong break to the extra light that might have been 
prominent in the progenitors. 
Despite not being obvious on 
visual inspection, however, the first and second derivatives of the profile show similar 
signatures, and in fact the deviations from a single profile fit (a Sersic or Nuker law) 
extrapolated outward from the central region (interior to the shoulder) 
are significant. Moreover, we find here (and in Figures~\ref{fig:demo.rem.appearance}-\ref{fig:demo})
that although the formal uncertainties may be  
larger when the profiles are smoothed in this manner,  
our two-component model is able to successfully fit them: \S~\ref{sec:profile.evol} 
demonstrates that even in systems with smoother profiles 
than these, our applied decompositions are robust.

\breaker
\begin{figure*}
    \centering
    \plotone{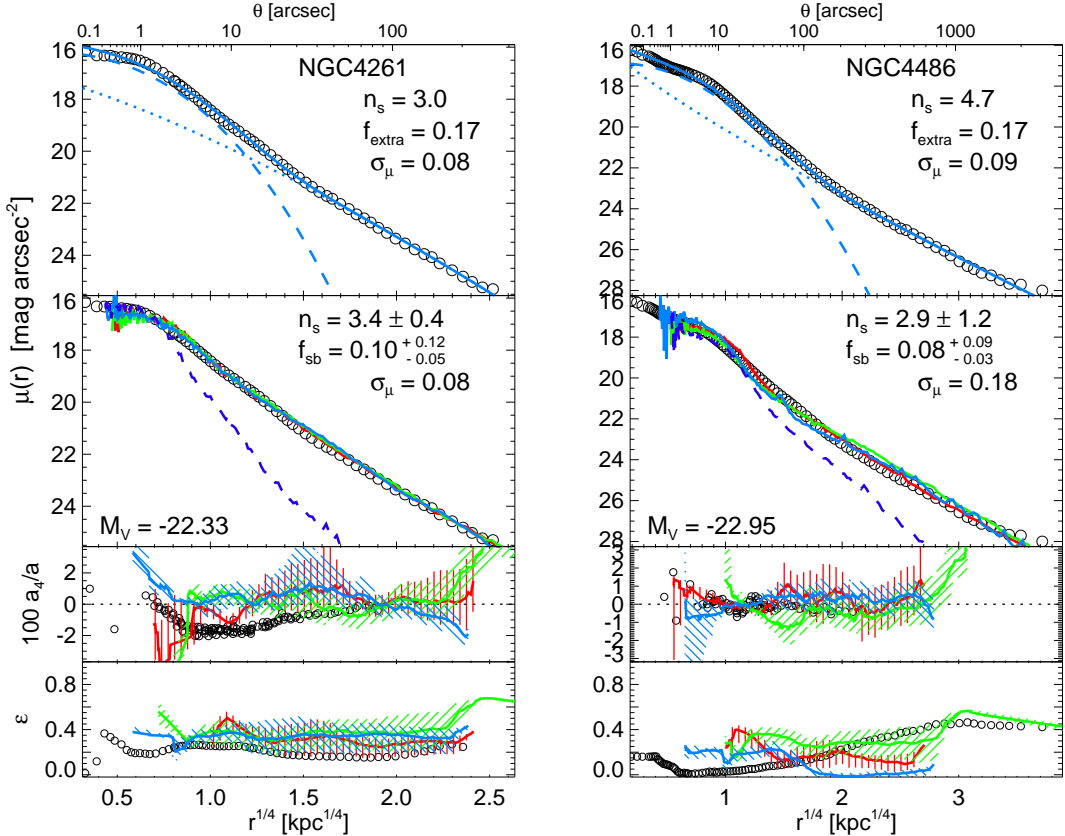}
    \caption{Surface brightness profiles -- decomposed into 
    dissipationless and dissipational/extra light components -- 
    are shown for core ellipticals in and around the 
    Virgo cluster$^{\ref{foot:4261}}$. Open circles show the observations, from \citet{jk:profiles}. 
    These are the highest-mass core ellipticals in Virgo
    ($\sim6-8\,\mstar$).
    {\em Upper:} Observed V-band surface brightness profile with our 
    two component best-fit model (solid, dashed, and dotted lines show the 
    total, inner/extra light component, and outer/pre-starburst component). 
    The best-fit outer Sersic index, extra light fraction, and rms residuals about the 
    fit are shown.
    {\em  Lower:} Colored lines show the corresponding surface brightness 
    profiles from the three simulations in our library which correspond 
    most closely to the observed system. Dashed line shows the 
    profile of the starburst light in the best-matching simulation. 
    The range of outer Sersic indices in the simulations (i.e.\ across sightlines for 
    these objects) and range of starburst mass fractions which match the 
    observed profile are shown, with the rms residuals of the observations about the 
    best-fit simulation$^{\ref{foot:explainfits}}$. 
    {\em Bottom:} Observed disky/boxy-ness ($a_{4}$) and ellipticity profiles, 
    with the median (solid) and $25-75\%$ range (shaded) corresponding profile 
    from the best-fitting simulations above. Note that these are not fitted for in any sense. 
    Figures~\ref{fig:jk2}-\ref{fig:jk5}
    show the other core ellipticals in the sample, ranked from most to least 
    massive$^{\ref{foot:loglog}}$.
    \label{fig:jk1}}
\end{figure*}
\begin{figure*}
    \centering
    \plotone{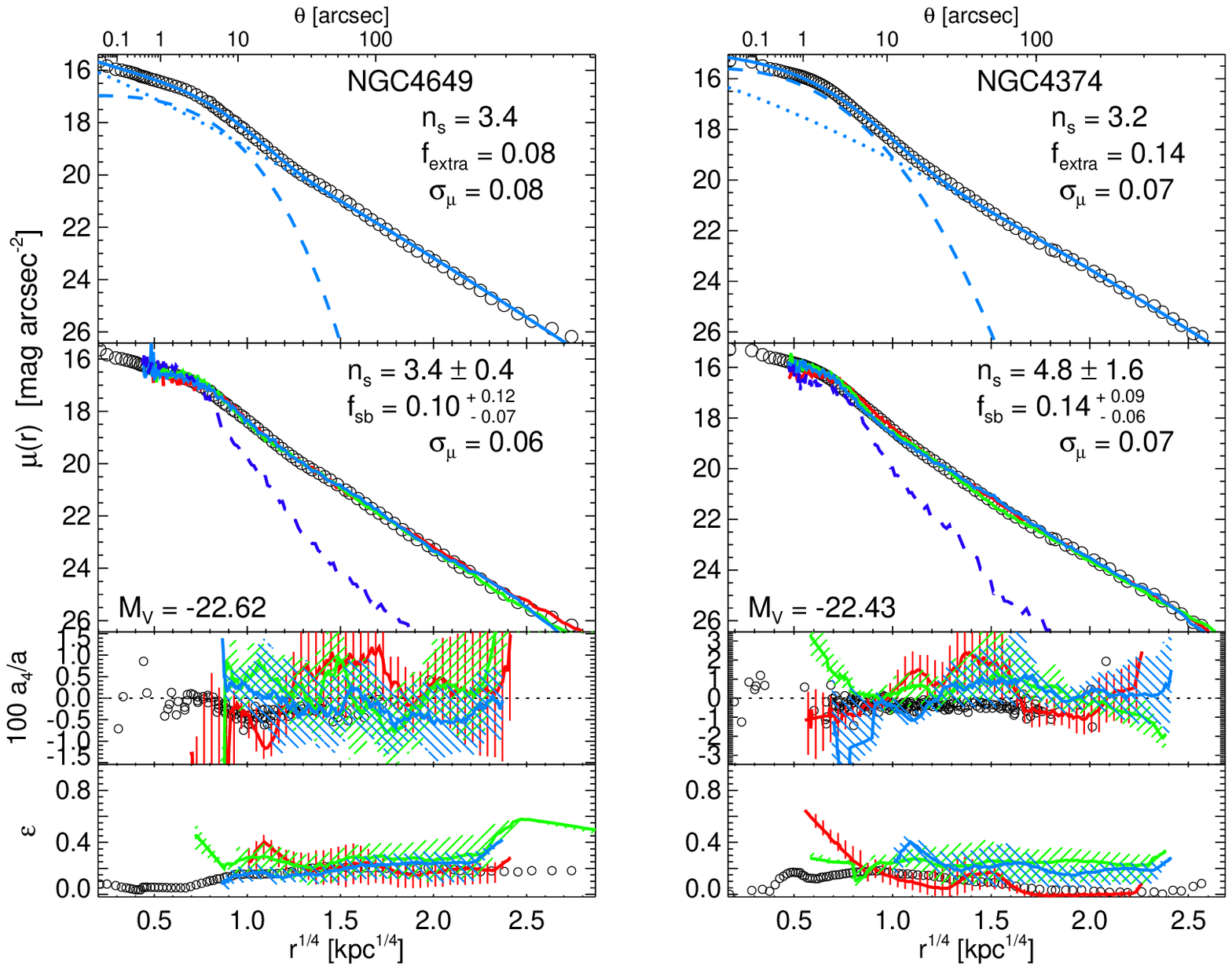}
    \caption{The next most massive core ellipticals ($\sim5-6\,\mstar$). 
    \label{fig:jk2}}
\end{figure*}
\begin{figure*}
    \centering
    \plotone{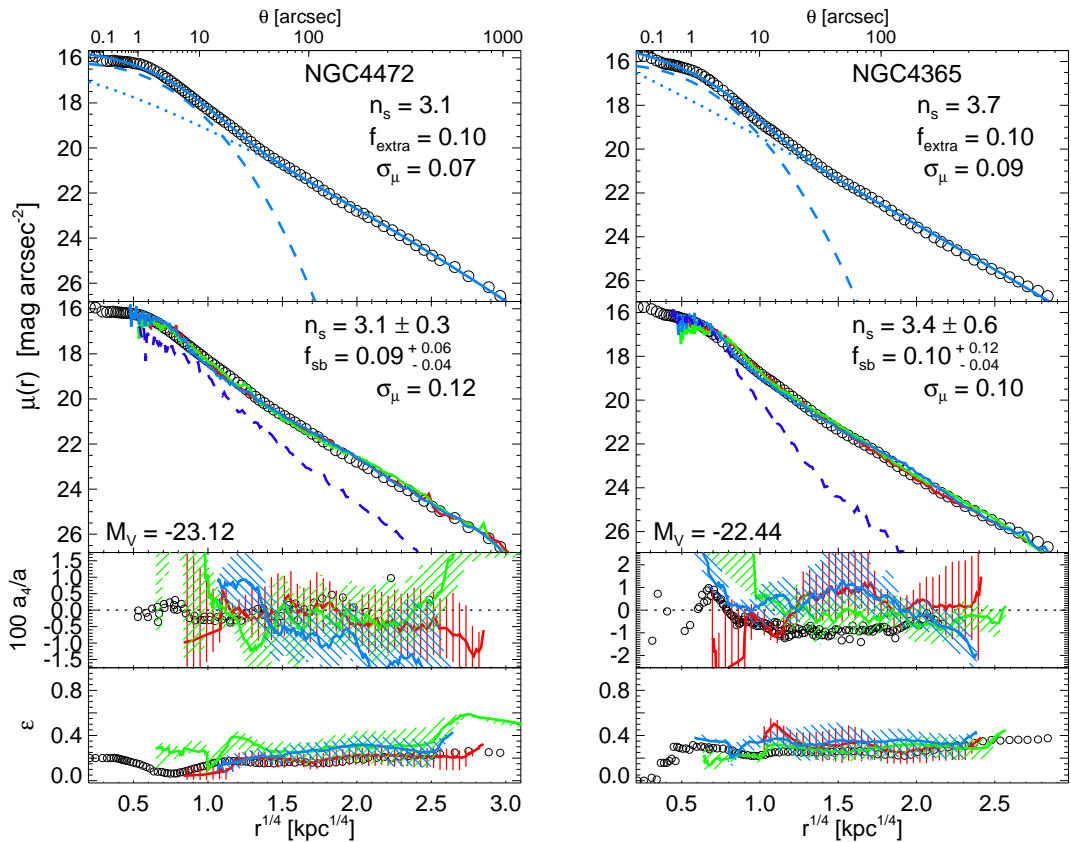}
    \caption{Similar massive core ellipticals ($\sim4-5\,\mstar$). 
    \label{fig:jk3}}
\end{figure*}
\begin{figure*}
    \centering
    \plotone{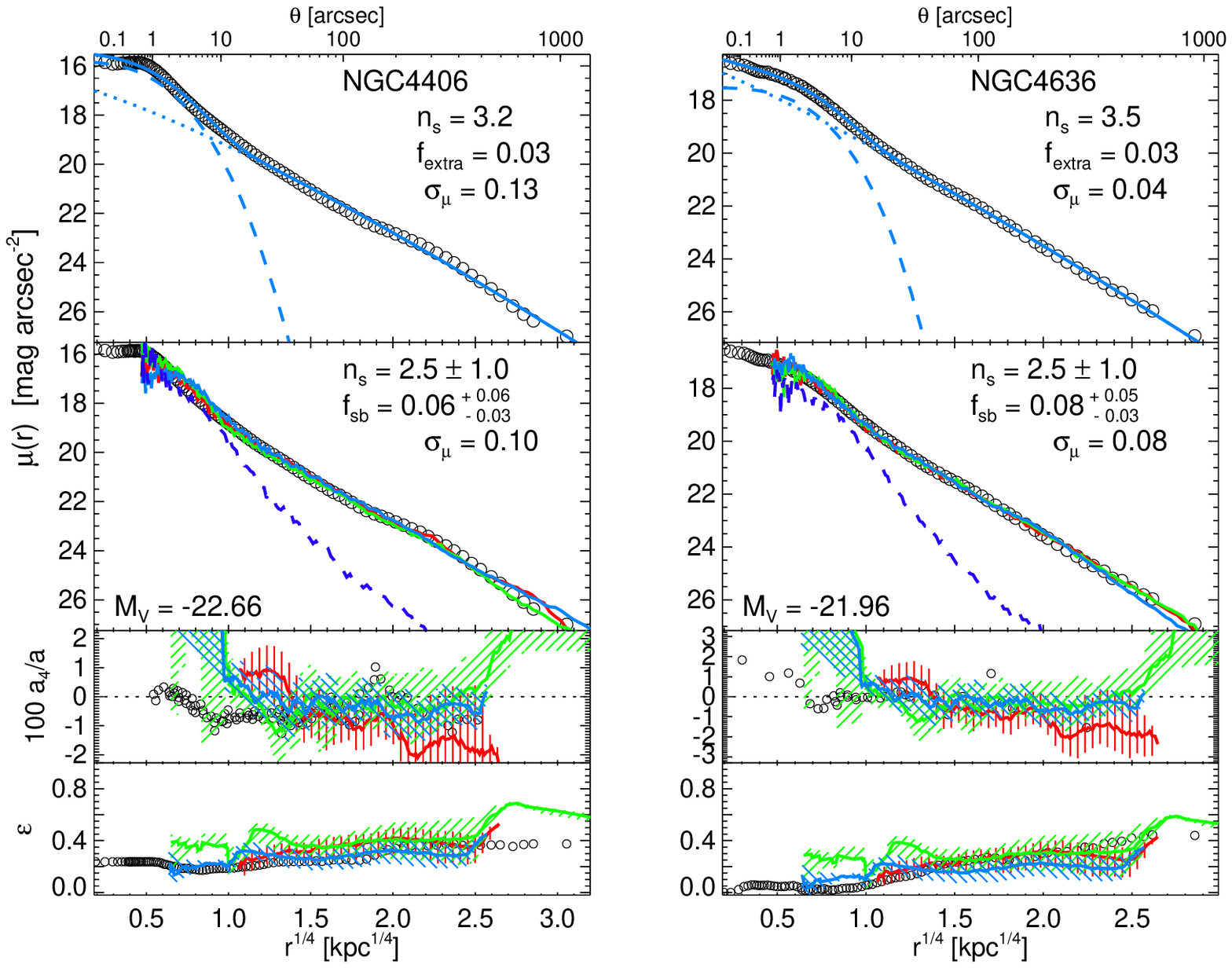}
    \caption{The next most massive core ellipticals ($\sim2-3\,\mstar$). 
    \label{fig:jk4}}
\end{figure*}
\begin{figure*}
    \centering
    \plotone{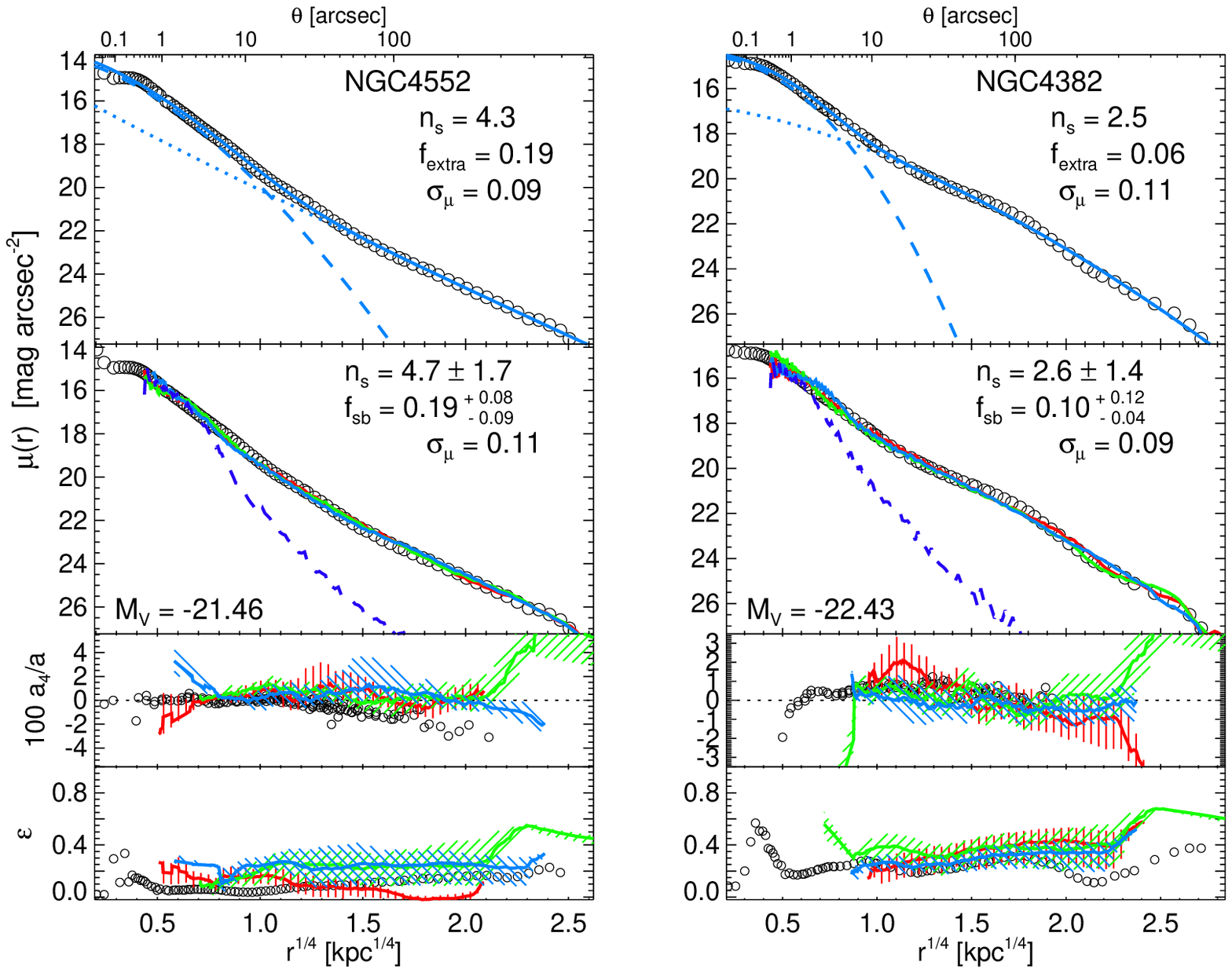}
    \caption{The least massive core ellipticals in the \citet{jk:profiles} 
    Virgo sample ($\sim1-2\,\mstar$). 
    \label{fig:jk5}}
\end{figure*}
\breaker

Together with the arguments in \S~\ref{sec:profile.evol}, 
this gives us reasonable confidence that we can 
extend our analysis to the compilation of observed ellipticals 
described in \S~\ref{sec:data}. 
Figures~\ref{fig:jk1}-\ref{fig:jk5} show surface brightness profiles
of objects in the Virgo core elliptical sample of \citet{jk:profiles}, in
order of most massive to least massive\footnote{\label{foot:4261}NGC 4261 is technically 
in the Virgo W cloud, but we 
show it because it is included in the Virgo samples 
from \citet{ferrarese:profiles} and \citet{jk:profiles}.}.
For each object, we plot the
surface brightness profile with the best-fit two component model, and
the corresponding fitted outer Sersic index and extra light fraction.
We refer to \paperone\ for the same comparisons
with local observed gas-rich merger remnants from \citet{rj:profiles}, 
and \papertwo\ for the comparisons with the cusp ellipticals in 
Virgo from \citet{jk:profiles}.
A complete list of fit parameters and compiled galaxy properties 
is included in Table~\ref{tbl:core.fits}.

We also compare each observed system with our library of 
re-merger simulations,
in a non-parametric fashion. We do this by allowing the normalization
of the simulated light profile to vary (within $\pm0.5$\,dex), and
quantifying the $\chi^{2}$ (variance of the observed points with
respect to the simulated light curve at
$>1$ gravitational softening length) of each simulation. We allow the
normalization to vary because we have a finite number of simulations
and therefore do not sample a continuum in e.g.\ total brightness, but
instead discretely sample at factor $\sim2$ intervals (we do not allow
the simulated profiles to vary by more than this amount, to avoid an
unphysical match to a simulation with very different total mass). We
do not allow any other parameters to vary -- i.e.\ we allow limited 
rescaling in the surface brightness of the simulated galaxies, but 
{\em not} their radii or other properties. 
Despite the allowed surface brightness rescaling, 
the best-fit simulations almost
always have similar total luminosities to the observed system, because
they must have a similar effective radius in order to be a good
match. Considering $\sim100$ sightlines to each of our simulations
(although, as noted in \paperone, the observed surface brightness
profile varies by a small amount sightline-to-sightline), we find the
best fit to each observed system.

We show in Figures~\ref{fig:jk1}-\ref{fig:jk5} the three simulations
which most closely match the observed light profile. For the best-fit
simulation, we also show the profile of the stars formed in the original,
central, merger-driven starburst, as described in
\S~\ref{sec:intro}. We show in the figures the outer Sersic indices
fitted to these simulations, along with the typical range both across
sightlines and across the best-fitting simulations (which together
give some rough approximation to the range of $n_{s}$ which might be
observed for these galaxies along different sightlines). We also show the
best-fit starburst mass fraction, along with the range across the
best-fitting simulations (described below), and the rms residuals of the
observed points with respect to the best fit\footnote{\label{foot:explainfits}
The values shown 
in Figures~\ref{fig:jk1}-\ref{fig:jk5} are based 
on comparison only to the profiles shown, from \citet{jk:profiles}. In 
Table~\ref{tbl:core.fits}, the values represent the results from all available 
data sets, including multiple different observations of the systems shown here, 
and so can be slightly different (however the differences are generally small).}.
In nearly every case, we easily 
find simulations which provide an excellent match to the observed
profiles, with variance $\dmu$ often less than even a
multi-component parameterized fit. The fits are good over the entire 
dynamic range from the largest observed radii ($\sim100\,$kpc) 
down to our resolution limits ($\sim50$\,pc)\footnote{\label{foot:loglog}
The dynamic range of the fits is somewhat difficult to discern in 
Figures~\ref{fig:jk1}-\ref{fig:jk5} owing to the plotting versus 
$r^{1/4}$; we therefore reproduce these figures plotting $\mu$ versus $r$ in 
Appendix~\ref{sec:appendix:jk}.}. 

In addition, for these simulations we show the isophotal shape and
ellipticity as a function of major axis radius, compared to that
observed.  Note that we do {\em not} fit these quantities, only the
surface brightness profile.  We show, for each simulation, the range
across sightlines in these quantities -- it is clear that these depend
much more strongly on sightline than the surface brightness profile
(this is primarily why we do not fit these quantities).  In every
case, there is a significant fraction of sightlines with shape and
ellipticity profiles roughly consistent with those observed, but the
simulations highlight the range of profile shapes for similar
spheroids to those observed.

Our simulations are also consistent with the observed kinematic 
($V$, $\sigma$) profiles of these systems, but we do not show an 
explicit comparison for individual objects because sightline-to-sightline 
variations are sufficiently large (see \S~\ref{sec:kinematics}) as to 
yield only weak constraints. For more robust quantities, such as e.g.\ the 
weighted rotation $\lambda_{R}$ defined in \citet{emsellem:sauron.rotation}, 
we find that our re-merger remnants yield good agreement with the 
(nearly universal) radial $\lambda_{R}(R)$ profile of ``slow rotators'' 
in the SAURON sample (see their Figure~2). The distribution of 
global rotation properties (e.g.\ $V/\sigma$) and correlations with 
isophotal shapes can be constraining (for a population): we 
consider this in \S~\ref{sec:kinematics}. There is, however, a great 
deal more  kinematic information than just a radial mean 
velocity profile: more detailed constraints for individual objects 
can, in principle, be obtained by comparison of simulations 
and high-resolution two dimensional velocity field maps (with higher-order 
velocity moments such as $h_{3}$ and $h_{4}$). Such a comparison 
is clearly warranted, but is outside the scope of this paper and will be the 
topic of future work \citep{cox:remerger.kinematics}.

\breaker
\begin{figure*}
    \centering
    \scaleup
    \plotter{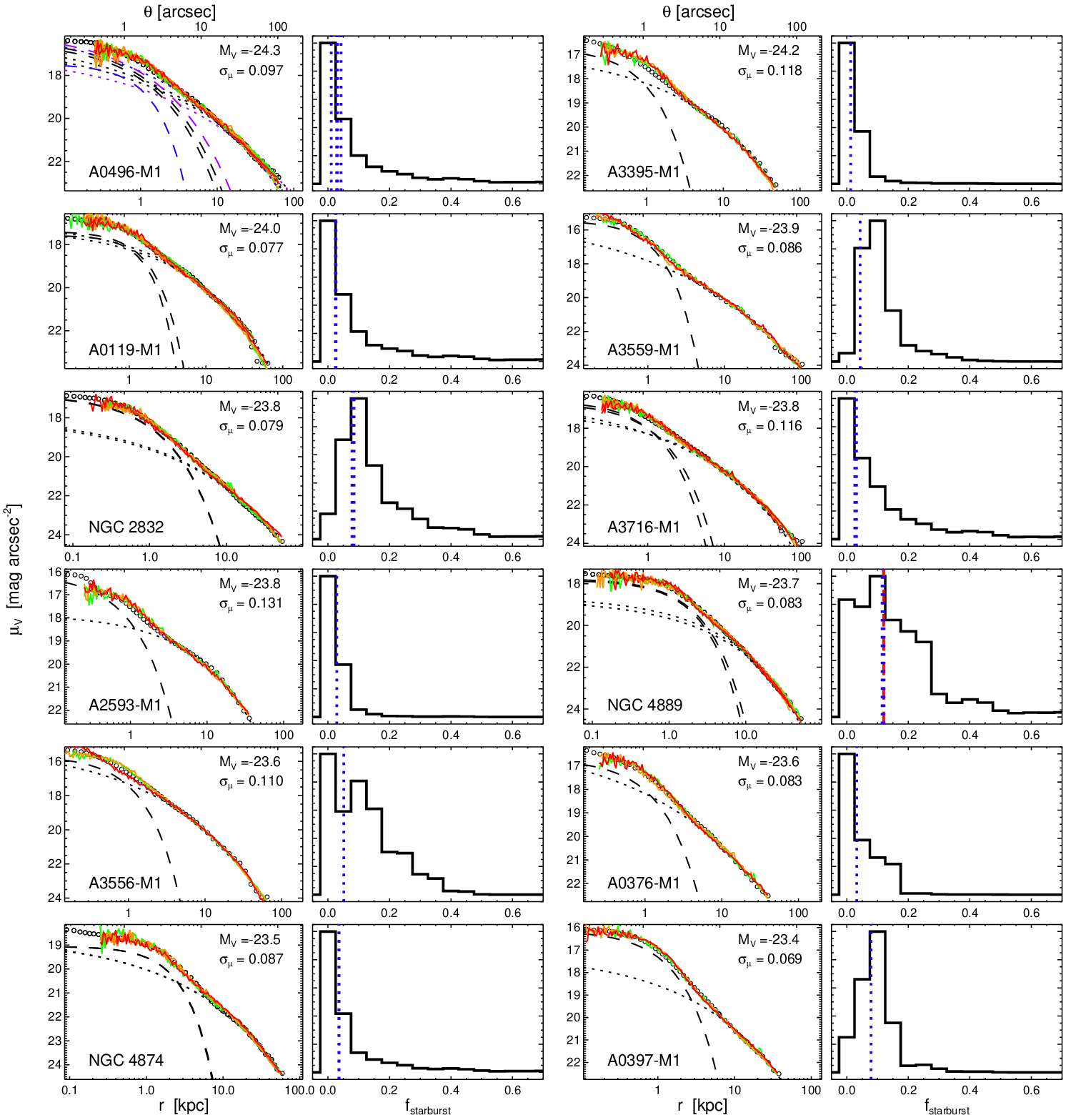}
    \caption{Observed surface brightness profiles of 
    a subset of the confirmed core 
    ellipticals from the sample of \citet{lauer:bimodal.profiles}, 
    with the best-matching two component 
    parameterized fit (dashed and dotted lines)
    and best-fitting simulations (red, orange, and green lines), as in Figure~\ref{fig:jk1}. 
    Where multiple sources of photometry are available, independent fits to each are 
    shown.
    The objects are ranked from brightest to faintest in $V$-band (as shown): 
    these are among the brightest galaxies in the sample; 
    Figures~\ref{fig:lauerpp2}-\ref{fig:lauerpp3} continue to fainter luminosities.
    Profiles are shown over a constant angular scale (top axis; bottom axis shows 
    physical radius in kpc).
    The corresponding ({\em right}) panel for each shows the distribution of physical 
    starburst fractions for the simulations which provide a good fit to the 
    observed profile (as described in the text), with the fitted (parameterized) 
    extra light fraction (blue dotted line; one for each source of 
    photometry) and observed secondary (recent) starburst 
    components (red dashed, where available). Note that our simulation resolution 
    limits do not extend within the central $\sim30-50\,$pc, and our fits 
    are not intended to describe these radii. 
    \label{fig:lauerpp1}}
\end{figure*}

\begin{figure*}
    \centering
    \scaleup
    \plotone{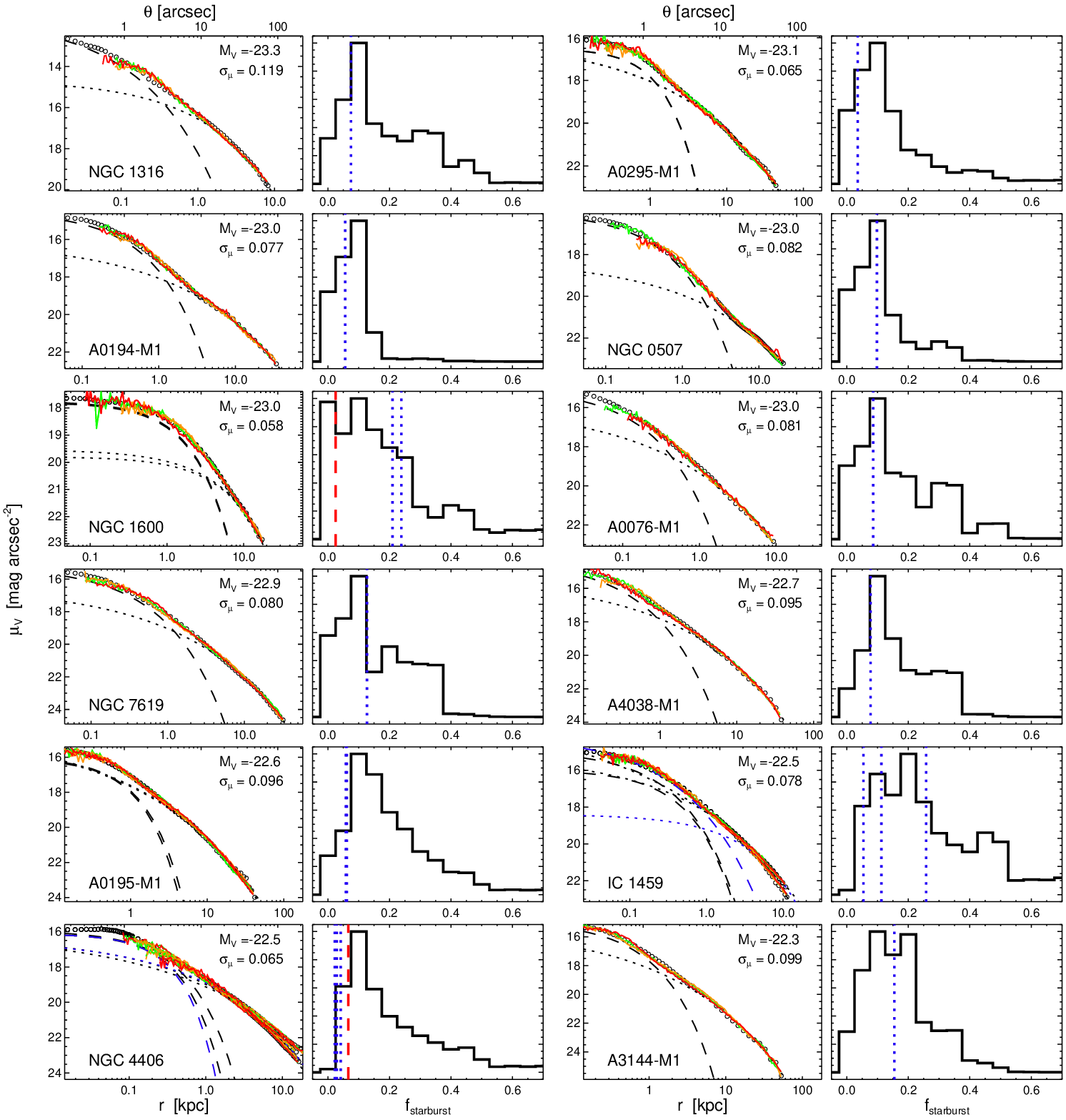}
    \caption{Figure~\ref{fig:lauerpp1}, continued.
    \label{fig:lauerpp2}}
\end{figure*}

\begin{figure*}
    \centering
    \scaleup
    \plotone{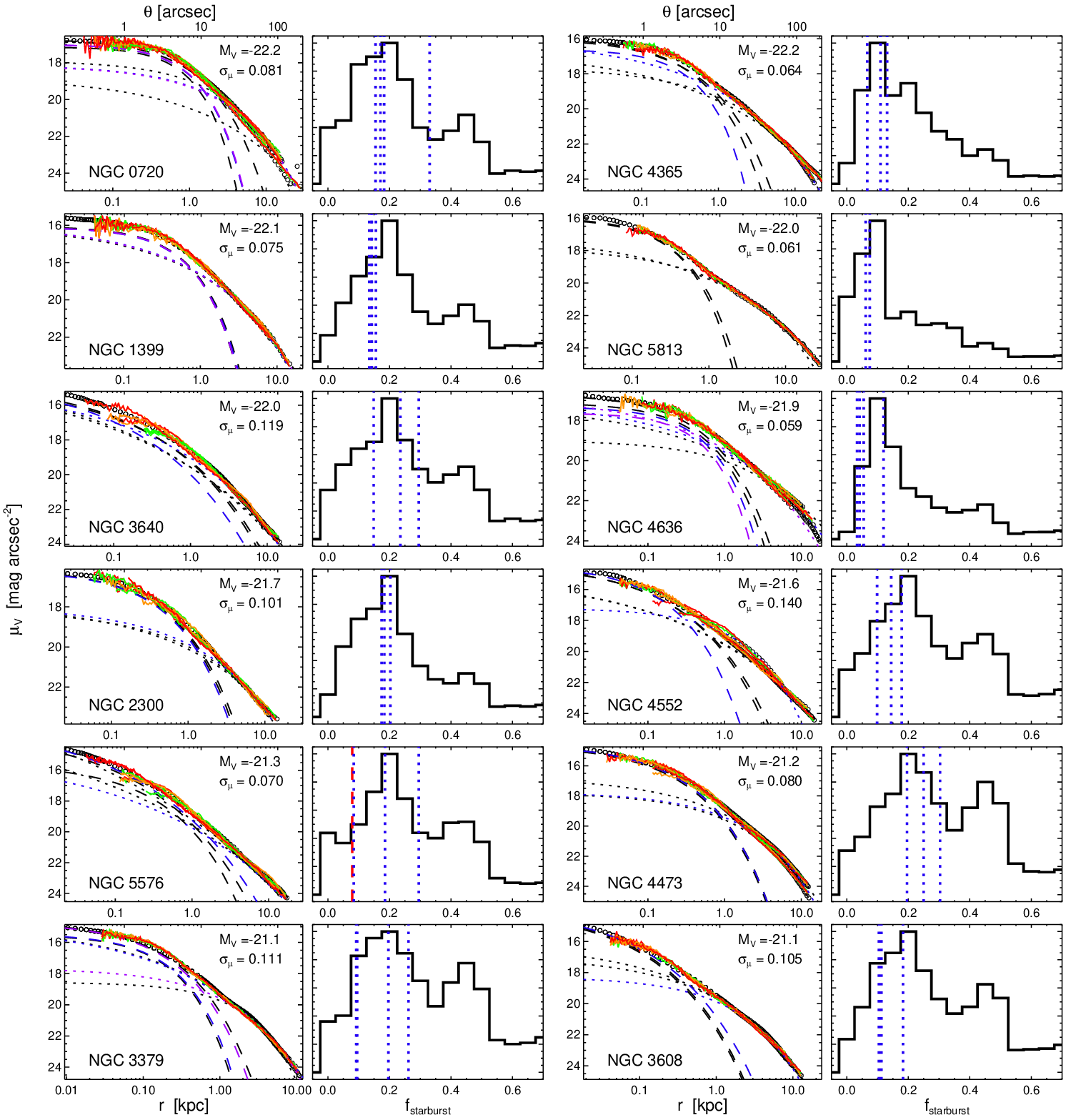}
    \caption{Figure~\ref{fig:lauerpp1}, continued.
    \label{fig:lauerpp3}}
\end{figure*}
\breaker

Figures~\ref{fig:lauerpp1}-\ref{fig:lauerpp3} again show the observed and best-fit simulated 
surface brightness profiles, for a subset of confirmed core ellipticals in 
the sample of \citet{lauer:bimodal.profiles}, in order of $V$-band magnitude from 
brightest to faintest. For each simulation in our library, we 
have a $\chi^{2}$ corresponding to its goodness of fit to the observed 
profile, and the genuine physical starburst mass fraction $\fsb$ 
from the original gas-rich merger. We can 
therefore construct a $\chi^{2}$-weighted distribution of $\fsb$ 
for each observed system -- essentially, the probability, across a 
uniform sample of initial conditions, that the observed profile 
was drawn from a simulation with the given starburst mass fraction. 
These are shown, and compared to the fitted extra light fraction for our 
two-component models. 
In general, the fitted extra light fraction corresponds well to 
the characteristic starburst mass fractions in simulations which produce 
similar light profiles.

\subsection{Independent Evidence for the Surviving Dissipational Component}
\label{sec:fitting:ssp.ev}

\begin{figure*}
    \centering
    \plotone{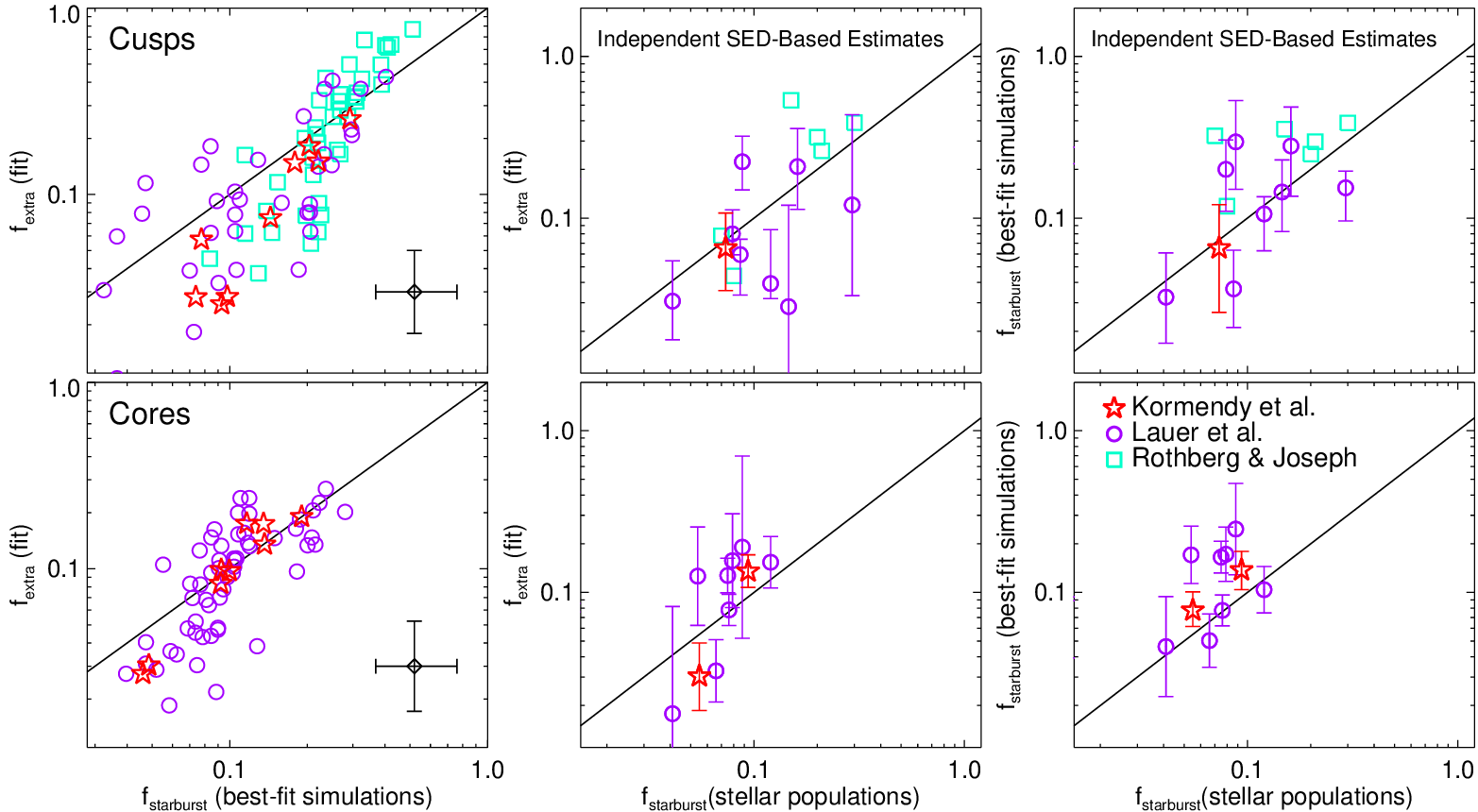}
    \caption{{\em Left:} Comparison of our estimated 
    mass in the fitted extra light component ($f_{\rm extra}$) versus 
    the starburst mass fraction in the best-fitting simulations ($f_{\rm sb}$; 
    as in Figures~\ref{fig:jk1}-\ref{fig:lauerpp1}). We show the results from 
    the core elliptical sample studied here ({\em bottom}; 
    red stars are systems from \citet{jk:profiles}, violet circles from \citet{lauer:bimodal.profiles})
    the observed gas-rich merger remnant and cusp elliptical 
    samples in \paperone\ and \papertwo\ ({\em top}; cyan squares are the 
    merger remnants from \citet{rj:profiles}). 
    The two estimates agree well, with a factor $\sim2-3$ scatter 
    in $f_{\rm extra}(f_{\rm sb})$ (similar to what we expect from our 
    simulations; see Figure~\ref{fig:rem.recovery}). Open point with 
    error bars shows the typical range in $f_{\rm extra}$ and $f_{\rm sb}$ 
    across different sources of photometry and in different wave bands. 
    {\em Center:} Fitted $f_{\rm extra}$ versus 
    independent observational estimates of 
    the mass fraction formed in a more recent starburst/star formation event, 
    from two-component stellar population model fits to the observed SEDs 
     \citep{schweizer96,titus:ssp.decomp,schweizer:7252,
     schweizer:ngc34.disk,reichardt:ssp.decomp,michard:ssp.decomp}. 
    {\em Right:} Same, but comparing the stellar population estimates 
    to $f_{\rm sb}$ from the best-fitting simulations. 
    More observations are needed to test our 
    estimates, but the stellar population data independently 
    suggest that our decompositions 
    are reasonable. 
    \label{fig:extra.vs.sb}}
\end{figure*}

\begin{figure*}
    \centering
    \scaleup
    \plotone{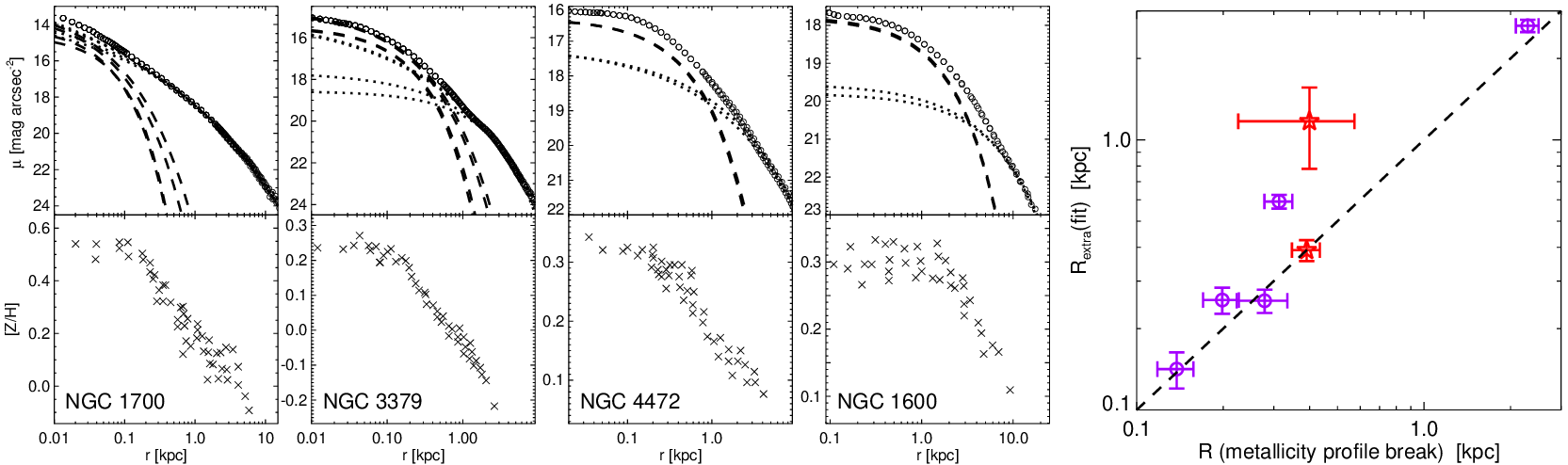}
    \caption{Independent tests of our two-component decompositions with 
    observed stellar population gradients in core ellipticals.    
    {\em Left:} Comparison of the surface brightness profile ({\em top}; points) 
    and two-component decomposition (dashed and dotted lines show the 
    fitted inner \&\ outer components for each source of photometry, 
    as Figure~\ref{fig:lauerpp1}) with the observed metallicity profile ({\em bottom}; 
    as discussed in text this is the most robust indicator of extra light)    
    for the cases observed in \citet{sanchezblazquez:ssp.gradients}. The comparison agrees with our 
    predictions in \papertwo\ and \S~\ref{sec:gradients}: the metallicity gradients are strong 
    at the radii of the transition from the outer (dissipationless) stellar populations 
    to the inner (dissipational, metal-enriched) population, and begin to flatten 
    within the radius of the inner component (since it is a similar population). 
    In each of these cases, we see a supporting (albeit weaker and more sensitive 
    to age effects, see \papertwo) trend towards younger stellar populations 
    at the same radii. 
    {\em Right:} Summary of these comparisons: radius 
    interior to which the fitted inner component dominates the light profile 
    in observed systems (points as Figure~\ref{fig:extra.vs.sb}), 
    versus the radius where the metallicity profile flattens below a power-law 
    inwards extrapolation larger radii (fitting the profiles to an outer power law 
    with an inner break to weakly changing ${\rm [Z/H]}$). 
    Metallicity profiles are from \citet{sanchezblazquez:ssp.gradients} and 
    \citet{mcdermid:sauron.profiles}. 
    The stellar population gradients independently support our two-component 
    decompositions, showing a more metal-rich, younger stellar population in the 
    center, with a characteristic radius that closely tracks our fitted radii for the 
    extra light. More observations are needed, 
    but with Figure~\ref{fig:extra.vs.sb} this lends confidence to our 
    two-component decompositions in core ellipticals. 
    \label{fig:z.grad.demo}}
\end{figure*}

Figure~\ref{fig:extra.vs.sb} compares our estimates of the dissipational 
mass fraction in the observed ellipticals: the 
directly fitted extra light fraction $f_{\rm extra}$ 
and inferred starburst mass fraction $f_{\rm sb}$ from the best-fitting simulations. 
For clarity we restrict to a sample where the range in $f_{\rm extra}$ 
from different sources of photometry is less than $\sim20\%$ (we find 
the same results for the entire sample, but with correspondingly larger 
scatter). We compare with the same quantities fit to cusp ellipticals 
and observed gas-rich merger remnants in \papertwo\ and \paperone, respectively. 
Our fitted decomposed extra 
light fraction reliably traces the inferred starburst 
mass fraction, with a factor $\sim2$ scatter similar to that predicted 
from our simulations (Figure~\ref{fig:rem.recovery}). This is 
true even for profiles without an obvious ``break'' or feature indicative 
of a transition to extra light. The starburst 
fraction $f_{\rm sb}$ itself must, in some sense, reflect the cold gas 
mass available in the disks just before the original gas-rich merger 
(and we show in \paperone\ that this is the case) -- in this 
physical sense (insofar as our simulations are reasonable qualitative proxies for 
observed profiles), our fitted $f_{\rm extra}$ and inferred $f_{\rm sb}$ 
are a robust reflection of the dissipational content of the progenitors. 

We can independently test these decompositions with 
other observations of stellar populations, shown in 
Figure~\ref{fig:extra.vs.sb}. Given sufficiently detailed observations 
and stellar population models, observers have estimated the 
mass fraction which formed in a more 
recent, central starburst (as opposed to the more extended quiescent
star formation history) -- completely 
independent of our analysis here -- for several of the observed 
systems \citep{schweizer96,titus:ssp.decomp,schweizer:7252,schweizer:ngc34.disk,
reichardt:ssp.decomp,michard:ssp.decomp}. Unfortunately, the observational 
expense required leaves us with samples of relatively limited 
size (there are only a few objects with sufficiently deep stellar population 
data to allow this comparison). Nevertheless, comparing our 
estimated $f_{\rm extra}$ or $f_{\rm sb}$ with these 
estimates for the mass fraction in the secondary (newly
formed/starburst) stellar populations, we find reasonable agreement. 
They all suggest that our fitted
extra light component is indeed a good proxy for the mass fraction
which was involved in the central, merger-driven starburst.
We find similar good agreement in these cases for both our 
core ellipticals here and the cusp ellipticals considered 
in \papertwo, and the merger remnants considered in \paperone, 
together constituting a sample of 
$\sim25$ objects.

Other independent 
lines of support further suggest that the decomposition is recovering a 
real physical difference in the populations. In Figures~\ref{fig:jk1}-\ref{fig:jk5}, 
there is often a change in isophotal shape and kinematics 
at the radii we associate with the dissipational remnant -- systems become 
slightly diskier, despite the fact that re-mergers will
smooth the isophotal shapes and destroy strong transitions in features 
such as the kinematics that might be present in cusp ellipticals 
at the extra light-outer light transition. 
We show in \S~\ref{sec:kinematics} that the distribution of 
global kinematic properties and isophotal shapes independently 
requires re-mergers with a relatively narrow range in starburst mass 
fractions, in agreement with our estimates from fitting the surface brightness 
profiles. 

Moreover, we show in \paperone\ 
that the extra light is associated with, and indeed drives, stellar 
population gradients in the remnant elliptical. In particular, we argue 
that metallicity gradients are most robust (age gradients are very 
weak once the system is old, and gradients in $\alpha$-enhancement 
are much more sensitive to initial conditions and the exact 
details of e.g.\ the starburst timing relative to the timescale for 
star formation in progenitor disks). We show in \S~\ref{sec:gradients} 
that these are expected to survive re-mergers relatively intact, 
and therefore should provide independent indicators of the 
dissipational versus dissipationless decomposition. 

There are a small number of systems 
in our observed core sample for which detailed stellar population 
gradients have been measured in \citet{sanchezblazquez:ssp.gradients} or 
two-dimensional stellar population maps have been constructed 
in \citet{mcdermid:sauron.profiles}. In almost every case, we see what is 
predicted: a strong rise in metallicity around the radii where 
our fits infer the transition between the dissipational and dissipationless 
components, with the metallicity gradient flattening 
within the radius where the extra light 
dominates the profile. Figure~\ref{fig:z.grad.demo} shows a few illustrative 
examples, comparing the observed metallicity gradients to our 
two-component surface brightness decompositions. 
If we attempt to quantify where the metallicity profile 
flattens (expected inside the radius where the extra light 
dominates, since the predicted gradient is largely driven by the 
transition from an outer, less metal-rich population into the 
more metal-rich dissipational 
population) by fitting it to an external power law with a break 
at some radius $R_{b}$, and compare this radius 
$R_{b}$ to the radius at which the fitted inner/starburst component 
dominates the profile (the outer radius where the 
fitted inner component surface brightness becomes larger than 
that of the outer component, or similarly the effective radius of the 
extra light component), we find a remarkably good correlation. 
In every case, the age gradients (while weak and less uniform, as predicted) 
also confirm that the central component we identify is younger 
than the outer profile, as predicted (the difference is usually 
small, $\sim1\,$Gyr, as expected in merger models with relatively 
early formation -- if the central components were instead formed 
by e.g.\ dissipation of gas from stellar mass loss \citep{ciottiostriker:recycling}, we 
might expect the central components to be much younger). 
More data are needed to test these decompositions and predictions 
over the entire dynamic range of interest, but our preliminary 
comparisons all lend some confidence that we are, in an average 
sense at least, recovering the appropriate physical separation 
between inner radii dominated by the remnants of extra light systems 
originally formed in dissipational starbursts and outer light 
profiles formed by violent relaxation acting on older, 
initially low-phase space 
density (disk) stars.

\breaker
\section{Properties and Scaling Laws of Dissipational Components}
\label{sec:scaling}

\subsection{Amount of Dissipation Versus Mass}
\label{sec:scaling:fsbvsmass}

The preceding analysis indicates that re-mergers do indeed have
a ``dissipational'' component (the surviving remnant of the original 
starburst that formed the progenitors' extra light), 
and that its mass can be estimated
from their surface brightness profiles.  Now, we examine how this
scales with other properties, specifically in reference to how the
extra light\footnote{In what follows, we use the phrase 
``extra light'' for the core ellipticals in 
the physically pertinent sense, to refer to the 
component that traces or reflects the original starburst mass, 
i.e.\ the centrally concentrated contribution to the light profile 
that is the remnant of the original gas-rich merger remnant 
extra light.} content was seen to scale with galaxy properties in cusp
ellipticals in \papertwo. 

\begin{figure*}
    \centering
    \scaleupp
    \plotone{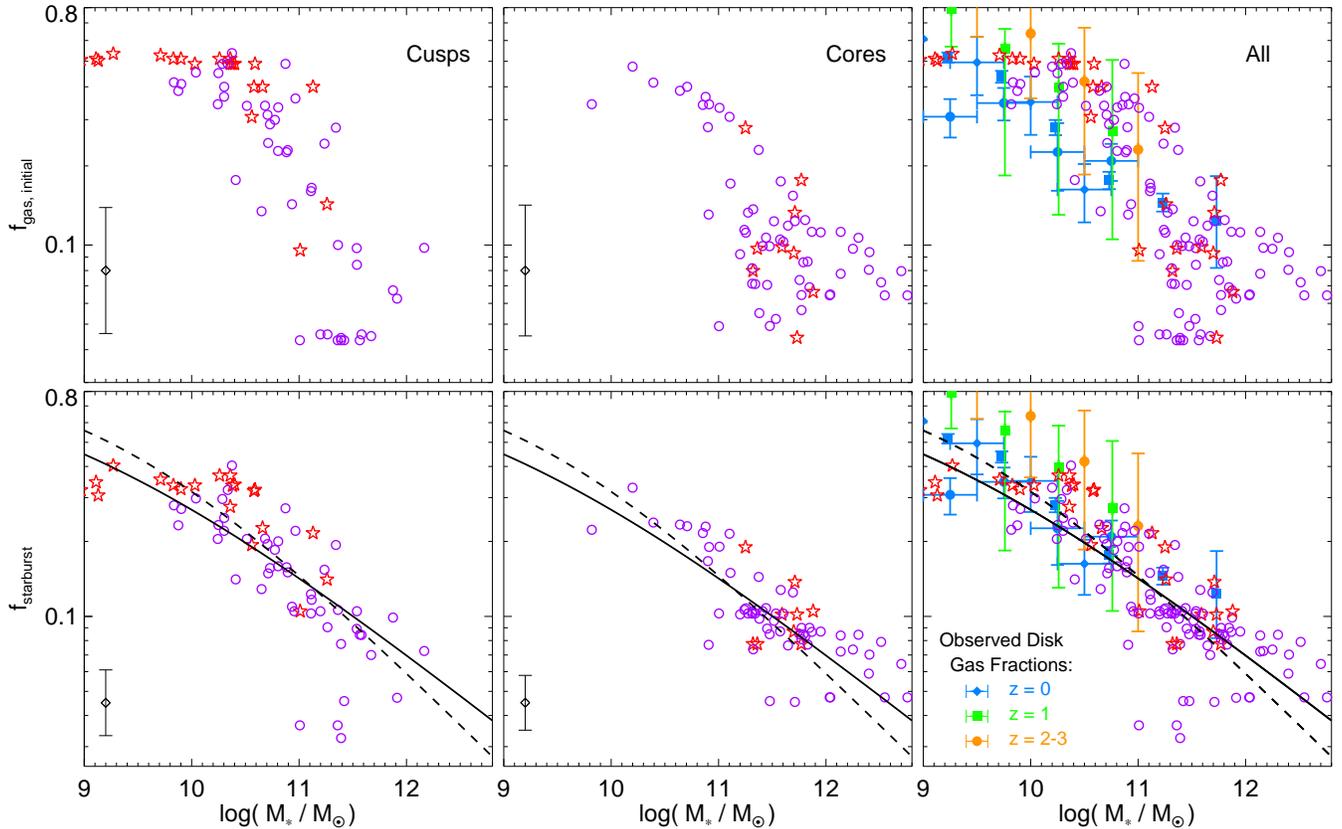}
    \caption{Inferred gas content (dissipational/starburst fraction) of 
    elliptical-producing mergers as a function of stellar mass. 
    Initial gas fraction ({\em top}) and physical final starburst mass 
    fraction ({\em bottom}) corresponding to the best-fit simulations to 
    each observed system in the samples of 
    \citet{lauer:bimodal.profiles} (circles) 
    and \citet{jk:profiles} (stars) are shown, 
    with the typical $25-75\%$ allowed range (error bar).
    We show results separately for cusp ellipticals ({\em left}), core ellipticals 
    ({\em center}), and both together ({\em right}). 
    Dashed (solid) line shows the fit to the data (Equation~\ref{eqn:fgas.m}) 
    in cusp (core) ellipticals. 
    Colored points with error bars indicate the mean (and $\pm1\,\sigma$ 
    range in) disk gas fractions at the same stellar mass, at 
    $z=0$ \citep[][blue diamonds, squares, and circles, respectively]
    {belldejong:tf,kannappan:gfs,mcgaugh:tf}, 
    $z=1$ \citep[][green squares]{shapley:z1.abundances}, and 
    $z=2$ \citep[][orange circles]{erb:lbg.gasmasses}. There is a clear trend of increasing 
    dissipation 
    required to explain elliptical profiles at lower masses 
    (significant at $>8\,\sigma$ in either core or cusp subsamples 
    separately), 
    in good agreement with the observed trend in progenitor disk 
    gas fractions over the redshift range where 
    cusp ellipticals are formed, and with what is invoked to explain  
    the observed densities and fundamental plane correlations of ellipticals 
    \citep[e.g.][]{kormendy:dissipation,hernquist:phasespace}.
    The best-fit trends in cusp and core populations are statistically 
    identical: i.e.\ the dissipational/extra light component is 
    preserved in re-mergers. 
    \label{fig:fgas.needed}}
\end{figure*}

Figure~\ref{fig:fgas.needed} plots the inferred starburst mass fraction 
$f_{\rm sb}$ for the observed systems as a function of stellar mass. In 
the same manner that we have defined a best-fit $f_{\rm sb}$ 
from the best-fit simulations, we can also define a best-fit 
``initial'' gas fraction (roughly the gas fraction $\sim1\,$Gyr before the 
final merger), and show this as well. We emphasize though that 
this is a much less robust quantity (for example, if we place the systems on a longer period 
orbit with the same initial gas fraction, then by the time they merge, more gas will 
have been consumed in the disks and the results will look like a shorter 
period orbit, lower 
``initial'' gas fraction simulation). In either case there is 
a clear trend of increasing dissipation 
(increasing fractional mass required in a dissipational starburst component) 
at lower masses. We compare this with the behavior found for 
observed cusp ellipticals in \papertwo. 
The significance of the correlation is unambiguous ($>8\,\sigma$, 
in each of the cusp and core samples). 
The two trends are 
statistically indistinguishable: fitting 
them as a function of stellar mass with the form 
\begin{equation}
\langle f_{\rm starburst} \rangle \approx 
{\Bigl[}1+{\Bigl(}\frac{M_{\ast}}{M_{0}}{\Bigr)}^{\alpha}{\Bigr]}^{-1}, 
\label{eqn:fgas.m}
\end{equation}
we obtain $(M_{0},\ \alpha)=(10^{9.2\pm0.2}\,\msun,\ 0.43\pm0.04)$ for the cusp sample and 
$(10^{8.8\pm0.3}\,\msun,\ 0.35\pm0.04)$ for the core sample, 
with roughly a constant factor $\sim2$ intrinsic scatter at each mass 
in both cases. 
It is worth noting that if identical systems begin on the cusp correlation and re-merge, they will 
conserve $f_{\rm starburst}$ but double $M_{\ast}$. To the extent that there is not a 
large offset between the two correlations, then, this suggests that at least the typical core 
ellipticals have not undergone a large number of re-mergers. A single dry re-merger, 
especially if it is of a more typical $\sim$1:3 mass ratio, is however consistent with 
what we see, within the scatter. 

In \papertwo\ we noted that the inferred gas fractions correspond well to 
observed disk galaxy gas fractions as a function of mass, from $z=0-2$.
We show these observations as well, and note that they are sufficient to explain the 
trend and scatter in the extra light fractions of both cusp and core ellipticals. 
In other words, the distribution in progenitor gas fractions implied
by the elliptical surface brightness profiles is, as a function of
mass, exactly what would be predicted if one assumes that the original 
progenitors were spirals, and that most of the systems were first formed by
a major merger somewhere between a redshift of $z\sim0-3$. 
In fact,
this is exactly what is inferred for the formation times of the
ellipticals in our sample and those of 
similar mass ($\lesssim$ a few $L_{\ast}$) 
from both observations of the early-type or red galaxy
mass functions \citep{bundy:mtrans,borch:mfs,fontana:highz.mfs,
hopkins:red.galaxies,hopkins:clustering,hopkins:groups.ell},
from direct stellar population synthesis studies
\citep{trager:ages,thomas05:ages, gallazzi:ssps}, and 
by association of elliptical galaxy formation with the 
triggering of quasar activity
\citep[e.g.][]{hopkins:qso.all,hopkins:bol.qlf,
hopkins:groups.qso}. 

There may be 
an interesting suggestion that the extremely massive, BCG tail of the 
core population (for which the various caveats described in \S~\ref{sec:profile.evol} apply) 
might asymptote to $\sim$ constant values of $f_{\rm sb}\sim0.05-0.1$ at 
$M_{\ast}\gtrsim 10^{12}\,M_{\ast}$. This would be expected if these systems 
all ``began'' at lower masses (where this gas fraction 
is typical of disks) and moved up
by a large number of dry re-mergers. Unfortunately, the data here are limited, 
but we stress that these systems do not significantly alter our comparison, 
and are not the systems we intend our modeling (applicable to those core 
ellipticals with a moderate re-merger history) to describe. 

\begin{figure}
    \centering
    \scaleup
    \plotter{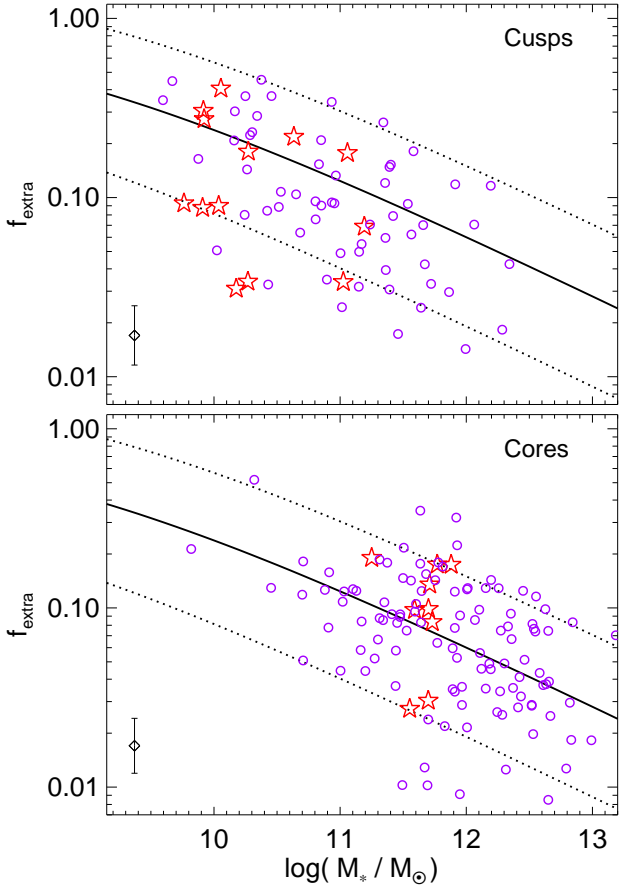}
    \caption{As Figure~\ref{fig:fgas.needed}, but showing 
    our empirically fitted $f_{\rm extra}$ as a function of 
    stellar mass. The trend of increasing dissipation 
    at lower masses is still clear and is consistent with 
    that in Figure~\ref{fig:fgas.needed}, but with an 
    expected extra factor $\sim2-3$ scatter 
    from the scatter in our purely empirical estimator. 
    Solid line shows the best-fit from Figure~\ref{fig:fgas.needed}, 
    dotted lines show the $\pm1\,\sigma$ scatter expected 
    based on the scatter in $f_{\rm extra}(f_{\rm sb})$ (see Figure~\ref{fig:extra.vs.sb}). 
    Fitting to just these data yield trends statistically consistent with the solid line, 
    ruling out no dependence of dissipation on mass at $>8\,\sigma$ confidence. 
    \label{fig:mass.vs.fgas}}
\end{figure}

The inference of $f_{\rm sb}$ requires some comparison with our simulations, 
but $f_{\rm extra}$ is a purely empirical quantity. We therefore 
repeat this exercise with the empirically fitted extra light component
$f_{\rm extra}$, and show the results in Figure~\ref{fig:mass.vs.fgas}. 
The trend seen in $f_{\rm extra}(M_{\ast})$ is completely consistent with 
that in $f_{\rm sb}$, but with a scatter larger by a factor $\sim2$ (exactly 
what we expect, based on the predicted and observed scatter 
in $f_{\rm extra}(f_{\rm sb})$). 
Moreover the trend we find in the core ellipticals is consistent with that 
in the cusp elliptical sample from \papertwo.
Considering just the data in Figure~\ref{fig:mass.vs.fgas}, 
even given its increased scatter, the trend of decreasing extra light fraction 
with mass is significant at $>6\,\sigma$ ($>8\,\sigma$, if we 
include both the cusp and core ellipticals). We have experimented with 
alternative, non-parametric (albeit less accurate) 
estimators based on e.g.\ the concentration 
indices or stellar populations 
of our simulations and observed systems, and obtain a similar answer. 
In short, even without reference to our simulations, 
however we derive an estimate of the
dissipational component, it is difficult to escape the conclusion that
it is more prominent in lower-mass ellipticals, even in core ellipticals 
that may have experienced re-mergers. 
The mass range
covered by the two samples is somewhat different, but where there is
overlap they are similar, and at the extremes the two populations appear to
be continuous extensions of the same trends.

\begin{figure*}
    \centering
    \plotone{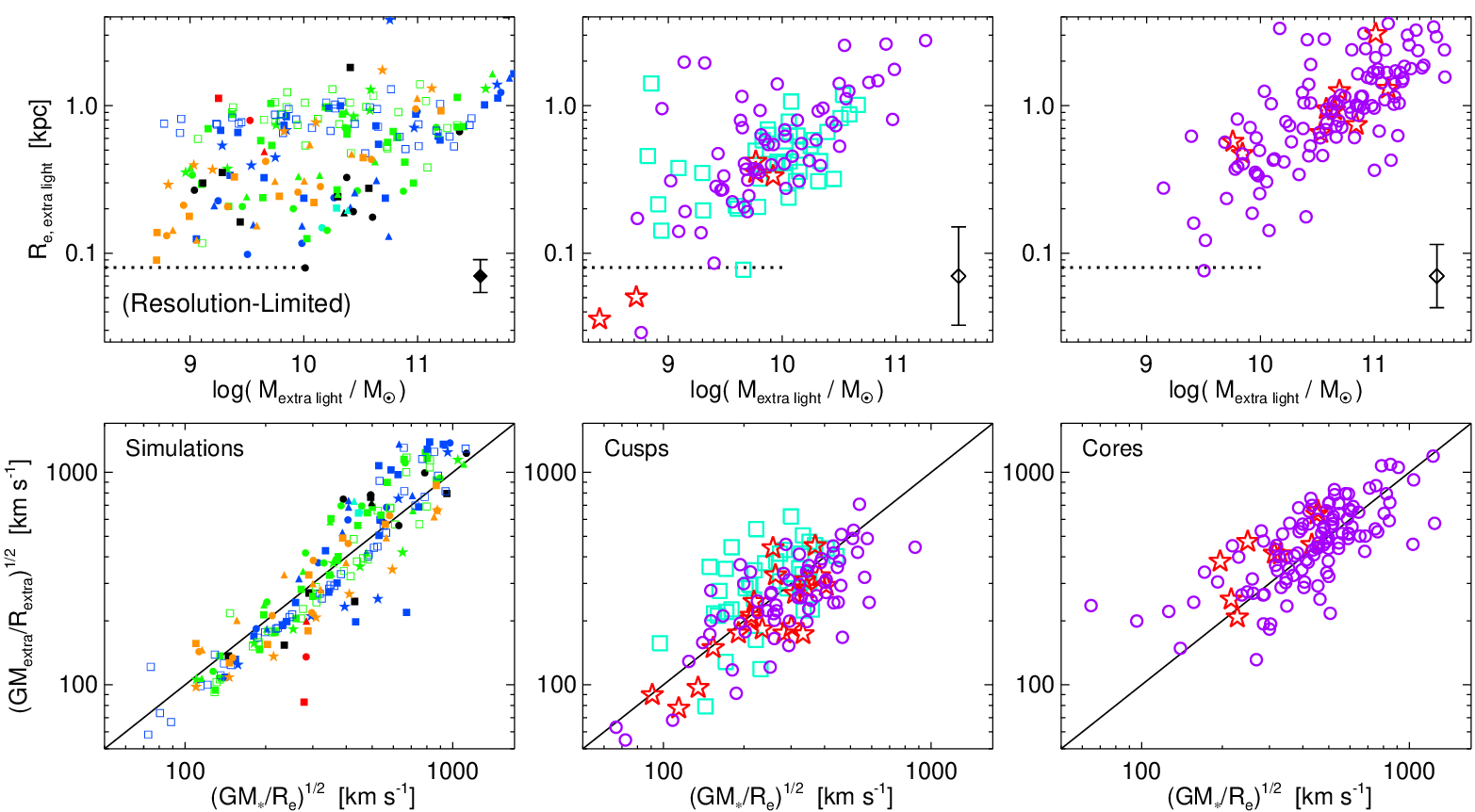}
    \caption{{\em Top:} Effective radius of the extra light component 
    ({\em not} equivalent to the radius where it breaks from the outer Sersic fit) 
    as a function of extra light mass
    (points as in Figure~\ref{fig:extra.vs.sb}). Simulations ({\em left}) and observed 
    cusp ({\em middle}) and core ({\em right}) ellipticals  
    are similar, especially 
    if we restrict to simulations with initial gas fractions $\sim0.2-0.4$. Resolution limits 
    (see \papertwo) prevent us from simulating systems with 
    $R_{\rm extra}\ll 100\,$pc, but this is only important for the few very lowest-mass 
    ellipticals ($L\lesssim 0.01\,L_{\ast}$).
    Filled diamond is typical sightline-to-sightline variance in the simulations, open 
    diamond the source-to-source (or band-to-band) scatter in observed profile fits. 
    {\em Bottom:} Effective velocity dispersion of the extra light component vs.\ that for the 
    whole galaxy. Solid line shows $(G\,M_{\rm extra}/R_{\rm extra})=(G\,M_{\ast}/R_{e})$ -- the 
    extra light collapses to the point where it is self-gravitating.
    The observed systems follow a trend which agrees 
    well with the simulations.
    \label{fig:sizes}}
\end{figure*}

\subsection{Sizes of the Extra Light/Dissipational Region}
\label{sec:scaling:xlsize}

The inferred dissipational components in simulations, cusp ellipticals, and core 
ellipticals also appear to follow a similar size-mass
relation (especially if we restrict to 
systems with similar $f_{\rm sb}$), shown in Figure~\ref{fig:sizes}. 
In \paperone\ we show that the size-mass relation is driven by the
condition that the gas collapsing into the central regions in the
final starburst becomes self-gravitating: $(G\,M_{\rm
extra}/R_{\rm extra})\sim(G\,M_{\ast}/R_{e})$. In Figure~\ref{fig:sizes} 
we show that the observed core ellipticals also follow this
constraint, with small scatter and dynamic range similar 
to that in our simulated mergers. 
This is expected even 
if these systems have experienced substantial re-merging, provided that
re-mergers 
puff up the system uniformly.

\begin{figure}
    \centering
    \plotter{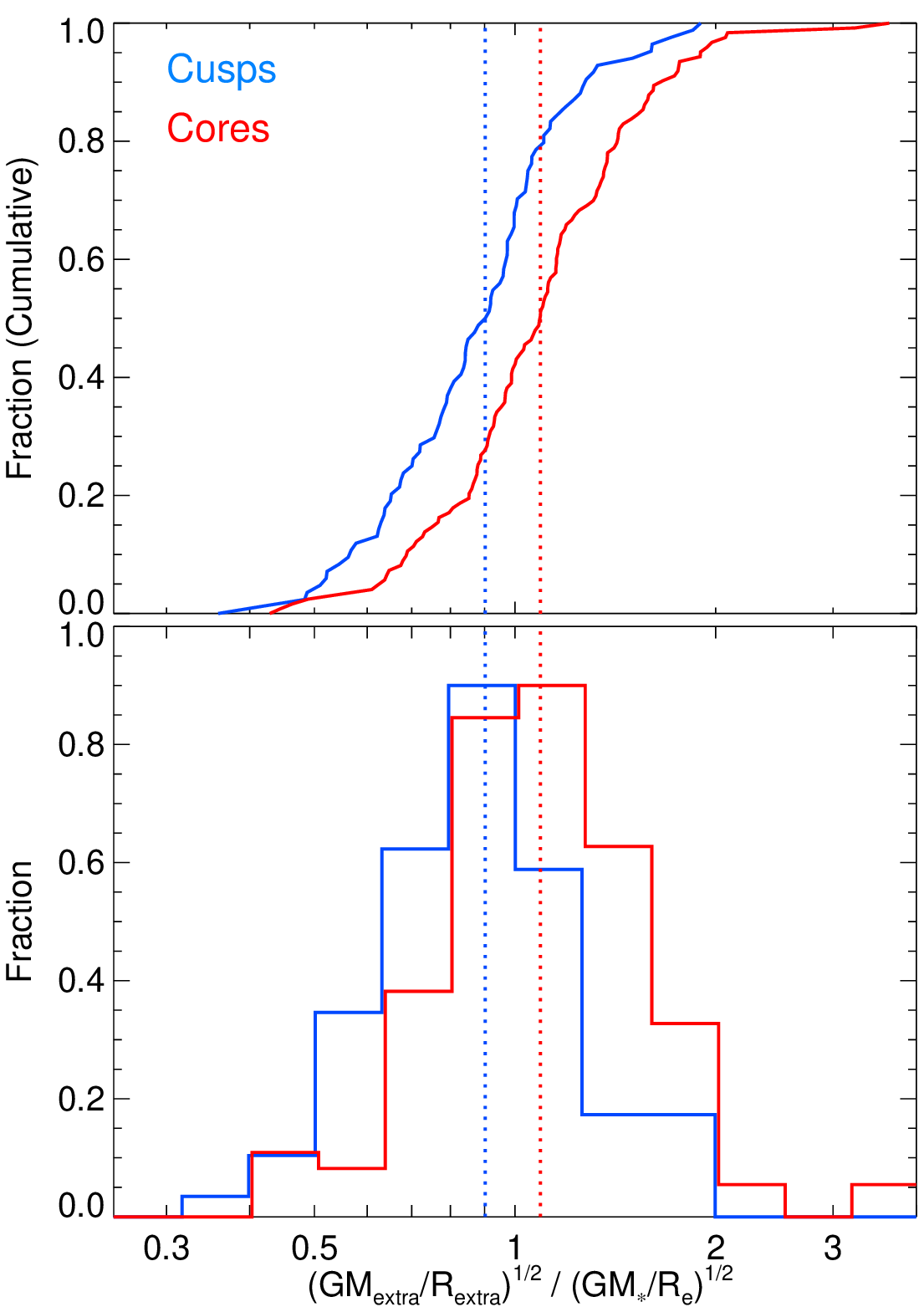}
    \caption{Cumulative ({\em top}) and differential ({\em bottom}) 
    distribution of the ratio of effective velocity dispersion of the inner light 
    component to that of the outer light component 
    ($(G\,M_{\rm extra}/R_{\rm extra})^{1/2}/(G\,M_{\ast}/R_{e})^{1/2}$), 
    for cusp ellipticals (blue) and core ellipticals (red). Dotted line 
    shows the median for each. Typical values are similar, but slightly 
    higher for core ellipticals (the difference in mean/median is significant 
    at $>3\,\sigma$), suggesting that the inner component is relatively less 
    puffed up by re-mergers than the outer component. 
    \label{fig:size.ratios}}
\end{figure}

To test, this, Figure~\ref{fig:size.ratios} plots the distribution in
the ratio of the effective velocity dispersion/self-gravity of the inner component
to that of the outer component, for the cusp and core elliptical
samples. Interestingly,
there is small offset, in the sense that the effective dispersions of
the core ellipticals are slightly higher than those of cusp
ellipticals. This is true even if we restrict the analysis to a narrow
mass bin -- i.e.\ the extra light in core ellipticals is (on average)
slightly more concentrated relative to the outer light component,
compared to cusp ellipticals.

It is unlikely that this is caused by scouring or whatever 
process might directly give rise to the core on $\lesssim50\,$pc scales, 
since the effect is seen even in the high-mass objects where 
the effective radius of the extra light component is $\gtrsim1\,$kpc. 
However, it is well established in numerical simulations that, when merging 
collisionless systems, angular momentum and orbital energy are preferentially transferred 
to less bound material, expanding their orbits while allowing the central regions to 
remain tightly bound \citep[see, e.g.][]{hernquist:bulgeless.mergers,hernquist:bulge.mergers}.
In practice, most of this less bound material is part of the outer halo -- 
so the gradient in terms of energy input and expansion of material across the
baryonic galaxy is weak (i.e.\ the profiles are stretched by a roughly constant 
factor in Figure~\ref{fig:rf.of.ri}). Nevertheless, a small effect 
of this nature would be expected in collisionless re-mergers. The observed 
offset is in fact weak, a factor $\sim1.5$ in $R_{\rm extra}/R_{\rm outer}$ (in the 
most extreme case of a single major 
re-merger where the inner profile was not affected by the merger and 
the outer profile doubled in radius, we would expect a factor of two offset), 
and more data would be needed to demonstrate that it is robust. However, 
if real, it provides further support to the idea that these have re-merged, but 
not destroyed their extra light content.

\begin{figure*}
    \centering
    \scaleup
    \plotone{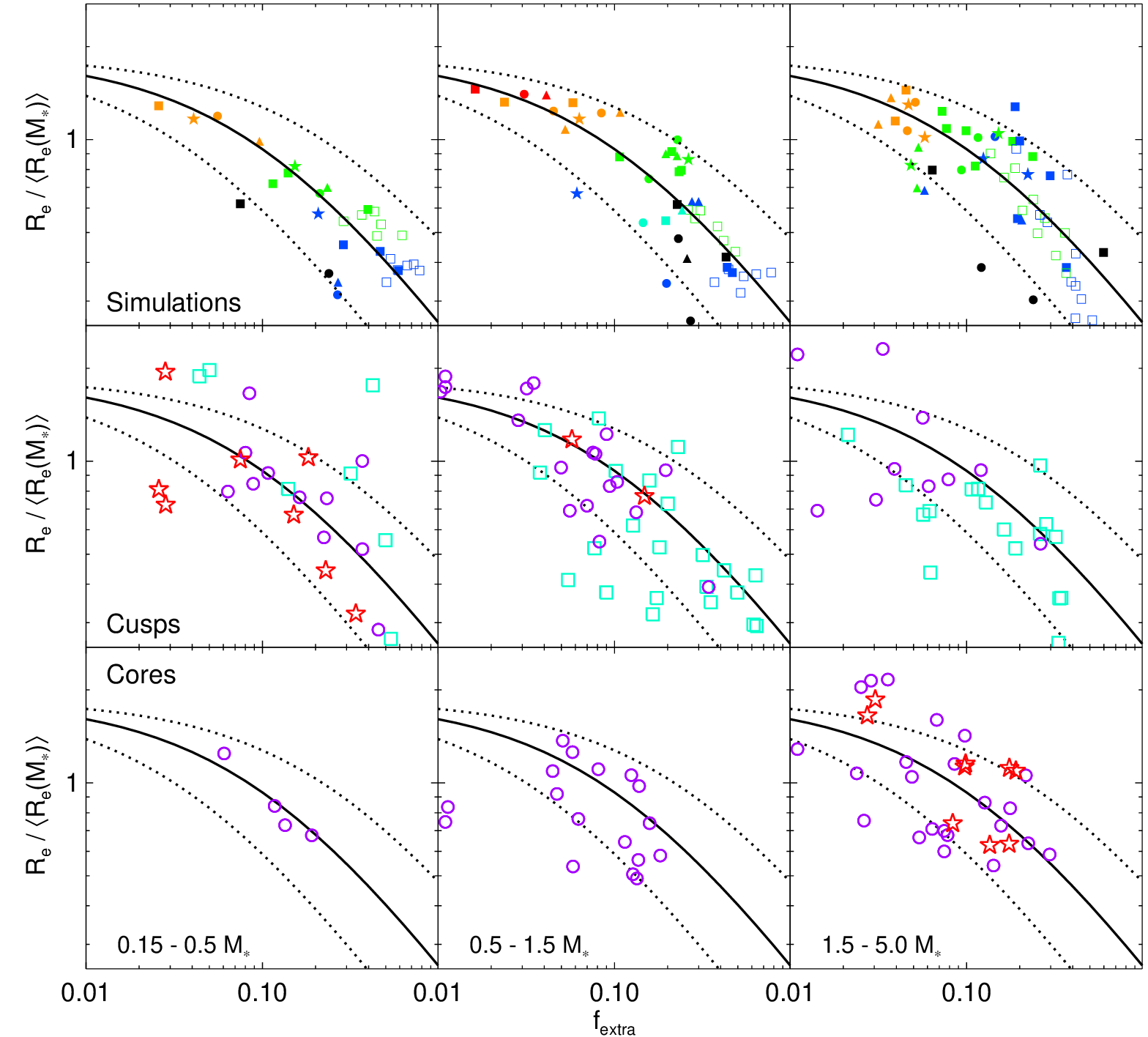}
    \caption{Effective radius $R_{e}$ relative to the 
    median value for all ellipticals of the same stellar mass, 
    as a function of our 
    fitted extra light fractions (the empirical 
    tracer of the dissipational/starburst mass fraction). 
    We compare simulated 
    merger remnants ({\em top}) with 
    observed cusp ellipticals and gas-rich merger remnants 
    ({\em middle}) and observed core 
    ellipticals ({\em bottom}), with points as in Figures~\ref{fig:rem.recovery} \&\ 
    \ref{fig:extra.vs.sb}.
    We show this in three bins of stellar mass
    (relative to $\mstar\approx10^{11}\,\msun$, or $M_{V}^{\ast}=-21$).
    Solid (dotted) lines show the mean ($\pm1\,\sigma$) correlation, 
    following the analytic solution for dissipational mergers and 
    fits to our simulation in \citet{covington:diss.size.expectation}. 
    Simulations and observations exhibit the same 
    behavior: systems with smaller $R_{e}$ at fixed mass have 
    systematically higher extra light fractions ($>6\,\sigma$ significance 
    in the observations). 
    This implies that, at fixed mass, 
    systems are driven along the fundamental plane by the relative amount of 
    dissipation involved in their formation.
    This behavior is true regardless of cusp/core 
    status, suggesting that (most) core
    ellipticals have not had a large number of re-mergers. 
    \label{fig:re.sigma.cusp}}
\end{figure*}

\subsection{Dissipation and the Sizes of Ellipticals}
\label{sec:scaling:sizes}

We demonstrated in \papertwo\ that the effective radii of cusp ellipticals scales with 
their extra light content -- systems with more dissipation are smaller, since more 
of the light comes from the compact starburst remnant. 
Figure~\ref{fig:re.sigma.cusp} 
extends this to the core ellipticals. 
Specifically, we determine $\langle{R_{e}(M_{\ast})}\rangle$
from the sample of \citet{shen:size.mass}, and take the ratio of 
the half mass radius of each system
(determined directly from the light profile,
or from the fits, it does not change the comparison) 
to that value.  Our mass bins are small enough,
however, that this makes little difference compared to just e.g.\
considering $R_{e}$ in a given bin. 
Again, we see a similar behavior in the cusp and core ellipticals: 
at a given stellar mass, systems with larger extra light have
systematically smaller $R_{e}$
(they also have slightly larger velocity dispersion $\sigma$, although the 
scatter is larger there in both simulations and observations). 
We showed that this could be predicted analytically, following 
\citet{covington:diss.size.expectation} given the impulse approximation to 
estimate the energy loss in the gaseous component, followed 
by collapse in a self-gravitating starburst. For typical conditions, this reduces 
to the simple approximation
\begin{equation}
R_{e} \approx \frac{R_{e}(\fsb=0)}{1+(f_{\rm sb}/f_{0})}, 
\end{equation}
where $f_{0}\approx0.25-0.30$ and $R_{e}(\fsb=0)$ 
is the radius expected for a gas-free remnant. 
In each case, the simulations and observed systems occupy a
similar locus to that predicted in this simple dissipational 
model. We can also construct this plot with the starburst mass
fraction $\fsb$ of the best-fitting simulation as the independent
variable, and find an even tighter correlation of the same nature. Core elliptical sizes 
depend on the relative strength of dissipation (evident in their profile 
shapes) in the same manner as in cusp ellipticals. 

This is expected: 
even if the systems with cores have expanded via re-mergers, they should (so long as 
there is not a wide range in number of re-mergers or systematic dependence 
of the number of re-mergers on starburst fraction at fixed mass) 
grow uniformly, and preserve these trends. 
We might expect some normalization offset: if the mean size-mass relation 
after a gas-rich merger is a power law 
$R_{e}\propto M_{\ast}^{\alpha}$, and 
two such systems with mass ratio $f$ (where $f$ is the mass ratio of the 
secondary to the primary) have a parabolic dry merger (and preserve profile shape), 
then the remnant will increase in radius by a factor 
$(1+f)^{2}/(1+f^{2-\alpha})$ relative to the primary. However, it has also increased 
in mass, so compared to ellipticals of the same (final) mass, its 
relative increase in size is only 
$(1+f)^{2-\alpha}/(1+f^{2-\alpha})$. For observationally suggested values 
$\alpha=0.56$ \citep{shen:size.mass}, this yields only a $30\%$ ($\sim0.1$\,dex) 
relative size increase for mass ratios of 1:3 through 1:1. This is 
easily dwarfed by the effects of dissipation seen in Figure~\ref{fig:re.sigma.cusp}, 
which can change the sizes of systems by nearly an order of magnitude. 
That there is not a substantial 
normalization offset between the trends for cusp and core ellipticals
in Figure~\ref{fig:re.sigma.cusp} therefore 
suggests that the number of major re-mergers for typical core ellipticals has been 
relatively moderate, but it is not a strong constraint.
Our simulations and the observations plotted 
suggest that the dominant factor regulating the size of an elliptical is the 
amount of dissipation, even for systems which have undergone re-mergers.

\breaker
\section{Dissipation and Re-Merger Kinematics/Shapes}
\label{sec:kinematics}

\begin{figure*}
    \centering
    \scaleup
    \plotterrr{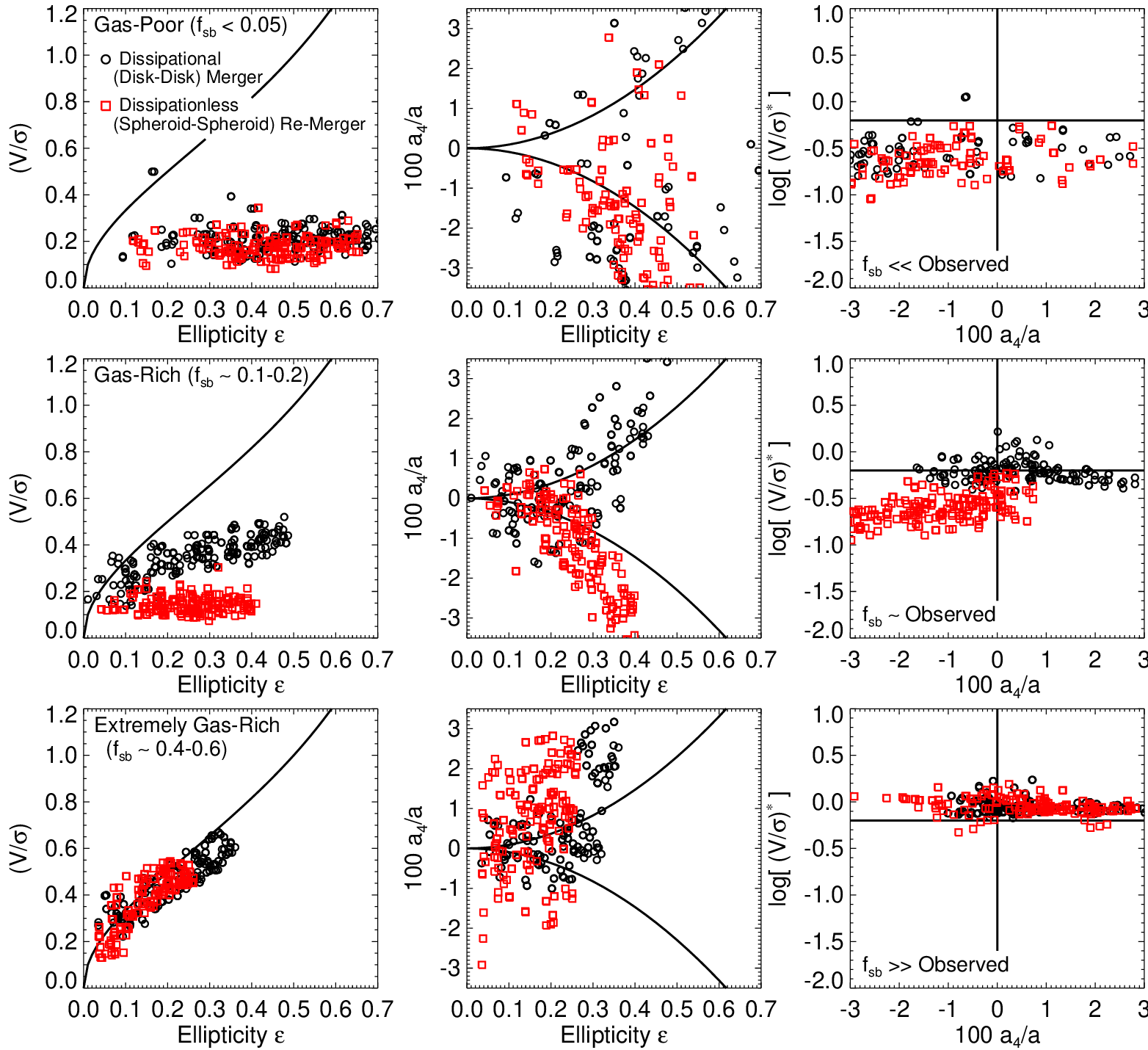}
    \caption{Rotation ($V/\sigma$), ellipticity $\epsilon$, and isophotal 
    boxy/diskiness ($a_{4}/a$) for a 1:1 identical disk-disk merger and 
    spheroid-spheroid 
    re-merger of that disk-disk remnant. {\em Top:} Case where the 
    initial disk-disk merger is gas-poor ($f_{\rm starburst}\approx 0.01$). 
    {\em Middle:} Case where the initial disk-disk merger is 
    gas-rich ($f_{\rm starburst}\sim0.1$, typical of $\sim1-5\,L_{\ast}$ 
    ellipticals). 
    {\em Bottom:} Case where the initial disk-disk merger is 
    extremely gas-rich ($f_{\rm starburst}\sim0.4-0.6$, larger 
    than we infer for any of the observed core ellipticals in our sample).     
    For each, we show a single 
    merger remnant, as viewed from $\sim100$ independent sightlines. 
    Gas-poor disk-disk mergers yield remnants that are slow rotators with 
    large ellipticity, without a correlation between $\epsilon$ and $a_{4}$, 
    and extending to very boxy isophotal shapes. This is relatively unchanged 
    in their re-mergers (simple energetic 
    concerns mean they cannot get significantly ``more round,'' and the 
    extension to ``too boxy'' isophotes and lack of $\epsilon-a_{4}$ correlation is 
    enhanced). Gas-rich disk-disk mergers are more round (owing to dissipation 
    building a central mass concentration), with kinematic subcomponents 
    that give rise to the rotation and disky isophotes observed in cusp ellipticals 
    (and their correlations). Their re-mergers destroy these subsystems, but 
    retain the central mass concentrations -- this enables the formation of 
    very round, but boxy and slowly-rotating systems 
    with the $\epsilon-a_{4}$ correlations appropriate for core ellipticals. 
    But systems which are significantly more gas-rich than we infer remain 
    disky and rapidly rotating even after re-mergers, in conflict  
    with the observed core population.
    \label{fig:kinematics.before.after}}
\end{figure*}

In \papertwo\ we showed that the existence and mass fraction in a 
dissipational component is related to the kinematics and shapes of 
cusp ellipticals. In particular, we confirmed what has been recognized in a growing 
number of simulations \citep[see e.g.][]{naab:gas,cox:kinematics,
jesseit:kinematics, burkert:anisotropy}, that disk-disk mergers without 
significant dissipational components (whether this component is created 
in the merger, as a gas-rich merger, or ``inherited'' from the progenitor disk 
bulges -- which formed it dissipationally in some earlier encounter/dissipational 
event) do not resemble any observed spheroids: they are slowly rotating and boxy, 
but highly elliptical. However, systems with the appropriate dissipational content 
for their stellar mass (as in e.g.\ Figure~\ref{fig:fgas.needed}) agree well with 
observed distributions in cusp ellipticals in all of these properties 
\citep[and the same is true for more sophisticated kinematic and shape 
measures, including their orbital isotropy, rotation, triaxiality, kinematic 
misalignments, and higher-order velocity moments 
$h_{3}$ and $h_{4}$;][]{cox:kinematics,jesseit:kinematics}. 
This is related to two aspects of the gas content in the mergers. First, 
the dissipational stellar component itself, by creating a dense central 
stellar mass concentration, acts for stars at $R_{e}$ almost as a central point 
mass with large stellar mass fraction $\sim10-20\%$ -- directly making the 
galaxy potential more spherical (less elliptical) 
and reducing the portion of phase space available to 
box orbits \citep{barnes.hernquist.91,barneshernquist96}.
Second, if more gas is available to form the starburst, more is 
likely to survive the merger. This gas will quickly re-form a disk (albeit often 
a very small, nuclear disk with relatively small mass fractions of order a 
couple percent, more often designated a nuclear subcomponent, if it can 
be recognized in observations), giving rise to more rapid rotation and diskier 
isophotal shapes. 

Given the connection between the dissipational component and the shapes/kinematics 
of the disk-disk merger remnant, it is worth examining how this compares 
after dissipationless spheroid-spheroid re-mergers (and how this scales with 
the original starburst/dissipational mass fraction). 
Figure~\ref{fig:kinematics.before.after} shows this comparison for 
representative simulations with different gas fractions. 
We consider the rotation ($V/\sigma$, where $V$ is the maximum 
rotational velocity from the projected stellar rotation curve and $\sigma$ 
the central dispersion), ellipticity $\epsilon$, and isophotal 
shape (boxy/diskiness, quantified as $a_{4}/a$ in standard fashion) 
for a 1:1 identical disk-disk merger and spheroid-spheroid re-merger of that remnant
\footnote{Specifically we show the 
rotation $(V/\sigma)$ or $(V/\sigma)^{\ast}$ measured within $\re$, and 
boxy/disky-ness $100\,a_{4}/a$, and 
ellipticity $\epsilon$ measured for the half-mass projected ($\re$) isophotal 
contour, for each of $\sim100$ lines-of-sight to the remnant uniformly 
sampling the unit sphere (i.e.\ representing the distribution across random 
viewing angles). The details of the fitting for our simulations are described in 
\citet{cox:kinematics}, and for the observations in \citet{bender:87.a4,
bender:88.shapes,bender:ell.kinematics,faber:catalogue,
simien:kinematics.1,simien:kinematics.6}.
We define rotation in the standard manner, relative to that 
of an oblate isotropic rotator, with the parameter $(V/\sigma)^{\ast}$, 
defined using the maximum circular velocity $V_{\rm c}$, 
velocity dispersion within $\re$, and ellipticity as \citep{kormendy:rotation.equation} 
\begin{equation}
(V/\sigma)^{\ast} = (V/\sigma) / \sqrt{\epsilon/(1-\epsilon)}.
\end{equation}
We exclude the coplanar merger simulations 
from our comparisons here: those 
simulations are idealized perfectly coplanar prograde orbits, and as such 
produce pathological orbit and phase structure (we do, however, include some  
more representative orbits below that are not far from coplanar).}.

In the case where the initial disk-disk merger is gas-poor ($f_{\rm starburst}\approx 0.01$), 
we recover the results in the papers described above: 
gas-poor disk-disk mergers yield remnants that are slow rotators with large ellipticity, without a correlation between $\epsilon$ and $a_{4}$, 
and extending to very boxy isophotal shapes, in stark disagreement with the 
properties of {\em either} cusp or core ellipticals (given similar 
resolution and seeing). This is relatively unchanged 
in re-mergers, and remains so regardless of the number of ``generations'' of 
re-mergers (we have experimented with series of several such re-mergers): 
without dissipation to build a central mass concentration (increase the 
central phase-space density), dissipationless processes cannot produce such a central 
density as required to make the potential sufficiently round and eliminate most of the 
phase space available to the most extreme boxy/triaxial orbital configurations 
(which are not observed in real systems). 
Regardless of re-merging, initially gas-free disk-disk mergers cannot ultimately 
reproduce the kinematics and shapes of core ellipticals. 

If instead the initial disk-disk merger is reasonably gas-rich 
($f_{\rm starburst}\sim0.1$), in the range that we have inferred 
from fitting observed surface brightness profiles of $\sim1-5\,L_{\ast}$ ellipticals 
(and in agreement with the observed gas fractions of disks in the 
relevant mass and redshift range), then the remnants are more round (owing to dissipation 
building a central mass concentration), with kinematic subcomponents 
that give rise to the rotation and disky isophotes 
in good agreement with the observed cusp population 
\citep[see e.g.][who show such subcomponents are ubiquitous 
in cusp ellipticals and drive much of this 
rotation]{krajnovic:cusps.have.subcomponents}. Re-mergers of 
these remnants destroy these kinematic subsystems (they are small disks/subsystems, 
and as such are violently relaxed in re-mergers), but 
retain the central mass concentrations (as we have argued they must), 
and this leads to the formation of 
sufficiently round, but boxy and slowly-rotating systems 
with the $\epsilon-a_{4}$ correlations appropriate for core ellipticals 
\citep[compare the observations in e.g.][]{bender:88.shapes,bender:ell.kinematics,
emsellem:sauron.rotation.data,emsellem:sauron.rotation}. The 
re-merger remnants of cusp ellipticals with the starburst mass 
fractions we have independently inferred from profile fitting appear to agree well 
with observed core elliptical kinematics and isophotal shapes. 

If the initial disk-disk merger is {\em too} gas-rich ($f_{\rm starburst}\sim0.4-0.6$), 
larger than that inferred for any of the observed core ellipticals in our sample, 
this presents an interesting case as well. These extremely gas-rich situations 
may apply to some of the most low-mass cusp ellipticals, but are clearly 
more gas-rich than expected at the masses of typical core ellipticals (and 
disk progenitors), 
even at high redshifts \citep{erb:lbg.gasmasses}. In these cases, we find that the 
initial disk-disk merger remnant is even more round, rapidly rotating, 
and disky. However, with such a large dissipational 
component, the kinematic subsystems are often quite massive -- in many 
cases they are more appropriately described as large surviving disks 
\citep[see e.g.][]{hopkins:disk.survival}. As a consequence, it becomes difficult to 
simply disrupt them in re-mergers. Moreover, conservation of 
the specific angular momentum 
in such large components makes it very difficult to achieve a slow rotator 
even if the subsystem is disrupted (as opposed to the case above, where the 
relevant subcomponents represent a small fraction of the galaxy mass); 
this has been seen when e.g.\ mock ellipticals with uniform rotation (as opposed to 
more realistic rotation associated with kinematic subcomponents) 
are re-merged \citep{naab:dry.mergers,burkert:anisotropy,jesseit:merger.rem.spin.vs.gas,
cox:remerger.kinematics}. As a consequence, the 
re-merger remnant looks like an oblate isotropic rapid rotator, not a slow rotator, 
and exhibits correlations between e.g.\ diskiness and ellipticity 
(and higher-order kinematic moments) that do not resemble either the 
cusp or core population.

\begin{figure*}
    \centering
    \scaleup
    \plotterrr{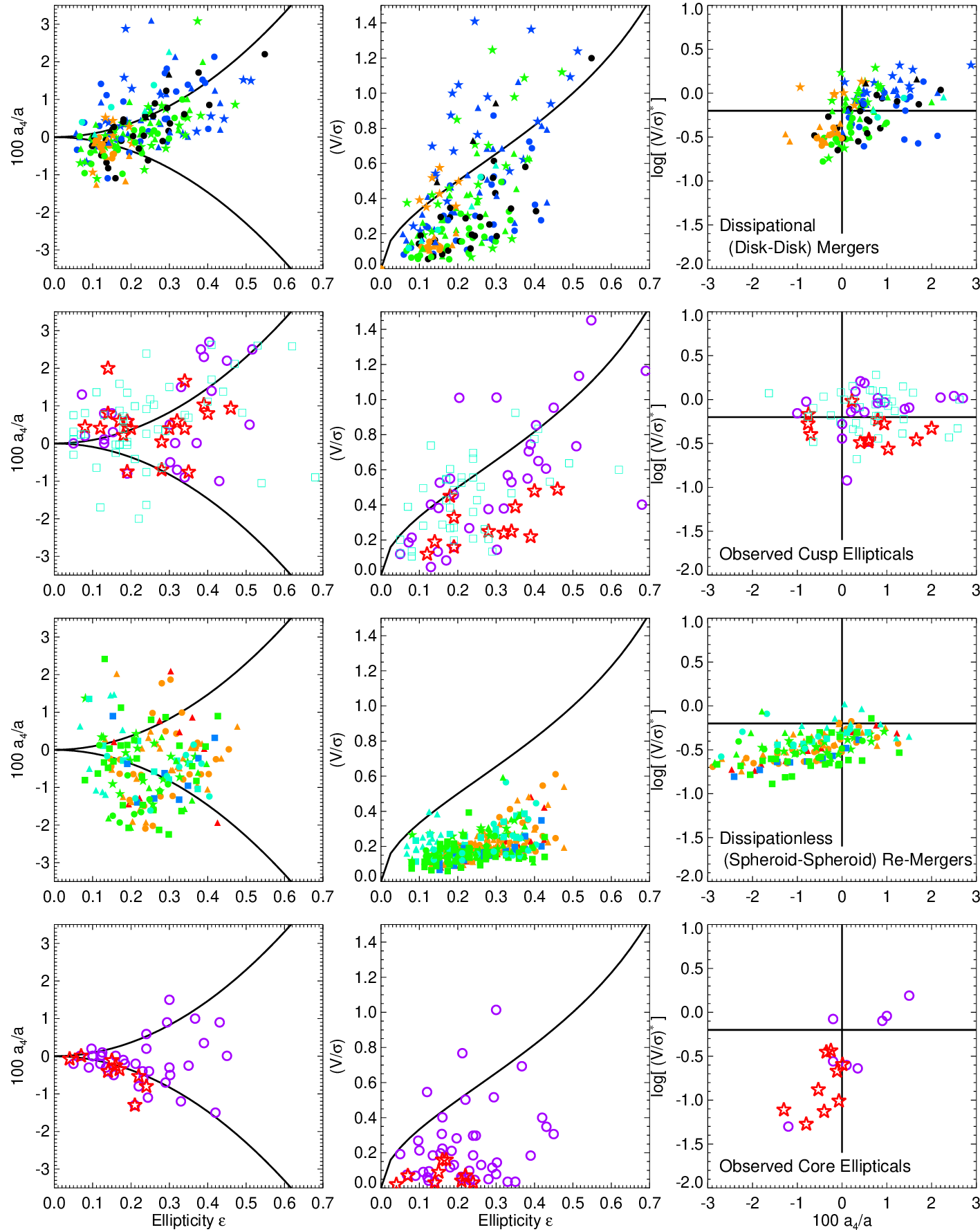}
    \caption{
    As Figure~\ref{fig:kinematics.before.after}, but comparing our library of simulations 
    to observed systems. 
    {\em Top:} Rotation ($V/\sigma$), ellipticity $\epsilon$, and isophotal 
    boxy/diskiness ($a_{4}/a$) for disk-disk merger remnants, with dissipational/starburst 
    fractions appropriate for their mass (matched to that inferred from the observed 
    systems; roughly $f_{\rm starburst}\sim0.1$ for most of the near $\sim L_{\ast}$ sample). 
    Each point is the median value for a given simulation, across $\sim100$ sightlines.
    Below, we compare observed cusp ellipticals and gas-rich merger remnants. 
    {\em Bottom:} Same, for spheroid-spheroid re-mergers of the original disk-disk merger 
    remnants, and observed core ellipticals. 
    The simulated and observed locii agree for both populations (despite being 
    distinct from one another). Comparing to Figure~\ref{fig:kinematics.before.after}, 
    it is clear that mergers and re-mergers of systems without 
    sufficient dissipation (less than that inferred from our surface brightness profile 
    fitting) do {\em not} agree with the observed cusp and core populations -- 
    significantly different dissipational fractions from our estimates are ruled out 
    by the kinematic data.
    \label{fig:kinematics.cuspcore}}
\end{figure*}

\begin{figure*}
    \centering
    \scaleup
    \plotter{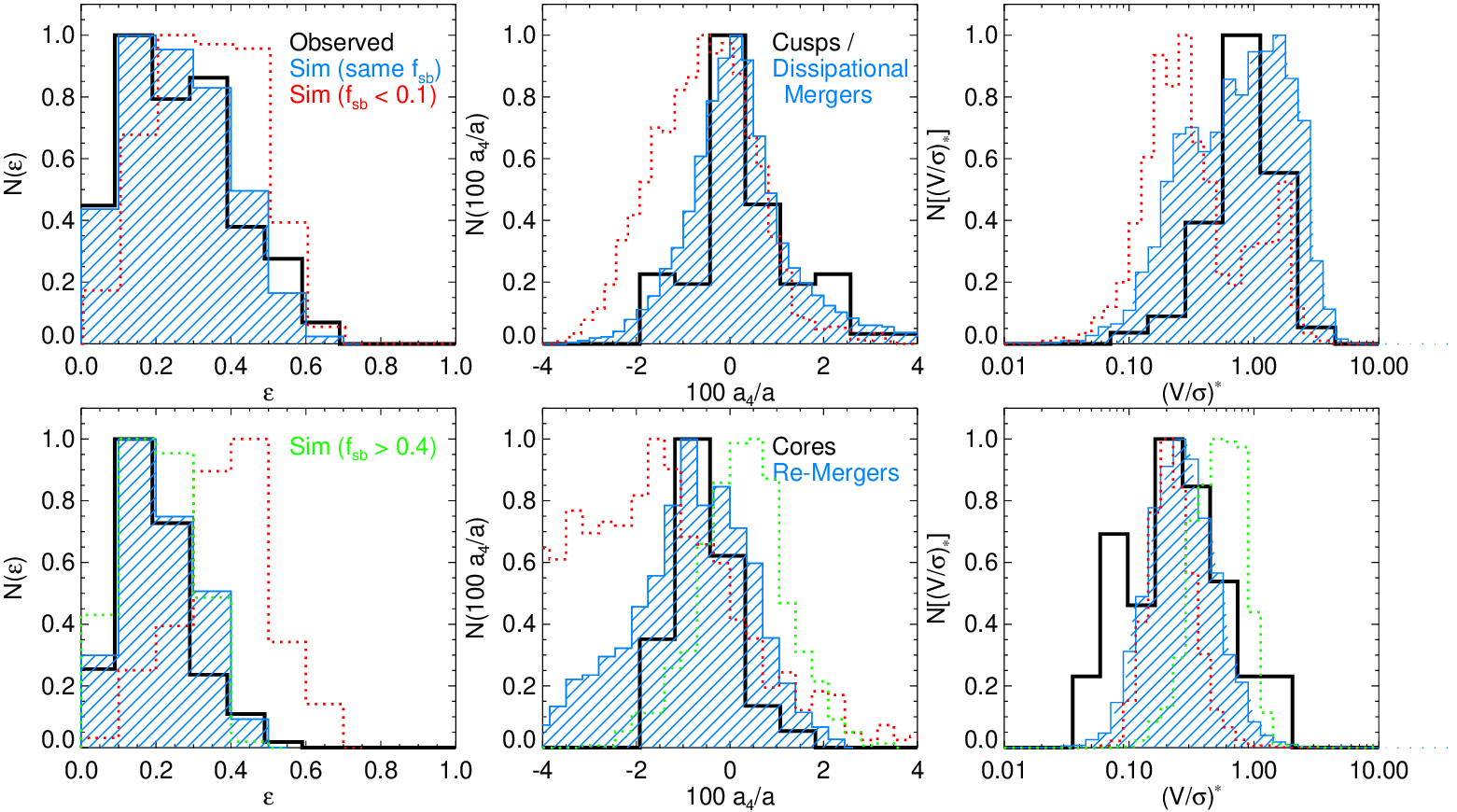}
    \caption{Histograms show the distribution in each property 
    from Figure~\ref{fig:kinematics.cuspcore} (ellipticity $\epsilon$, 
    isphotal shape boxy/diskiness $a_{4}/a$, and rotation $(V/\sigma)^{\ast}$), 
    for observed systems and simulations. 
    {\em Top:} Observed cusp ellipticals (black line) compared to simulated 
    disk-disk merger remnants 
    which have similar moderate extra light fractions (blue shaded),  
    and gas-poor (nearly dissipationless) disk-disk mergers ($\fsb<0.1$; red dotted). 
    The simulation distributions uniformly 
    sample each simulation in solid angle over $\sim100$ lines-of-sight -- sightline-to-sightline
    and object-to-object differences contribute comparably to the scatter. 
    {\em Bottom:} Same, for observed core ellipticals (black line) compared to 
    simulated re-mergers of the disk-disk merger remnants with appropriate 
    extra light fractions (dissipational fractions in the original disk-disk merger; blue shaded) 
    and with much lower (red dotted) or higher (green dotted) dissipational content. 
    The simulations, when they have the appropriate extra light/dissipational content, 
    agree with the observed distributions. Their is some ``memory'' of the amount of dissipation 
    retained in the kinematics: systems without sufficient extra light are too elliptical and 
    have a much broader $a_{4}$ distribution extending to extremely boxy isophotal shapes; systems 
    with too much extra light remain too disky and rapidly rotating even after re-mergers.  
    \label{fig:kinematics.histograms}}
\end{figure*}

Some of the observed differences between cusp and core distributions 
owe to the fact that core ellipticals (because they predominate at high masses) tend towards 
lower $f_{\rm sb}$ and cusp ellipticals (dominant at low masses) towards 
higher $f_{\rm sb}$. However, 
at the same mass, the two populations evidence similar 
$f_{\rm sb}$ (Figure~\ref{fig:fgas.needed}), but show differences in their kinematics. 
Figure~\ref{fig:kinematics.before.after} demonstrates that at fixed 
$f_{\rm sb}$ the kinematics also depend 
significantly -- and in a different manner -- on merger history. As a consequence, 
the well-established observational differences between these populations 
\citep[e.g.][]{bender89,faber:ell.centers} are non-trivial -- 
there is no population with a large number of 
dissipationless re-mergers 
(at any $f_{\rm sb}$) that matches the observed kinematic/shape distributions of 
cusp ellipticals, and no population of disk-disk mergers (at any $f_{\rm sb}$) 
that matches the observed distributions of core ellipticals\footnote{As discussed 
below, this statement is not true for any {\em single} object, but applies to the 
populations as a whole.}.

Figures~\ref{fig:kinematics.cuspcore} \&\ \ref{fig:kinematics.histograms} 
show how the rotation and isophotal shapes 
of our simulated disk-disk and re-merger remnants compare 
with observed cusp and core elliptical populations. 
We specifically consider simulations with similar dissipational 
fractions to the observations (we only compare simulations that have values of $f_{\rm sb}$ 
in broad agreement with the locus we observationally infer from fitting the observed 
systems in Figure~\ref{fig:fgas.needed}; equivalently but more strictly, we can consider 
only the set of simulations each of which is a best-fit match to one of the observed 
surface brightness profiles, 
as in Figure~\ref{fig:jk1}). Despite these simulations being constrained to 
match the observed systems only in their averaged surface brightness profile, they 
occupy a similar locus to the observations in each sense 
in Figures~\ref{fig:kinematics.cuspcore} \&\ \ref{fig:kinematics.histograms}. 

Figure~\ref{fig:kinematics.histograms} illustrates that the distribution in $\epsilon$, $a_{4}/a$ and 
$(V/\sigma)^{\ast}$ for core ellipticals constrains not only their merger history, but 
also the dissipational content of the {\em original} gas-rich mergers, 
(even if these systems are the product of spheroid-spheroid re-mergers). 
This constraint agrees well with our estimates from their 
surface brightness profiles: 
systems which originally had too much or too little gas, relative 
to that we infer from the surface brightness profiles, fail to 
reproduce the observed shapes and kinematics (with or without re-mergers). 

Unlike the case of surface brightness profiles, however, such a constraint 
can only be derived from a statistical population of ellipticals; as in 
Figure~\ref{fig:kinematics.before.after}, the sightline-to-sightline scatter 
(even the object-to-object scatter in median properties seen in 
Figure~\ref{fig:kinematics.cuspcore}) is sufficiently large that the isophotal 
shape and/or rotation of any {\em individual} elliptical, observed 
from only the single available viewing angle, does not present a 
strong constraint on the gas-richness of the original spheroid-forming merger 
(especially after allowing for the dependence on merger history in a 
re-merged population). However, given the distribution observed in a 
statistical population, it is possible to make a strong statement of independent 
support for typical dissipational fractions similar to that inferred from the individual 
surface brightness profile fits. 

Finally, we note that one population that our simulations may 
have difficulty reproducing are the most extreme slow-rotators, 
with $(V/\sigma)^{\ast} < 0.1$. This may be a by-product of 
observational uncertainties or 
differences between 
our simulation versus observational fitting methodologies. At this level, 
both simulations and observations are sensitive to resolution and to the outer 
radii within which velocities are defined \citep[with velocities, even in slow rotators, 
rising significantly at larger radii; see e.g.][]{emsellem:sauron.rotation}; 
many of the observed objects in this category 
have $V/\sigma$ measured within relatively small radii $\lesssim0.6\,R_{e}$. 
We have broadly matched the simulation resolution and radii where kinematic 
quantities are defined ($=R_{e}$) to most of the observations shown, 
but a more rigorous comparison should carefully match the sampling 
radii for given sub-classes of objects and consider full kinematic profiles, 
rather than the integrated quantities here. 

The difference between our simulations and 
these slowest rotators may also reflect a real physical difference: 
probably our simplification of realistic merger histories. As demonstrated in 
e.g.\ \citet{boylankolchin:mergers.fp,boylankolchin:dry.mergers}, 
the rotation properties of spheroid-spheroid 
merger remnants can be significantly dependent on orbital parameters. We do 
not find as strong a dependence as those authors (namely because their model 
ellipticals did not include a central mass concentration as do the systems here; 
the presence of this component leads to more regular tube and loop orbits as 
opposed to the long box orbits that support the most extreme ellipticity/rotation 
configurations seen in that paper); however, we still expect that a systematic 
bias towards certain orbital parameters (for example, a preference for 
particularly radial orbits in the formation of the most massive ellipticals, 
if they accrete galaxies along filaments) can change $(V/\sigma)^{\ast}$ 
in a corresponding systematic manner. 

Also, the systems with such low 
rotation are typically the most massive systems, with early stellar formation 
times \citep[see e.g.][]{bender:ell.kinematics,bender:ell.kinematics.a4,
ferrarese:type12,kuntschner:line.strength.maps,
emsellem:sauron.rotation,cappellari:anisotropy,jk:profiles}; these are most likely to 
have undergone a mix of major and minor mergers, with e.g.\ five 1:5 mergers 
becoming a more likely channel for mass doubling in dry mergers than a single 1:1 merger 
for the most massive systems $L\gg L_{\ast}$. Although this will not change our 
analysis of the extra light components in the surface brightness profiles 
(provided these are as robust to 
minor mergers as they are to major mergers), 
it has long been proposed as a means to yield systems with 
low net angular momentum and such little rotation \citep{gallagherostriker72,
ostrikertremaine75,hausman:mergers,
weil94:multiple.merger.kinematics,weil96:multiple.merger.scalings,naab:etg.formation}. 
The formation of these objects, therefore, requires further study, with more 
cosmologically representative accretion and merger histories.

\breaker
\section{Outer Sersic Indices of Ellipticals with Cores}
\label{sec:outer.sersic}

Given the issues surrounding profile fitting 
described in \S~\ref{sec:profile.evol}, we 
have to be somewhat careful when we attempt to describe the 
typical outer Sersic indices of core ellipticals. 
We can (and do) obtain different answers depending on how we 
approach our fits, even for data sets of excellent quality and large dynamic 
range. However, we still find that the distribution of Sersic indices in 
core ellipticals is relatively narrow, with a characteristic value somewhat 
higher than those in cusp galaxies (as expected if they are indeed 
re-merger remnants). 

\begin{figure*}
    \centering
    \scaleup
    \plotone{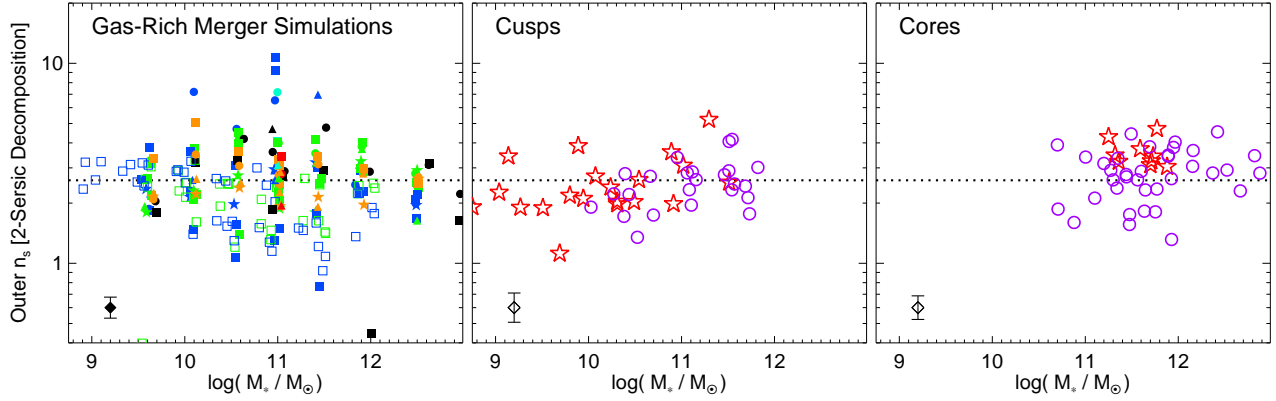}
    \caption{Outer Sersic indices (from our two-component 
    decomposition) in ellipticals and simulated gas-rich merger 
    remnants (point style as in Figure~\ref{fig:extra.vs.sb}). 
    Filled diamond is typical sightline-to-sightline variation in the simulations, open 
    diamond the source-to-source (or band-to-band) scatter in observed profile fits. 
    Gas-rich merger remnants and cusp 
    ellipticals have characteristic 
    $n_{s}\sim2-3$, without a strong systematic dependence on mass or other properties 
    (dotted line is the median for cusp ellipticals and gas-rich merger remnants, 
    $n_{s}=2.7$); 
    core ellipticals are the same, 
    but with slightly higher characteristic $n_{s}\sim3-5$.
    \label{fig:ns.mass}}
\end{figure*}

Figure~\ref{fig:ns.mass} compares the Sersic indices of the outer component 
in our standard two-component decompositions fitted 
to our gas-rich merger 
simulations, cusp and core ellipticals. 
Here, the same fitting procedure has been applied over a self-consistent 
dynamic range, regardless of the cusp or core status (profile at $\lesssim50\,$pc) 
of the system \citep[for a detailed discussion of the dynamic range 
requirements for reliable Sersic index fits, see][]{jk:profiles}. 
As we demonstrated in \papertwo, for cusp ellipticals and 
gas-rich merger remnants considered in this fashion, there is no dependence 
of outer Sersic index on stellar mass or other galaxy properties. The cusps and 
gas-rich mergers have a relatively narrow Sersic index distribution centered 
around $n_{s}\sim2.6-2.7$ with a width of $\pm0.75$. 

Applying our two-component models to the core ellipticals, we obtain a similar result. 
There is no evidence for a significant dependence of the fitted outer Sersic index 
on mass or any other property within the core elliptical sample. There is an offset 
between the Sersic index distribution of this sample and that of the cusp 
ellipticals, roughly corresponding to an increase in Sersic index of 
$\Delta n_{s}\approx 1$ (i.e.\ a median $n_{s}\sim3.6$ and dispersion $\pm1.0$). 

\begin{figure}
    \centering
    \plotter{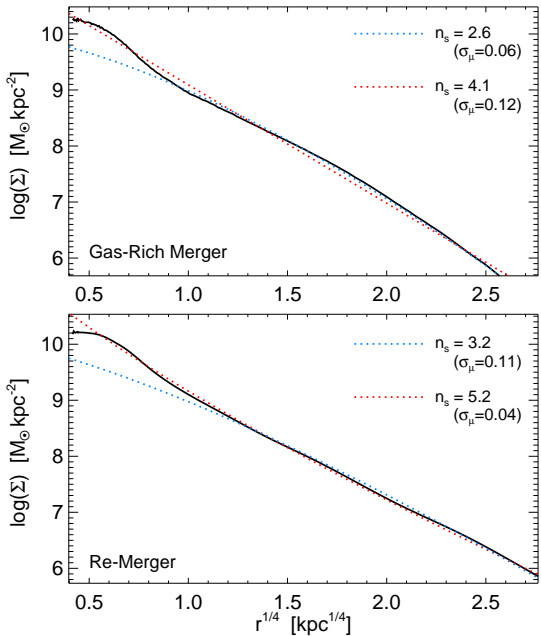}
    \caption{Illustration of how profile shapes change in a re-merger, as a 
    consequence of the scattering of stars. {\em Top:} Gas-rich merger remnant. 
    Dotted blue line shows the outer Sersic index from our typical 
    two-component fits ($n_{s}=2.6$; concave down in this projection) 
    with the rms deviations of the 
    profile from the best-fit two-component model (the rise above 
    this curve accurately traces the physical starburst component). 
    Dotted red line shows the result 
    of fitting the entire profile to a core-Sersic profile (or a 
    Sersic profile with some downward break at small radii; $n_{s}=4.1$). 
    Both methods yield reasonably good fits, but the meaning and values of $n_{s}$ 
    are different (the $n_{s}$ obtained fitting the profile in the latter manner reflects 
    some combination of the dissipational and dissipationless components). 
    For gas-rich merger remnants, the sharper ``break'' in the profile also yields 
    significant residuals about the cored-Sersic fit. 
    {\em Bottom:} Same, after a gas-poor re-merger. 
    The two-component decomposition is just as successful at recovering the 
    physical dissipational and dissipationless components. However, 
    the scattering of stars 
    to larger radii and smoothing of the transition to the starburst component 
    yields a situation where other possible fitting functions -- in 
    this case a core-Sersic fit, with a higher outer 
    index ($n_{s}=5.2$; concave up here) and central deficit -- can be excellent fits 
    in a formal statistical sense. 
    \label{fig:highlownsdemo}}
\end{figure}

As described in \S~\ref{sec:profile.evol}, 
the change in the outer profile shape from cusp to core ellipticals 
is expected to be a real consequence of re-mergers. Figure~\ref{fig:highlownsdemo} 
illustrates the difference between typical profiles before and after a re-merger. 
To lowest order, the profile is simply puffed out by a uniform factor (which 
has no effect on the Sersic index). 
However, 
there is significant scatter in final radii at fixed initial radii -- in other words, 
more stellar light will be scattered to large radii with a re-merger 
\citep[and, roughly speaking, this fraction will increase weakly as the amount of 
violence, in terms of the orbital energy, of the re-mergers is increased; see 
e.g.][]{boylankolchin:dry.mergers}. 
This will slightly raise the outer Sersic index, 
and help build up the extended envelopes seen around 
many massive galaxies and BCGs. 
Exactly how much this outer profile is raised depends somewhat on how we 
define our profile fits -- in a more robust physical sense, it reflects a lognormal 
scattering of stars by $\sim0.4$\,dex (Figure~\ref{fig:rf.of.ri}). 

Since there was no strong dependence on mass in the cusp population, we 
expect no strong dependence in the core population, so long as there is 
not a dramatic dependence of merger history (within this 
sub-population) on mass. To the extent that we see these effects and no substantial 
mass dependence, it implies that most of our core ellipticals have undergone a 
relatively small number of re-mergers, and 
are therefore reasonable analogs of our re-merger simulations 
(where we see this change -- where we see 
large envelopes that imply very high Sersic indices for the outer 
components -- is at the extreme masses of  $\gtrsim$ a few $L_{\ast}$ BCGs, 
where we have emphasized our models should not be extrapolated without 
caution). 

This relatively modest increase in Sersic index may seem surprising, as 
are the (still relatively low) Sersic indices fitted in this manner, given 
that massive ellipticals are often fitted to high 
Sersic indices $n_{s}\gtrsim6$. However, we caution that those results 
are based on fitting those profiles to either single Sersic or core-Sersic type laws 
(or similarly an outer Sersic law allowing for some inner ``deficit'' or core), 
and are {\em not} directly comparable to the results fitting the two-component 
models adopted here. For comparison, we therefore consider the results obtained 
with these fitting functions and how they relate to the galactic components 
inferred from our two-component decompositions. 

\begin{figure*}
    \centering
    \scaleup
    \plotone{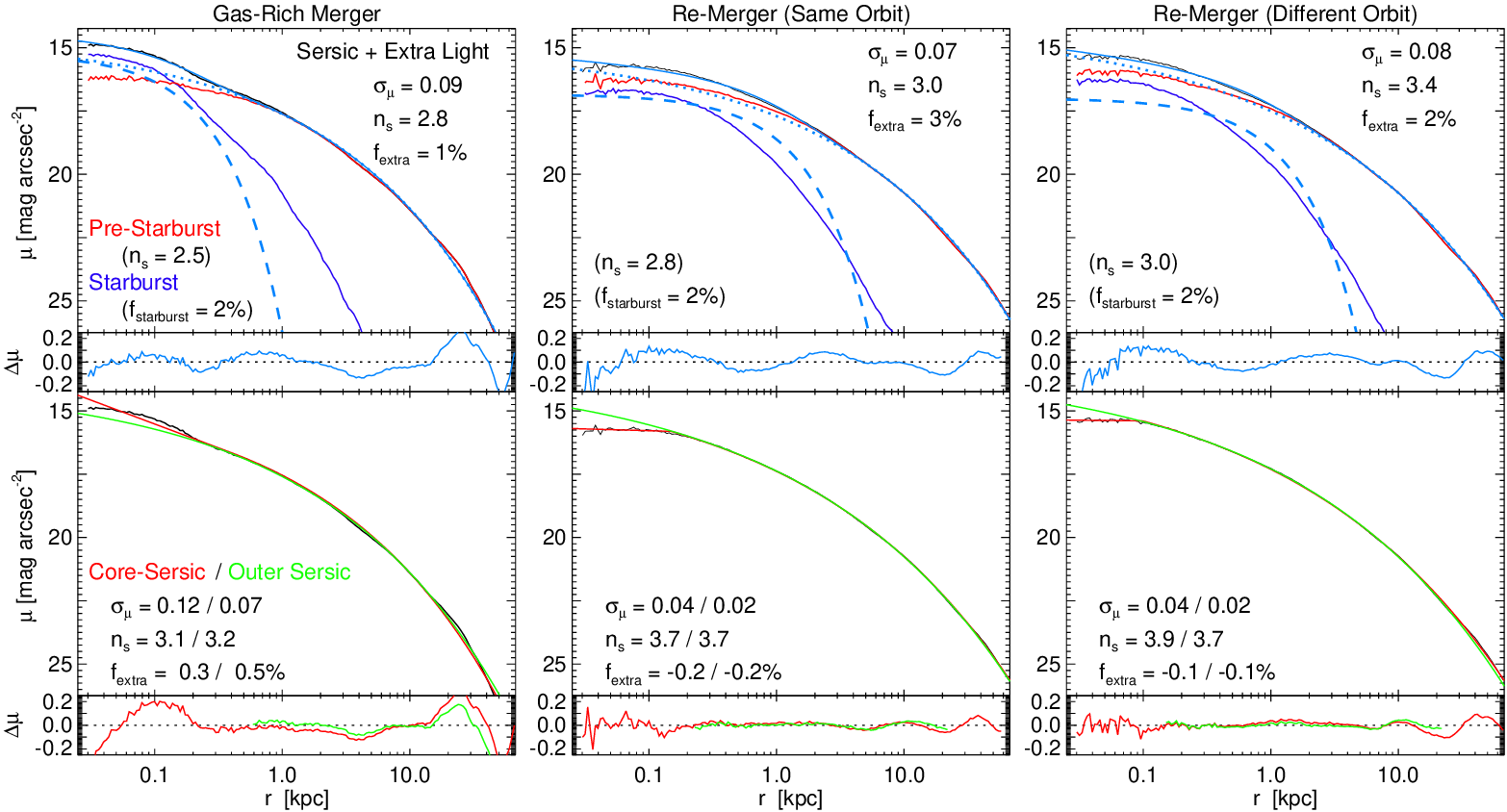}
    \caption{{\em Top:} Surface brightness profiles from the smooth extra light 
    simulations shown in Figure~\ref{fig:demo.rem.appearance}, including 
    the original gas-rich merger remnant ({\em left}), a major re-merger 
    remnant ({\em center}) and another re-merger with different orbital 
    parameters ({\em right}). We show the physical decomposition into 
    pre-starburst and starburst stars (tracking the starburst component 
    after re-mergers), with the mass fraction in the starburst and Sersic index directly 
    fitted to the pre-starburst component. We compare with our two-component 
    fit, with corresponding extra light mass fraction, outer Sersic index, 
    and rms residuals $\dmu$ (in ${\rm mag\, arcsec^{-2}}$), and show the fit 
    residuals ($\Delta\mu$). As before, the two-component decomposition 
    reliably recovers the physical decomposition of the light. 
    {\em Bottom:} Same, but fitted with a core-Sersic law profile (red) or 
    a single Sersic law fitted only to radii outside of where the central 
    profile begins to deviate from the Sersic law (green). 
    In the original gas-rich merger remnant, these methods all show some 
    extra light, and the core-Sersic model performs poorly. 
    After a re-merger, 
    the core-Sersic model formally fits the profile well with a higher Sersic index 
    that reflects the {\em combination} of the outer profile 
    and the inner extra light profile. 
    If we were to simply take $f_{\rm extra}$ as the integrated light relative to that 
    from extrapolation of the Sersic fit to all radii (instead of using our 
    calibrated two-component decomposition), we would infer 
    much less ``extra'' light in the gas-rich remnant and some 
    ``deficit'' ({\em missing} light) in the re-merger remnants -- 
    the starburst component is still there (with the mass fraction correctly inferred 
    by our two-component fits), but these parameterized fits reflect it only 
    indirectly (with their higher Sersic indices) -- inferring the ``extra'' or dissipational 
    component mass from these alternative 
    fitting functions is not physically motivated, even if they are better statistical 
    fits to the light profile. 
    \label{fig:demo.ml.appearance}}
\end{figure*}

Figures~\ref{fig:highlownsdemo} \&\ \ref{fig:demo.ml.appearance} 
compare the results of our two-component fits with those obtained fitting 
our simulated re-mergers to a ``core-Sersic'' law of the form 
\begin{equation}
I = I'\,{\Bigl[}1+{\Bigl(}\frac{r_{b}}{r}{\Bigr)}^{\alpha}{\Bigr]}^{\gamma/\alpha}\,\exp{{\Bigl[}
-b_{n}\,{\Bigl(}\frac{r^{\alpha}+r_{b}^{\alpha}}{r_{e}^{\alpha}}{\Bigr)}^{1/(\alpha\,n)}{\Bigr]}}
\end{equation}
\citep[e.g.][]{graham:core.sersic}, where $r_{b}$ is the core break
radius within which the profile shifts to a power law of slope
$\gamma$, $r_{e}$ is the effective radius and $n$ the Sersic index of
the outer light profile, and $\alpha$ a parameter describing how
rapidly the break occurs. 
Whether we free $\alpha$ or set $\alpha\rightarrow\infty$ (the limit of a 
sharp break into the profile core) we obtain similar results (although the fit 
quality $\chi^{2}/\nu$ is slightly improved with a free $\alpha$). 

We also consider fitting a single Sersic index to the outer profile, i.e.\ the profile 
outside of the central core \citep[following the 
methodology of][]{jk:profiles}. In practice, we fit a core-Sersic law 
to identify the radius 
where the inner profile breaks from the outer Sersic law, and then re-fit the 
radii larger than $\sim1-2$ times this radius to a single Sersic law (we 
can also in more detail fit steadily inwards from larger radii until the fit residuals 
begin to rapidly grow, which yields similar results). 

In the case of the direct gas-rich merger remnants, where the breaks between 
the inner and outer components are sharp (and where the outer component 
Sersic indices are relatively low), the results are more or less similar. 
The two-component fits not only match the physical decompositions 
in the simulations, but they also happen to be excellent fits in a formal statistical 
sense. The functional form of the Sersic distribution is such that fitting just the 
outer profiles to a Sersic function -- in this case giving low $n_{s}$ -- 
yields a shallow inwards extrapolation, similar to that in the progenitor ($n_{s}\sim1$) 
disks and therefore the dissipationless component; so the extra light again 
appears as a characteristic central excess (although the nature of 
fitting inwards ``as far as possible'' yields a mild bias towards some underestimation 
of the extra light content). 
The core-Sersic distribution (or equivalently, forcing a 
single Sersic function to fit the entire profile, inner and outer), 
designed to describe ellipticals with higher 
outer $n_{s}$ and a downward break at small radii, performs 
reasonably well fitting a higher $n_{s}$ Sersic distribution to the entire profile (including 
the extra light), but the relatively sharp break between the components yields 
noticeably larger residuals. 

If re-mergers exactly preserved profile shape, the behavior would, of course, 
be identical. 
In reality however, a re-merger scatters stars out to large radii (in addition 
to the overall ``puffing up'' of the profile; Figure~\ref{fig:rf.of.ri}), which raises 
the characteristic Sersic index $n_{s}$ describing the outer profile shape, 
and smooths any sharp breaks in the inner profile. As a consequence, 
we obtain rather different results applying alternative fitting functions to 
re-merger remnants. In Figure~\ref{fig:highlownsdemo}, 
for example, fitting from the outer radii inwards yields a 
high outer Sersic index $n_{s}\sim5$; the nature of the Sersic distribution is 
such that higher-$n_{s}$ outer profiles extrapolate to relatively 
large central densities, so the fitted distribution maps onto or attaches smoothly to 
the high central density that is the extra light component. 
Eventually, the inward extrapolation of the high-$n_{s}$ profiles rises 
above the light profile (and this will be even more pronounced in the 
presence of black hole scouring flattening the nuclear profile slope), 
so the profile appears to ``breaks downwards'' 
relative to the outer Sersic fit. Together with the smoothing of 
sharp features near the transition between the dissipational 
and dissipationless components, this means that one can -- 
after a re-merger -- obtain 
very good statistical profile fits with simple cored-Sersic models 
($\dmu\lesssim0.04\,{\rm mag\,arcsec^{-2}}$), 
with considerably higher outer $n_{s}$. 

Figure~\ref{fig:demo.ml.appearance} demonstrates this in a more 
rigorous manner: we compare the results of these different fitting 
methods with the known physical decompositions in the simulations. 
This emphasizes an important point: the good fits obtained with e.g.\ 
cored Sersic models (that do not include an explicit ``extra'' component) 
do {\em not} necessarily mean that the extra light (as we physically 
define it: stars formed originally in a dissipational starburst) is 
not still manifest in the light profile. 
This stellar material survives the dry
spheroid-spheroid re-mergers we simulate, 
and the rank order of particles in radius is sufficiently preserved 
that it still contributes a substantial component of the light 
within $\sim$kpc scales in Figure~\ref{fig:demo.ml.appearance}. 
In fact, the high {\em central} densities implied by the large 
Sersic indices of the fitted cored Sersic models 
\citep[typical $n_{s}\sim6-8$ in observed cored systems; see e.g.][]{
prugniel:fp.non-homology,trujillo:sersic.fits,ferrarese:profiles,jk:profiles} 
can only be part of the fit if the extra light is included -- scattering of stars 
cannot, by Louisville's theorem, raise the central density of the 
elliptical. If there were no dissipational component in core ellipticals, 
then their outer portions would resemble a high-$n_{s}$ envelope, 
but their inner portions ($\lesssim 2-3\,R_{e}$) would resemble a 
low-$n_{s}$, low density disk (which would mean, among other things, 
that they could not be well-fit over a significant dynamic range 
by {\em any} Sersic function). Note that the apparent ``deficit'' 
seen in these simulations (i.e.\ shortfall of the inner light profile 
relative to the inwards extrapolation of the outer Sersic fit) 
does not reflect a physical deficit (since the high-$n_{s}$ outer 
profile is much steeper in the central regions than the profiles 
of the actual progenitors); this lends some caution to the 
approach of using such comparisons to infer 
the mass ``scoured'' by a nuclear binary black hole 
(which we discuss further in \S~\ref{sec:missing.light}).

\begin{figure}
    \centering
    \plotone{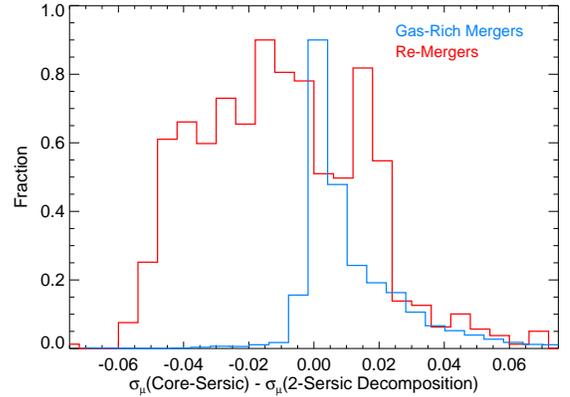}
    \caption{Summary of the difference in fitting single gas-rich merger remnants (blue) 
    and re-merger remnants (red). We show the distribution for both samples of simulations 
    in relative fit quality of a core-Sersic law and our two-component decomposition -- 
    i.e.\ the difference in $\dmu$ (in ${\rm mag\,arcsec^{-2}}$) between 
    the best-fit core-Sersic law and our two-component fit (note the two forms have 
    the same number of degrees of freedom). Negative values 
    mean the core-Sersic law is formally a better fit. The gas-rich mergers are poorly 
    fitted by a core-Sersic law -- they almost all prefer the two-component 
    decomposition, or little difference (that being in cases where there is only a small 
    fraction of extra light). A large fraction of re-merger remnants are better 
    fit by a core-Sersic law.
    \label{fig:dif.fit.qual.hists}}
\end{figure}

This general behavior is quite common; 
Figure~\ref{fig:dif.fit.qual.hists} summarizes these results for our library of 
simulations -- we plot the distribution of relative fit quality (for which we adopt the 
proxy of the difference in $\dmu$ between a core-Sersic fit and 
our two-component decomposition), for both single gas-rich merger 
remnants and re-merger remnants. 
These examples illustrate the importance of 
combining models and observations in interpreting the parameters obtained 
from galaxy surface brightness profile fits: these different functions 
are fitting the same profiles, with broadly similar quality (and with 
varying relative quality in populations with different merger histories), 
but the Sersic indices and other parameters obtained from the fits 
have different physical interpretations in the context of our simulations. 
Our proposed decompositions, 
although physically motivated, are not required by the observations: in a purely 
statistical sense, the two-component decomposition is not only non-unique as a 
parameterization, it is not even necessarily the best-fitting choice in all cases. 
Our intention here is not to argue that the proposed decomposition is a 
better fit in a formal sense to observed profiles, 
but rather that it is useful because it tends, 
when adopted in numerical experiments, 
to recover the correct physical decomposition between the 
dense inner (high surface-brightness) component of the galaxy originally 
formed by dissipation of gas into the central regions of the system, and 
the less dense outer component originally formed by violent relaxation of 
stellar disks. If we were to 
interpret the extra or excess light to be that light in the real profile
above the prediction of the best-fit single Sersic or core-Sersic model, 
in the same manner that we do relative to the outer components in 
our two-component fits, we would (incorrectly) infer 
effectively no extra light in most of our simulated 
re-merger remnants. Clearly, calibrating a given method against 
simulations and models is important for assigning it a specific 
physical meaning. 

Interestingly, we find that only 
our re-merger remnants are better fit (in a purely formal sense) 
by a core-Sersic law, as opposed 
to a two component model, and we find this in general becomes more true 
as we repeatedly re-merge systems. As a consequence, although the parameters 
of the core-Sersic or single Sersic fit may not be directly analogous to those 
obtained from our two-component fits, the fact that a galaxy can be formally better fit by one of 
those functions is predicted to be a consequence of dry re-mergers which 
will smooth the profile in the manner described above. 
Indeed, a tight association is observed 
between galaxies with nuclear cores and galaxies better fit by a single Sersic 
or core-Sersic law, despite differences in the spatial scales where the 
two are determined \citep{ferrarese:profiles,jk:profiles}. 
Our models provide a natural explanation for this, which not only reconciles 
such profile fits with the survival of dissipational/extra light components 
from cusp elliptical progenitors, but in fact requires such components 
to explain the highest observed Sersic indices from single Sersic or 
core-Sersic fits. This lends further 
confidence to our modeling of profile shapes, even at the level of 
smooth profiles apparently better fit by functional forms different from 
the convenient two-component decomposition we adopt in this paper.

\begin{figure*}
    \centering
    \scaleup
    \plotone{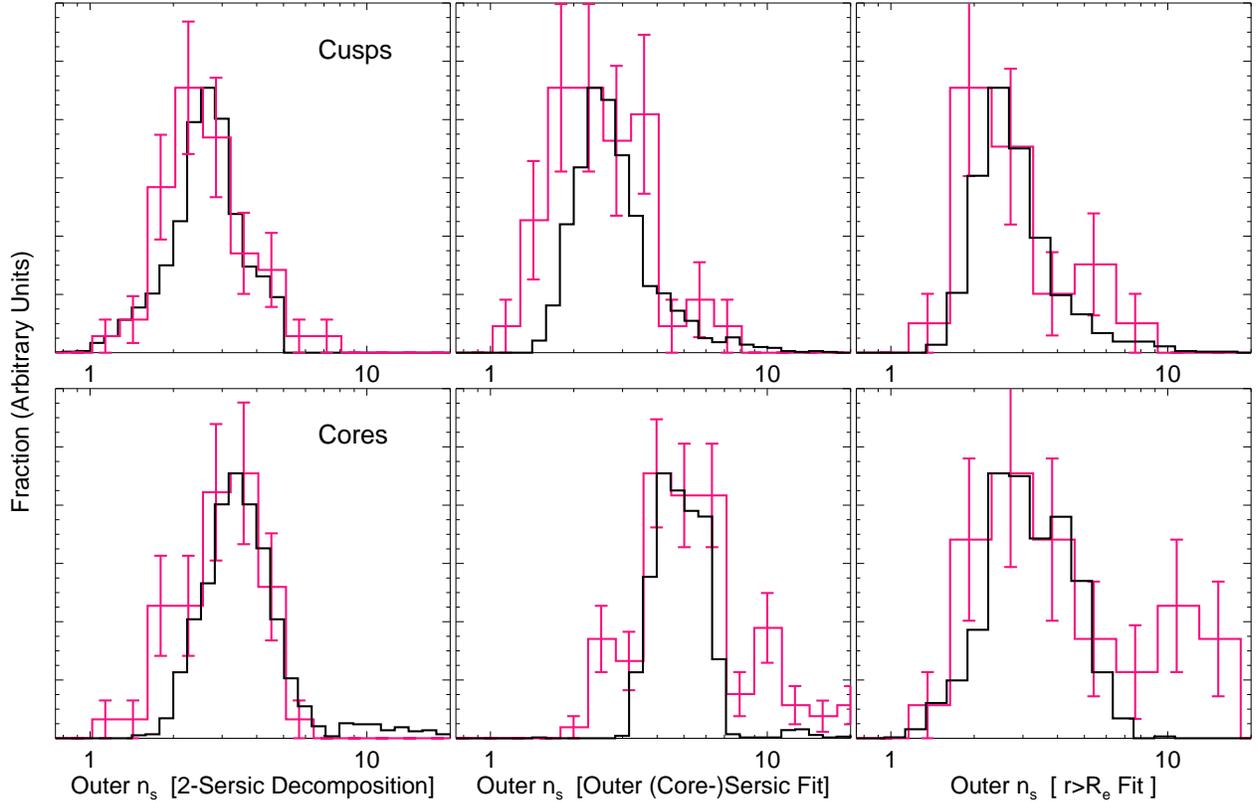}
    \caption{Distribution of outer Sersic indices in ellipticals, using 
    different methods for determining $n_{s}$, for cusp ellipticals and gas-rich merger remnant 
    simulations ({\em top}) and core ellipticals and spheroid-spheroid 
    re-merger simulations ({\em bottom}). Solid black lines show the results 
    for our entire sample of simulations (each across $\sim100$ sightlines). Colored lines show the 
    results for the combined observed samples of \citet{lauer:bimodal.profiles} and \citet{jk:profiles}, 
    with Poisson error bars. 
    {\em Left:} Outer Sersic index determined using our standard two-component decomposition. 
    {\em Center:} Outer Sersic index determined 
    following \citet{jk:profiles} or \citet{ferrarese:profiles} (fitting an ``outer'' Sersic index inwards, 
    including an extra component if there is some small-scale rise above it or allowing 
    a deficit or core-Sersic fit when there is a core or high outer $n_{s}$). 
    {\em Right:} Outer Sersic index determined by re-fitting only the observed profile data 
    at $r>R_{e}$ to a Sersic law. 
    Cuspy ellipticals 
    have a fairly narrow range of $n_{s}\sim2.7\pm0.8$, in good agreement 
    with gas-rich merger simulations, and relatively insensitive to the parameterization. 
    Core ellipticals also have a narrow range with a higher  $n\sim3.5-4\pm1.0$, in 
    good agreement with re-merger simulations, 
    when parameterized with our two component fit or fitted directly at large radii -- however, 
    the issues in the choice of fitting function described in \S~\ref{sec:missing.light} 
    make the estimated $n_{s}$ distribution more sensitive to the parameterization.
    \label{fig:ns.distrib}}
\end{figure*}

Given these different choices that can be made when estimating an observational 
Sersic index, we compare in Figure~\ref{fig:ns.distrib} the distribution of Sersic 
indices for cusp and core ellipticals, obtained with different methods. 
First, we show the distribution of Sersic indices measured from 
our two-component decomposition, 
in both the observed samples and our simulations. We compare 
the distribution extracted from our gas-rich merger simulations to the distribution 
fitted to the observed 
cusp elliptical profiles from \citet{jk:profiles} and \citet{lauer:bimodal.profiles}, 
and compare the distribution extracted from our re-merger 
simulations to that fitted to the observed core ellipticals from the same samples 
(with the same method applied). Because (as demonstrated 
in \papertwo) the fitted outer Sersic indices can be quite sensitive to the 
photometric depth and systematics between different observations, we restrict ourselves to a 
robust subsample described in \papertwo: specifically, those ellipticals for 
which we have $>3$ independent sources of photometry that agree 
to within $\sim20\%$ on the fitted value of $n_{s}$. We note that we 
obtain identical results (but with larger scatter) if we consider our 
entire samples. 

We compare these distributions with what we would obtain using two other methods 
to estimate the outer Sersic index. 
First, we follow \citet{jk:profiles} and fit a single outer Sersic profile inwards 
from large radii until the 
central profile begins to break upwards or downwards from this extrapolation, 
as described above (the comparison in the core ellipticals is similar if we fit the 
entire profile with a core-Sersic distribution). 
Second, 
we attempt to directly fit just the outer radii -- we first fit a single $r^{1/4}$ law to estimate 
the effective radius, then consider only the profile at radii larger than $\sim0.5-2.0\,R_{e}$ 
(within this range, our results are not dramatically sensitive to the exact cut). 
We then fit a single Sersic law just to these points. 
This limits our 
observational samples to  
only those systems 
with a large number of observed points at $r\gg R_{e}$ (where such a fit is possible). 

Applied to the cusp ellipticals, 
these methods yield similar distributions, all with median 
$n_{s}\sim2.5$ and similar scatter ($\pm0.7-0.9$), as expected from 
our previous comparisons. Applied to core ellipticals, the 
Sersic index distribution is, as expected, more 
sensitive to the method adopted. Fitting directly only to 
large radii yields a roughly similar distribution in $n_{s}$ to our two-component 
model, with typical $n_{s}\sim3-4$ (perhaps somewhat larger). 
There is more scatter (dispersion $\Delta n_{s}\sim1-1.5$) -- this relates 
to two effects. First, our restriction to radii $r\gtrsim R_{e}$ limits the number of 
data points and dynamic range of the fits, increasing the associated uncertainties. 
Second, the fact (as noted above) that the rise in outer $n_{s}$ 
(in our two-component model) in a re-merger 
is primarily driven at large radii from scattered stars, not by any increased 
density at small radii (which still behaves like a lower-$n_{s}$ system). 
Fitting only the large radii is therefore not restrained by the curvature and lack 
of sharply rising density at small $r$, and is more sensitive to the 
increased scattering of stars into the envelope at large radii. 
Adopting the method of fitting a single Sersic profile with a downward break 
from \citet{jk:profiles} or fitting core-Sersic models \citep{graham:core.sersic,
ferrarese:profiles} 
gives a higher typical $n_{s}\sim4-5$ and 
scatter $\Delta n_{s}\sim2-3$, reflecting the physical behavior outlined above. 

In any case, our simulated distributions and the observed Sersic index distributions 
match well when the same methods are applied to determine 
$n_{s}$. There is a significant (albeit not dramatic) and physically meaningful offset in the 
Sersic indices of gas-rich merger remnants and re-merger remnants (again, 
regardless of methodology used to estimate $n_{s}$), 
which appears to be reflected in the Sersic index distributions of 
cusp and core ellipticals. This 
supports the notion that there is some difference in the populations, 
in the sense expected for a relatively small number of major dry re-mergers.

\breaker
\section{The Appearance of ``Missing'' Light: A Caution}
\label{sec:missing.light}

As we discussed in \S~\ref{sec:profile.evol} and \S~\ref{sec:outer.sersic}, 
core ellipticals are sometimes referred to as having 
``missing light'' or ``mass deficits'' at small radii 
(generally below the size scales of the extra light of interest 
to us here), which has been associated with the action of 
black hole ``scouring'' scattering stars from the nucleus. 
We have argued that such scouring should not make a difference 
to the scales of interest for our inferences of the dissipational content in these 
objects. 
Having said that, however, the co-existence of ``extra'' and ``missing'' light 
(i.e.\ the presence of cores in the central profiles of ellipticals dominated, 
at their center, by the dense remnants of dissipational star formation) 
can mean that some additional care is needed when observers attempt to 
estimate the amount of ``missing'' light (i.e.\ how much stellar mass 
has been kicked out from small radii in order to form the central core). 
The problem is that one desires 
a method to empirically determine where (and by how much) the profile 
has been modified by scouring. Of course, any such estimate requires some 
knowledge of what the profile shape was {\em before} the scouring took 
place (roughly, the pre-``dry merger'' profile of the galaxy). If 
the galaxy profile were simply one-component, then provided some physical 
model for the profile or robust constraints on the profile shapes of 
progenitor (non-scoured) galaxies, this would be straightforward. 
But the profile of the progenitors is in fact multi-component -- so the question 
arises, can (and if so, by how much does) the amount of ``extra light'' or 
dissipation in the galaxy bias the estimators of scoured/``missing'' light?

Our comparisons in \S~\ref{sec:outer.sersic} should lend 
some insight to this discussion. 
In Figure~\ref{fig:demo.ml.appearance}, for example, we consider the 
profiles of re-merger remnants fit to both our two-component 
decomposition and to (instead) a core-Sersic distribution or an outer 
single Sersic distribution down to radii including the outer portions of the 
dissipational/extra light component. 
If we were to assume that the true ``pre dry-merger'' profile of 
the galaxies was given by the inwards extrapolation of whatever fitting law 
we chose (or, in the case of the core-Sersic law, by the inwards extrapolation of 
the ``Sersic'' part of the law, without the central flattening implied by the ``core''), 
then the application of the core-Sersic law to the 
simulations in Figure~\ref{fig:demo.ml.appearance} yields some 
apparent ``missing light'' in their nuclei (i.e.\ they fall below the inward extrapolation 
of the outer Sersic indices from these particular fitting functions). 
This happens not because of 
scouring in our simulations (recall, we do not resolve the scales where initial 
cusps would be transformed into cores), but because the outer Sersic indices 
of the core-Sersic (or limited single Sersic) fits are quite large, and large Sersic 
indices extrapolate (as $r\rightarrow0$) to very high central densities -- 
even above the true densities of the dissipational central components (which 
themselves drive the fits towards high $n_{s}$ values, as described 
in \S~\ref{sec:outer.sersic}). In short, 
the estimate of the ``pre-merger'' (or ``un-cored'') profile at small radii is biased to higher 
values than the true ``pre-merger'' profile. 

Physically, two things contribute to this bias (see \S~\ref{sec:outer.sersic}): 
the re-merger smooths the difference 
between the extra light (which rises to higher surface densities at $r<R_{e}$) 
and dissipationless/outer component, leading to some of it (potentially) being 
included in the fit; also, the re-merger scatters more stars out 
to large radii (by $\sim0.4$\,dex), 
broadening the overall profile and raising the outer $n_{s}$. 

This can yield a bias because of the nature of the Sersic profile -- 
if we fix the profile normalization at $R_{e}$, then changing $n_{s}$ from, for example, 
$n_{s}=3$ to 
$n_{s}=6$ yields relatively little change at moderate radii (amounting to a factor 
$\sim2-3$ higher surface brightness at $r=0.1\,R_{e}$ and $r=5\,R_{e}$), 
but as $r\rightarrow0$ the profiles diverge strongly. An 
$n_{s}=3$ profile contains $\sim0.1-0.2\%$ of its mass within $0.01\,R_{e}$,
but an $n_{s}=6$ profile contains $\sim1-2\%$, an order of magnitude difference. 
If the nuclear profile shape within $\sim0.01\,R_{e}$ were flat in both cases, 
one would therefore infer an order of magnitude more 
``missing light'' in the higher-$n_{s}$ case. 
In fact, all the observed cases 
where the ``missing light'' fraction is 
estimated to be as high as $\gtrsim1\%$ ($\sim10-100\,M_{\rm BH}$) 
have high outer Sersic indices $n_{s}\sim6-10$ 
\citep[see][]{ferrarese:profiles,cote:virgo,jk:profiles}. The physical concern is that 
these Sersic indices are much higher than those of typical ``cusp'' ellipticals 
(with slopes corresponding to $n_{s}\sim2-4$ in their central regions) -- 
in other words, the assumed ``pre-scouring'' profile in these cases is rising 
much more steeply as $r\rightarrow0$ than is actually observed in the 
presumed progenitors of core ellipticals. 

This is not to say that the presence of central light deficits is 
meaningless -- simply that, with this methodology, the inferred amount of missing light at 
small scales ($\sim R_{\rm BH}$) can be 
biased (generally towards higher values) by features on large scales ($\sim R_{e}$). 
A more robust null hypothesis (i.e.\ estimator of what the 
nuclear profile would be in the absence of scouring) may be obtained if 
we assume the ``pre-merger'' nuclear profile rises about as steeply as observed in typical 
cusp ellipticals. 

For example, if we examine the profile of just the inner dominant component 
(the ``extra light'') in cusp ellipticals, we expect that the median non-scoured distribution 
would rise about as steeply as an $n_{s}\sim3.6$ profile (or a 
Nuker profile with inner logarithmic slope $\gamma\sim0.7-0.8$) once well interior 
to the effective radius of the extra light. In core ellipticals, the observed nuclear profile 
at $r\ll R_{\rm extra}$ is flatter, more like median $n_{s}\sim1.6$ (or Nuker slope $\gamma\sim0.2$) 
-- this, of course, is why they are cores.  
Taking the typical nuclear shape in cusp ellipticals as our null hypothesis, 
we obtain characteristic missing light fractions $\sim 0.5-3\times10^{-3} M_{\rm gal}$, 
comparable to the black hole mass and 
expectations from scouring models. Alternatively, 
\citet{lauer:massive.bhs} estimate the scoured 
mass as that within $r_{\gamma}$ (defined as the radius where the 
logarithmic slope $\gamma=1/2$, which is reasonably representative of 
the radius in which scouring has significantly flattened the profile), 
giving a median value $\sim2.4\,M_{\rm BH}$. 

This is important because the high values sometimes inferred for 
the scoured mass fraction or ``missing light''  ($\gtrsim0.01\,M_{\rm gal}$) 
would actually present a challenge to the black hole scouring models -- 
those models predict a scoured core mass $\sim M_{\rm BH}$, but 
in some cases these empirical estimators yield scoured masses $\sim10-100\,M_{\rm BH}$. 
Our simulations demonstrate that these extreme observational estimates can 
probably be explained as artifacts of the fitting functions chosen, in cases where 
the outer $n_{s}$ is large. Qualitatively, this is not expected to alter 
the classification of cusp and core ellipticals; but if a robust {\em quantitative} estimate of the 
scoured or core mass is desired, then care is needed in empirically evaluating these cases. 
We find that if we adopt 
estimators less sensitive to the profile at large radii, 
these high values come down, more in line 
with the expectations from scouring models.

\breaker
\section{Discussion}
\label{sec:discuss}

In \paperone\ and \papertwo, we 
demonstrated that ``extra light,'' in the sense of stellar populations 
formed from dissipation in e.g.\ a
gas-rich merger induced starburst, is ubiquitous in cusp ellipticals 
and gas-rich merger remnants, and studied how its properties 
are related to and (in some cases) drive those of the galaxy. 
Here, we show that this dissipational component is expected to survive 
subsequent re-mergers of those ellipticals, even major gas-poor 
spheroid-spheroid mergers, in the sense that it will continue to 
contribute substantially to the central light profile and can be empirically 
recovered. We apply this to a large observed sample of 
ellipticals with central cores (i.e.\ flattening of their light profiles 
within the central $\sim30-50\,$pc), and show that 
they are consistent with a surviving 
dissipational ``relic extra light'' component 
which our adopted empirical fitting machinery and comparison 
with simulations allow us to recover and interpret. 

\subsection{Comparing Simulations and Observations:
Empirical Decomposition of Light Profiles}
\label{sec:discuss:separating}

In \papertwo, we argued that stars in cuspy ellipticals/gas-rich merger remnants 
should be separated into at least two distinct populations. 
First, stars which are formed in the disks (or otherwise in extended distributions in 
progenitor galaxies) before the final merger and coalescence 
of the progenitors. The final merger scatters these stellar orbits and they 
undergo violent relaxation. They dominate the light, even in highly gas-rich 
merger remnants, outside of $\sim0.5-1\,$kpc, and form a Sersic-law profile 
owing to their partial violent relaxation. Second, 
the starburst or dissipational population, 
formed in the central gas concentration in the final merger. 
This component is compact, and 
dominates the light inside a small radius $\lesssim0.5-1\,$kpc. 
These stars {\em do not} undergo 
significant violent relaxation, but form in a nearly fixed background potential 
set by the dissipationless component of the galaxy.
\footnote{There is also a third 
component present in simulations and important for 
observed kinematics but not prominent in light profile fitting: 
gas moved to large radii temporarily either by feedback or tidal effects, which 
settles into the relaxed remnant and re-forms small rotationally supported 
components \citep[embedded disks, kinematically decoupled cores, etc.; e.g.][]{hernquist:kinematic.subsystems,hopkins:disk.survival}.}
We developed, tested, and studied in detail 
in \paperone\ a two-component fit decomposition that, in simulations, 
could reliably extract the properties of these two physically distinct components.

Here, we demonstrate that, to lowest order, the two primary components 
structuring the surface brightness profile are expected to survive in 
dissipationless spheroid-spheroid re-mergers, in the 
sense that the physical starburst component still forms a more compact distribution 
that dominates or contributes significantly to the profile at small radii. 
Re-mergers will generally expand or puff up the system by a factor $\sim2$ 
in size, and smooth out the profile by mixing stellar populations and scattering 
stars by a typical factor $\sim0.4\,$dex in radius (see Figure~\ref{fig:rf.of.ri}) -- 
this can make fitting the 
profile more sensitive to the prescription adopted, 
and may smooth an obvious break in the profile around the transition 
from dissipational to dissipationless components 
(Figures~\ref{fig:highlownsdemo}-\ref{fig:demo.ml.appearance}), 
but does not 
fundamentally remove the extra light in a physical sense. 

We apply our same fitting procedure to simulated re-merger remnants, and 
find that, despite the re-merger, it is still able to statistically
recover the physically meaningful components of the original, 
spheroid-forming gas-rich merger (i.e.\ the original dissipational or starburst 
component and the original dissipationless or pre-starburst component; 
see Figures~\ref{fig:demo.rem.appearance}-\ref{fig:rem.recovery}). 
In other words, even after a re-merger, the surface brightness profiles of 
ellipticals retain information about the gas fractions and starburst mass 
fractions of their original gas-rich mergers. Our parametric fitting form is 
simple.  We consider the sum of two Sersic laws: an inner extra light component, 
for which a fixed $n_{s}=1$ works best in a mean sense when the data are 
not especially constraining in the central regions, but for which $n_{s}$ 
can be freed if the data are of sufficient quality; and an outer 
component with a free Sersic index. We explicitly demonstrate that this 
approach is successful even when (as is common in re-merger 
remnants) the profile is smoothed by re-mergers and 
obvious breaks might not be present in the profile 
(Figures~\ref{fig:highlownsdemo}-\ref{fig:dif.fit.qual.hists}).

We apply this to a large sample of ``core'' ellipticals, which (it is often 
argued) have been modified by re-mergers, and find that it is reliable and 
implies significant ``extra light'' (by which we mean 
the remains of the original 
gas-rich component) in almost all cases 
(Figures~\ref{fig:shoulders}-\ref{fig:lauerpp3}). 
We also match each of the observed profiles to our library of
simulations (of both gas-rich mergers and re-mergers) 
-- i.e.\ directly find the simulation mass profile which
most closely resembles that observed, and consider non-parametric
estimators of the mass in a central starburst component. We find that
in all cases we have simulations which provide good matches to the observed
systems, comparable to the typical point-to-point variance inherent
in the simulation surface brightness profiles
($\dmu\lesssim0.1$). The physical starburst
components in these best-fitting simulations are closely related to
those that we fit directly to the observed profiles, lending further
support to our attempt to physically decompose the profiles (Figure~\ref{fig:extra.vs.sb}). 
Where available, stellar population models including extended 
star formation and a subsequent burst independently 
support our inferred starburst mass fractions (Figure~\ref{fig:extra.vs.sb}). 
Likewise, metallicity, age, and abundance gradients, where 
available, support our decompositions, typically demonstrating a 
smooth transition to a younger, more metal rich population 
at the radii where the dissipational component begins to dominate 
the profile, as predicted in \papertwo\ (Figure~\ref{fig:z.grad.demo}). 
A complete list of fit parameters and compiled galaxy properties 
is included in Table~\ref{tbl:core.fits}.

\subsection{Predictions and Observations}
\label{sec:discuss:predictions}

Fundamentally, we argue that 
{\em all ellipticals -- including those with central cores -- 
are ``extra light'' ellipticals (dissipational systems)}, insofar 
as ``extra light'' refers to a component originally formed 
dissipationally on top of a background of dissipationlessly scattered 
envelope (a physical two-component nature).  Ellipticals with cores
show just as much dissipation, at a given mass, as cuspy or power-law ellipticals. 
Over the core population, 
the mass contributed by a central starburst or dissipational population can 
vary substantially, but in a physical sense, it always exists in addition to 
the pre-starburst or violently relaxed stellar populations. 
In terms of their dissipational or starburst population properties, we 
demonstrate that core ellipticals appear to be, for the most part, a continuous 
extension of the cusp population. 
The correlations obeyed by the inferred dissipational component itself are continuous and 
agree where there is overlap in galaxy properties between the 
cusp and core populations, as expected in physical 
models where core ellipticals are formed by re-mergers of cusp elliptical 
progenitors. 

Specifically, we find: 
\\

{\bf (1)} {\em The mass fraction in the dissipational or starburst component 
of both cusp and core ellipticals is a strongly decreasing function of 
mass} (Figures~\ref{fig:fgas.needed}-\ref{fig:mass.vs.fgas}). 
In detail, the mean starburst mass fractions can be approximated 
as Equation~(\ref{eqn:fgas.m}): 
\begin{eqnarray}
\nonumber & & 
\langle f_{\rm starburst} \rangle \sim {\Bigl[}1+{\Bigl(}\frac{M_{\ast}}{10^{9.15}\,\msun}
{\Bigr)}^{0.4}{\Bigr]}^{-1}, 
\end{eqnarray}
with a factor $\sim2$ scatter at each mass. The trend is similar for 
both cusp and core ellipticals, and the gas fractions needed 
span a range bracketed by the typical observed
gas fractions of spiral galaxies at the same mass, at $z=0$
(bracketing the low end of the required gas fractions) and $z\sim2-3$
(bracketing the high end). Core ellipticals have not only preserved their 
dissipational content, but its fraction reflects that of their ultimate progenitor 
disks. 
\\

{\bf (2)} At each mass, the degree of dissipation strongly affects the 
sizes of the remnants. In both observations and simulations {\em we
demonstrate a tight correlation between effective (half-light) radius
at a given stellar mass and the inferred dissipational/extra light
fraction} (Figure~\ref{fig:re.sigma.cusp}). 
This owes to the compact nature of the central dissipational
component -- increasing the 
mass fraction in this component means that the half-light radius must be
smaller.  The correlations obeyed between dissipational content and size 
at fixed mass are similar for cusp and core ellipticals. 

Despite the fact that re-mergers puff up remnants by a factor $\sim2$, 
their masses also double, so 
this only moves them by about $\sim0.1\,$dex off the mean size-stellar 
mass relation obeyed by gas-rich merger remnants (which has an 
intrinsic scatter $\sim0.3\,$dex). On the other hand, we demonstrate that 
changes in the original starburst or extra light fraction can alter 
the remnant size by nearly an order of magnitude at fixed mass. 
Thus, even if core ellipticals have undergone a moderate number of 
re-mergers, we expect that the degree of dissipation in their original 
formation should still be the most important factor setting their 
sizes today, and we demonstrate this in the observations. 
\\

{\bf (3)} Re-mergers roughly preserve the size-mass relation 
of the dissipational/starburst component 
itself, approximately set by the radius at which the starburst component 
becomes self-gravitating ($G\,M_{\rm extra}/R_{\rm extra} \approx G\,M_{\ast}/R_{e}$); 
and we do find that the core ellipticals and cusp ellipticals 
obey nearly the same correlation (Figure~\ref{fig:sizes}). 
If re-mergers preferentially puff out 
low binding energy material (i.e.\ material at large initial radii), then 
we might expect that the quantity 
($G\,M_{\rm extra}/R_{\rm extra} \approx G\,M_{\ast}/R_{e}$) should 
increase slightly in a re-merger (by a maximum factor $\sim2$ in a 
1:1 merger, if the inner component 
is not expanded at all, and the outer component expands by a factor of $2$). 
We see tentative evidence for such an offset, but it is small (a factor $\sim1.4-1.5$; 
Figure~\ref{fig:size.ratios}). 
\\

{\bf (4)} In \papertwo\ we demonstrated that the extra light component
gives rise to stellar population gradients in the 
remnant. We show here that these gradients are only weakly affected 
by re-mergers, in agreement with observations which find that cusp and core 
ellipticals do not show significant differences (at fixed galaxy properties) 
in their gradient strengths (Figure~\ref{fig:grad.fx.demo}-\ref{fig:grad.correlations}). 
To the extent that both radius and gradient strength 
scale with the contribution of the original dissipational component at fixed mass, 
we find that our prediction from \papertwo\ should also hold for core ellipticals 
(namely that at fixed mass, smaller ellipticals should, on average, exhibit 
stronger metallicity gradients), and early observational evidence appears to 
support this \citep[e.g.][see also Figure~\ref{fig:z.grad.demo}]{mehlert:ssp.gradients,
sanchezblazquez:ssp.gradients,reda:ssp.gradients}. 
There are other lines of evidence for the survival of extra light: 
\citet{lauer:centers} find that 
nuclear color gradients, ellipticities, and isophotal twists 
in the centers of cusp and core 
galaxies are continuous and trace similar distributions 
as a function of mass and luminosity. 
Analogous correlations should hold with integrated stellar 
populations, which we discuss in \papertwo, 
provided that these are correlated with the gas fractions of 
the pre-merger disks. 
\\

{\bf (5)} Given the appropriate dissipational mass fraction inferred 
from fits to the observed surface brightness profiles, simulated re-merger 
remnants reproduce the global kinematics and isophotal shape distributions 
of core ellipticals (specifically rotation $V/\sigma$, ellipticity, and boxyness $a_{4}/a$), 
and their mutual correlations (Figures~\ref{fig:kinematics.cuspcore}-\ref{fig:kinematics.histograms}). 
These distributions are {\em only} reproduced by simulations with matched 
dissipational mass fractions: systems with too little dissipation (lacking a 
high central density to make the potential more round and disrupt box orbits) 
are too elliptical, and systems with too much dissipation remain too disky and 
rapidly rotating even after several re-mergers (Figures~\ref{fig:kinematics.before.after} \&\ 
\ref{fig:kinematics.histograms}). More detailed comparison of simulations with 
full two-dimensional velocity fields, along with higher-order measures of the velocity 
field at a given point, are needed to robustly discern whether scenarios as simple 
as a single re-merger are viable formation mechanisms for slowly-rotating, boxy 
ellipticals, but similar dissipational content appears to be a basic prerequisite. 
\\

{\bf (6)} We predict and find 
that the outer Sersic indices of ellipticals 
with cores (in our two-component decompositions)  
trace a roughly constant distribution with median $n_{s}\sim3-4$ and scatter 
$\Delta n_{s} \sim1$, without a strong dependence on galaxy mass, 
effective radius, or other properties (Figure~\ref{fig:ns.mass}). 
We found similar results in \papertwo\ for the cusp elliptical population, 
but with a lower median $n_{s}\sim2-3$. The difference arises naturally 
in our re-merger simulations, as the scattering of stars in subsequent mergers 
will tend to broaden the outer regions of the light profile, leaving a less 
steep falloff and giving rise to a higher Sersic index. That there is 
an offset in the outer Sersic index distribution of these populations suggests 
that indeed some sizable fraction of core ellipticals have experienced 
a re-merger, but that the offset is relatively small suggests that there have 
not been a large number of such re-mergers. 

We emphasize that these Sersic indices are not directly comparable 
to those in previous studies, which fit different functional forms to the 
light profile or fit e.g.\ a Sersic profile to the entire light profile (including the 
extra light). If we fit our simulations to a single Sersic index or core-Sersic 
profile where it is the formal best fit, we recover a stronger dependence 
of Sersic index on galaxy mass, luminosity, or effective radius, similar to 
earlier claims 
\citep{caon:sersic.fits,prugniel:fp.non-homology,
graham:bulges,trujillo:sersic.fits,ferrarese:profiles}: our models 
are consistent with the results derived using these methods, 
however this does not strictly 
trace the physical dissipationless component of the galaxy. In fact, 
we show in \S~\ref{sec:outer.sersic} 
(Figures~\ref{fig:highlownsdemo}-\ref{fig:dif.fit.qual.hists}) 
that re-merger 
remnants are often formally well-fit by core-Sersic laws (Sersic 
profiles with a central deficit), in agreement with observations of 
core ellipticals. These cases obtain high Sersic indices, driven 
in part by the presence of a dissipational component at small radii, 
raising the central surface brightness (Figure~\ref{fig:highlownsdemo}). 
These functional forms are a precise formal parameterization 
of the light profile, but the best-fit parameters determined in this manner 
reflect the combination of the dissipational and dissipationless 
outer components, and do not trivially translate to physical 
descriptions of the components of the galaxy. 

Indeed, as demonstrated in \papertwo, 
many of the results
in studies using these fitting approaches 
are actually driven by a dependence of extra light on galaxy 
properties. 
The meaningful dependence of outer Sersic index 
as we quantify it (tracing the dissipationless component) 
on mass owes to the dependence of typical merger history on 
mass (dwarf spheroidals and pseudobulges will have lower 
typical $n_{s}$ than gas-rich merger remnants, which 
themselves have lower typical $n_{s}$ than re-merger 
remnants, and these populations, on average form an 
increasing sequence in mass). 
For a broadly similar merger history (in terms of e.g.\ number of 
major mergers), the profile of the true dissipationless 
component is expected to be self-similar, as we 
recover (for both the cusp and core populations), since it is determined 
purely by gravity.
\\

These and other correlations argue that core ellipticals should be 
properly thought of as containing ``extra'' or dissipational components 
in the same physical (although not necessarily observational) manner as 
cusp ellipticals; their central, high density regions formed 
via dissipational processes. 
We therefore refer to \papertwo\ for a number of other proposed 
observational tests of the models herein, 
which should hold for core ellipticals as well. 
This is a critical test of the merger hypothesis 
and supports the notion that even massive, slowly rotating, boxy 
ellipticals with cores were originally formed in gas-rich major mergers, 
albeit potentially modified by subsequent re-mergers. 

Points {(1)} and {(2)} emphasize that, even if a system has experienced 
re-mergers, and even in observed massive core ellipticals, dissipation 
is the key driver of galaxies along the fundamental plane (in terms 
of galaxy stellar mass). The degree of 
dissipation in the original gas-rich merger is the most important factor 
determining the size of the object and the ratio of baryonic to dark matter 
within the effective radius of the stellar light, and the correlations 
obeyed by the observed systems agree well with our simulations. 

We show that this degree of dissipation is a 
systematic function of mass, without a significant offset between 
cusp and core elliptical populations (as expected if it is set by the 
gas fractions of the progenitor disks, and there have not been a large 
number of re-mergers of individual objects). In other words, we 
are able to demonstrate that the amount of 
dissipation expected based on known disk gas fractions as a function of 
mass is precisely that needed to explain the extra light in the surface brightness 
profiles and to reconcile the densities and radii of disks and ellipticals 
as a function of mass. We investigate this further in \citet{hopkins:cusps.fp}, 
and show how it gives rise to the fundamental plane scalings and ``tilt.''
Once on the fundamental plane, our simulations and other experiments 
\citep{boylankolchin:mergers.fp,robertson:fp} find that 
re-mergers typically move more or less parallel to the fundamental plane, 
emphasizing that the tilt must arise in the initial, gas-rich 
spheroid-forming merger.

Although some of these effects will become increasingly
scattered or smeared out by a large number of re-mergers, multiple lines of 
evidence above suggest that the typical core elliptical has undergone 
relatively limited dry re-merging. 
If re-mergers are indeed the mechanism 
by which cores are formed, our comparisons suggest a small number 
$\sim1$\,major re-mergers per object since its formation in a gas-rich merger. 

This is in line with other observational indications from e.g.\ the mass function 
evolution of ellipticals and red galaxies \citep[e.g.][]{bundy:mfs,borch:mfs,
pannella:mfs,franceschini:mfs,fontana:highz.mfs} and direct observational estimates of the dry 
merger rate \citep{lin:merger.fraction,lin:mergers.by.type,
vandokkum:dry.mergers,bell:dry.mergers}. 
It is also generally expected in 
cosmological models \citep{delucia:ell.formation,zheng:hod.evolution,hopkins:groups.ell}, 
where only the 
most massive BCGs with $M_{\ast}\gg10^{12}\,\msun$ or so are expected 
to have a large number of major re-mergers. Even in these cases, it is not clear 
whether such mergers actually proceed efficiently, or whether the secondary 
is tidally destroyed before the merger and added as part of an extended envelope 
or intercluster light \citep{gallagherostriker72,monaco:sat.destruction,
conroy:sat.destruction,purcell:sat.destruction}. At these masses, the growth of the system 
may also be dominated by a large number of minor mergers, rather than 
a few major mergers \citep[major mergers 
being the dominant ``dry'' growth mode 
at $\lesssim$ a few $\lstar$; see][]{maller:sph.merger.rates,masjedi:cross.correlations}. 

Our constraints on this extreme of 
the population are, unfortunately, much weaker. Moreover, we do not intend our 
modeling to be extrapolated to this regime. Such systems formed, for the most 
part, at early times from potentially very different progenitors than those 
we model, with more complex subsequent merger histories our simulations do 
not capture
\citep[see, e.g.][]{li:z6.quasar}.
More detailed analysis of individual objects, and more detailed 
models incorporating a fully cosmological context are called for to study 
the formation history of galaxies in this regime. However, such systems 
constitute only a small fraction ($\lesssim10-20\%$) of the mass 
density in even the core 
elliptical population (and just $\sim3-5\%$ of the total mass density in 
ellipticals), so the vast majority of the stellar populations of 
spheroids should be reasonably represented by our modeling.

\subsection{Profile Fitting and Tests of Core Creation Models}
\label{sec:discuss:missing}

The existence of nuclear cores in these ellipticals is commonly attributed 
to ``scouring'' by a binary black hole in gas-poor re-mergers. This will 
eject stars in close encounters with the binary, flattening the initial 
steeply-rising power-law cusps into the observed shallow cores. To the 
extent that the stars in this initial cusp are ejected to larger orbits, leaving a 
core, these remnants can be thought of as having ``missing'' light 
in their centers.
We emphasize that this is completely consistent with there 
also being ``extra light,'' as we define it, in these galaxies. The ``extra light'' 
we refer to is 
the remnant of stars formed in dissipational central star formation events, 
and dominates the profile within rather large radii $\sim$kpc, blending 
smoothly onto the outer dissipationless profile. Scouring 
will flatten the nuclear peak or cusp of this dissipational remnant, 
but will not be likely to remove the $\sim10\%$ of the stellar mass 
that constitutes the dissipational starburst 
relic. Despite the terminology, ``extra'' and ``missing'' 
light are not mutually exclusive. In a strict physical sense, all ellipticals 
are dissipational/``extra light'' ellipticals -- 
they have all experienced some dissipational 
star formation. Of these dissipational ellipticals, there are  
``un-cored'' or ``un-scoured'' (``power-law,'' ``cusp,'' or ``cuspy core'') ellipticals 
(presumably objects whose last major merger was gas-rich, allowing the 
black holes to merge quickly, and forming the central cusp via new star formation) and 
``cored'' or ``scoured'' ellipticals (presumably the remnants of subsequent 
spheroid-spheroid ``dry'' re-mergers).  The ``missing light'' is a core -- a 
flattened nuclear profile as opposed to a 
continued steeply rising cusp -- within the very center of the dense dissipational 
component. 

Our methodology is robust to re-mergers and scouring: our two-component 
profiles still recover the dissipational component in simulations allowing for any 
reasonable model of nuclear scouring. However, we demonstrate that the 
same profile can be well-fitted by different assumed functional forms, 
and the parameters obtained from these fits -- particularly in regards to the 
``extra'' and ``missing'' light -- are not trivially related and require careful 
interpretation. 
In particular, observations find that core ellipticals can be reasonably
well-fit in a formal statistical sense by cored Sersic laws (an outer Sersic profile 
with an inner ``flattening'' or deficit), generally with high 
Sersic indices $n_{s}\gtrsim5$, and we obtain consistent results 
fitting our re-merger simulations with these methods. 
There is no explicit ``extra'' component in these fits: rather, 
in our simulations, we find that the high $n_{s}$ values -- which 
extrapolate to high central densities before the flattening or deficit -- 
implicitly reflect the extra light (that being the central, dense component 
that enables the high-$n_{s}$ fit; otherwise the system would have a 
high outer $n_{s}$ but a low central density, not resembling any Sersic profile). 
Because the re-merger has smeared out the light profile, removing 
characteristic features (or kinematic subsystems) 
that might have (in the original gas-rich merger 
remnant) been a more obvious indication of the transition to radii 
where stars formed dissipationally dominate the profile, 
the physical breakdown of these systems is not obvious from a 
strictly empirical standpoint. By calibrating different fitting methods 
with our simulations, we have attempting to provide an interpretive 
context and physical motivation for specific interpretations of 
these fits. 

In a related manner, this raises some cautions regarding 
the purely empirical methods sometimes used to 
estimate how a profile has been modified by core scouring -- 
i.e.\ how much mass is ``missing'' from the nuclear region 
(relative to the steep nuclear rise of the progenitor cusps). 
This is a subtle issue. There may of course be no light actually missing 
after scouring -- rather, stars are scattered from small radii to 
large, flattening the central profile (what is really desired is an estimate of 
how much stellar mass has been scattered -- i.e.\ how much stellar mass 
must be moved from the nuclear region to larger radii to explain the 
difference in profiles). So any such estimate is sensitive to the model 
for the nuclear profiles of the progenitors -- equivalently, what the profile 
would be in the absence of scouring. 

We explicitly consider one such (commonly adopted) model, where the 
implicitly assumed progenitor profile is based on the 
inwards extrapolation of an outer Sersic law (as in e.g.\ core-Sersic 
fits). In these fits, the effects described above (effects on the profile 
at larger radii than the core itself) can lead to a very large 
outer Sersic index in the formal best fit; in the high-$n_{s}$ regime ($n_{s}\gtrsim6$), 
the nature of the Sersic profile is such that these profiles then rise
steeply at small radii (more steeply even than the power-law nuclear 
profiles of cusp ellipticals), and a small change in the outer profile (raising $n_{s}$) 
can substantially raise the extrapolated (and assumed ``pre-merger'') inner profile. 
This can bias the estimate of the ``missing light'' fraction 
towards very high values ($\sim1-5\%$) and literally interpreted can make it appear 
as if the ``core'' extends to radii $\sim$kpc. In fact, we find that adopting 
this methodology in such cases, we would infer ``missing light'' at this 
level even in simulations where there is no 
physical scouring (i.e.\ there is no actual ``missing light''). Moreover, 
if real, these values would imply scoured masses $\sim10-50\,M_{\rm BH}$,
in troubling disagreement with models of scouring which predict 
scoured mass deficits $\sim0.1-1\,M_{\rm BH}$ per major re-merger 
\citep[see e.g.][]{milosavljevic:core.mass,
merritt:mass.deficit,sesana:binary.bh.mergers}. If stars from centrophilic orbits 
in triaxial potentials allow the binary to coalesce rapidly (providing 
stars from near $\sim R_{e}$, where the loss of even 
$\sim 10\,M_{\rm BH}$ makes no difference to the profile, to harden the binary), 
as suggested in some idealized calculations 
\citep{berczik:triaxial.bh.mergers,holley:triaxial.loss.cone} and 
merger remnant simulations \citep{hoffman:prep}, the difficulty for the 
models in explaining such large ``missing masses'' grows. 

Given the steep dependence of the implicitly assumed central progenitor 
profile on Sersic index in this regime, and the appearance of ``missing light'' 
even in simulations without scouring given a literal interpretation of this 
fitting methodology, it is likely that the ``problems'' for the scouring models 
in these extreme cases reflect more the observational uncertainty regarding 
the appropriate ``un-scoured'' profile, rather than fundamental uncertainties 
in the physics of scouring. 
Future work, as observations and models improve, should attempt to 
carefully model the progenitor profiles and develop estimators 
that are less sensitive to the profile shape at radii much larger than the core 
itself. An important test for any such 
estimators should be that re-merged progenitors with initially 
cuspy profiles (down to the desired resolution limits), without scouring included, 
should typically yield little or no ``missing light.'' 
We briefly consider a couple of possibilities for such estimators of the 
scoured mass (less sensitive to the large scale radii), and 
obtain (re-analyzing the observations) typical scoured mass 
estimates of $\sim0.5-3\,M_{\rm BH}$, in better agreement with 
scouring models. 

Further study, in particular observation of the nuclear regions $\sim R_{\rm BH}$ 
is needed to test models of scouring and test the accuracy of different 
estimators for the scoured stellar mass. For example, 
scouring is expected to preferentially eliminate stars 
on radial orbits and leave a bias for tangential orbits 
within the radius affected
\citep[e.g.][]{quinlan:bh.binary.tang.orbit.bias}. \citet{gebhardt:nuclear.anisotropies} see 
tentative evidence for this in a limited sample of ellipticals; the 
major-axis radii within which the effect appears are generally 
$\sim0.5-3\,R_{\rm BH}$, as expected in scouring theories 
and our revised inferences from core profiles. A preliminary 
comparison supports our conclusions (and caveats) here, but the number of 
relevant ellipticals is small. If similar observations can 
be obtained for a sample of ellipticals with alternative estimates of the missing mass 
fraction, it can be determined whether some estimators are biased towards 
putting too much or too little of the profile into the scoured component.

\subsection{Summary}
\label{sec:discuss:summary}

We have developed a paradigm to understand the
structure of both cusp and core ellipticals, in which there are
fundamentally two stellar components: a dissipational central
starburst component and a more extended violently relaxed component.
We have shown that the separation 
between these components can be inferred with observations
of sufficient quality, and used to understand the formation history of
ellipticals as a function of a wide range of properties. This allows
us to demonstrate that dissipation is critical to
understanding the properties of ellipticals, including (but not
limited to) the structure of their surface brightness profiles, their
sizes, ellipticities, isophotal shapes and rotation, age, color, and
metallicity gradients (and their evolution), and the gas content and
properties of their progenitors. 

In particular, we argue here 
that this remains true for ellipticals with cores and re-merger remnants -- 
in other words, {\em all ellipticals, including core ellipticals, are 
fundamentally ``extra light'' ellipticals}. The core ellipticals form 
a continuous family with the cusp ellipticals in terms of their 
dissipational content, as expected if these cusp ellipticals are their progenitors 
and if the number of re-mergers (if important) has not been 
large ($\sim1-3$ major re-mergers) for the typical core elliptical. 
We demonstrate that, despite the possibility that these systems 
have expanded via re-mergers, the degree of dissipation in the 
original gas-rich merger (the memory of which is retained in the 
surface brightness profile) remains the most important factor determining 
the size, gradient strength, and other properties of the remnant. 

We have studied the properties and identified robust trends 
of dissipational stellar remnants in the nuclei of elliptical galaxies with cores and 
remnants of gas-poor, spheroid-spheroid re-mergers, 
across a 
large set of simulations, in which we vary e.g.\ the galaxy masses, initial gas 
fractions, concentrations, halo masses, presence or absence of bulges, presence or 
absence of black holes, feedback parameters from supernovae and stellar winds, 
orbital parameters and disk inclinations, and mass ratios of the merging galaxies. 
This range of parameters allows us to identify the most important physics. 

As we found in \paperone\ and \papertwo, the most important factor determining 
the structure of the remnant (insofar as the properties we consider are concerned) 
is how much mass is in the original (in the original or last spheroid forming, 
gas-rich merger) dissipationless (violently relaxed) component, versus 
the mass fraction in the dissipational (starburst) component. 
Orbital parameters and initial galaxy structure can, in principle, 
affect the remnant surface brightness profile 
significantly, but only indirectly, insofar as these help to set
the amount of gas which will be available at the time of the final coalescence of the 
galaxy nuclei (i.e.\ how much mass ends up in the starburst component, as opposed to 
being violently relaxed in this final merger). 

Re-mergers will expand 
the original gas-rich merger remnants, and smooth the profile (scattering stars 
substantially about their mean final radii), but, as it is well established that 
dissipationless mergers conserve (in a mean sense) particle rank order in binding 
energy and therefore radius \citep{barnes:disk.halo.mergers}, they will preserve these 
components. In other words, the product of a re-merger is, to lowest order, 
the sum of the two progenitor spheroid dissipationless components (constituting 
the nearly self-similar outer violently relaxed component) and 
their inner dissipational/starburst components (constituting the 
central dissipational/starburst remnant component of the final re-merger 
remnant, despite the fact that there might be little or no new dissipation in the 
re-merger). 

We have demonstrated that this makes predictions for how fundamental
plane scalings arise, which we study further in \citet{hopkins:cusps.fp}.  We
make a wide range of new predictions for the distributions of these
properties and how they scale with the degree of dissipation, and how
they should scale with each other and various other observational
proxies for this degree of dissipation (which we define herein).
We have predicted and shown (given these proxies) that dissipation is
indeed more important (contributing a larger mass fraction) in
low-mass ellipticals, in line with expectations based on how gas
fractions are known to scale with disk mass.  Testing all of these
with better observations should be possible in the near future, with
well-defined samples of ellipticals and continued improvements in
mapping e.g.\ the surface brightness profiles, stellar populations and
their gradients, and structural properties of ellipticals over a wide
dynamic range.  

Given our decompositions, we observationally
confirm the long-standing prediction of the merger hypothesis, that
sufficient dissipation should have occurred in the inner regions of
ellipticals to explain the discrepancy between their central densities
and those of their progenitor spirals, a confirmation that fits well
in line with what is now well-established in gas-rich merger
simulations and is also directly seen in progress in ongoing/recent
mergers, which have (through clear recent central star formation)
raised their phase space densities to be comparable to ellipticals
\citep{kormendysanders92,Doyon94,
Genzel01,rothberg.joseph:kinematics,
tacconi:ulirgs.sb.profiles,dasyra:pg.qso.dynamics}. We 
show that this is true for the core elliptical population just as 
we demonstrated it for merger remnants in \paperone\ and 
cusp ellipticals in \papertwo. In other words, the {\em same} 
mechanism can explain the central densities in both cusp 
and core ellipticals -- while core ellipticals may be modified 
by subsequent re-mergers, no alternative formation mechanism 
for them (in the sense of formation of their spheroid progenitors) 
is required. 

We demonstrate some important caveats in 
observational studies of core ellipticals. 
These ellipticals are multi-component (dissipational plus dissipationless, in 
the physical sense above) galaxies, but 
they have shallow nuclear profiles (central cores) 
instead of steeply rising central cusps, and can therefore be thought of 
as ``missing light'' galaxies in the sense that some process (e.g.\ 
scattering of stars from a black hole binary) has flattened 
the nuclear profile slope to create the core. This is fundamentally different from 
the processes involved in the evolution of the ``extra light'' 
(by which we mean the remnant of the starburst/dissipational 
star formation) and they should not 
be confused -- whatever process forms cores likely acts on small scales and involves 
a relatively small fraction of the galaxy mass, 
much less than the $\sim$kpc scales and $\sim10\%$ of the galaxy mass 
that are characteristic of the dissipational component. 
Owing to the smoothing of the outer profile 
in a re-merger, re-merger remnants can often be better fit (in a purely formal sense) 
by e.g.\ single Sersic profiles with central deficits or core-Sersic laws: where 
observed, we argue that this is an indication of such re-mergers, but we stress 
that it does not mean the galaxies are not, in fact, two-component objects 
in a physical sense (simply that the fitted function reflects some combination of the 
two components). For this reason, 
the two-component profiles adopted here lend 
themselves to more direct physical interpretation and comparison with 
galaxy formation models. 
Furthermore, we demonstrate that care is needed when using such 
core-Sersic fits as an estimator of 
the ``scoured'' nuclear mass -- when the outer Sersic index is large, the 
inferred ``missing mass'' can be quite sensitive to the details of the 
profile at large radii. This may explain estimates of scoured masses 
which are much larger than those predicted by models of core creation. 

There are, of course, other changes to galaxy properties in re-mergers. 
As discussed in \paperone\ and \papertwo, gas which survives the original 
spheroid-forming gas-rich merger will quiescently settle into 
the galaxy and form kinematic sub-components (in particular, 
embedded disks and kinematically decoupled nuclear 
components). We demonstrated that these 
do not contribute significantly to the surface brightness profile, 
and therefore they are not evident in our analysis. However, 
\citet{cox:kinematics} and other numerical studies of 
gas-rich mergers \citep{naab:gas,burkert:anisotropy} 
have demonstrated that these subsystems can contribute strongly to 
the isophotal shapes (in particular driving the diskyness 
of the remnant) and kinematics, yielding 
distributions of shape and kinematic properties 
(including ellipticity, isophotal shape 
$a_{4}/a$, rotation $(v/\sigma)^{\ast}$, 
anisotropy, triaxiality, and 
kinematic misalignments) in good agreement with 
observed cusp or disky, rapidly rotating $\sim\lstar$ ellipticals. 

The evolution of these sub-systems in re-mergers 
is not readily apparent in the surface brightness profiles of the 
remnants -- their contribution to the profile is much less than the 
smoothing effects and variations introduced by the re-merger 
(shown in \S~\ref{sec:profile.evol}). However, 
their evolution in re-mergers is briefly 
discussed in \S~\ref{sec:kinematics} and will be studied in detail in 
\citet{cox:remerger.kinematics}, who show that they are 
generically destroyed by major spheroid-spheroid re-mergers. 
This explains the significant effects on the global kinematic 
properties of re-merged ellipticals shown in \S~\ref{sec:kinematics} -- making 
them rounder, boxier, and more slowly rotating, in 
agreement with the observed properties of massive 
core or boxy, slowly rotating ellipticals. But effects on the velocity field 
in detail (as a function of radius, at higher order in the velocity moments, 
and in spatial distribution rather than just azimuthal average) 
will be more pronounced, and present opportunities not just to test whether 
or not re-mergers are good analogs to boxy/slowly-rotating ellipticals, 
but to distinguish the extent to which substructure and kinematic subcomponents 
contribute to these rotation/isophotal shape properties, as opposed to 
e.g.\ global angular momentum content or features of the spheroid potential. 
Kinematic misalignments, kinematically decoupled subsystems, 
triaxiality, and trends of isotropy with radius are also 
more sensitive to orbital parameters 
\citep[see e.g.][]{cox:kinematics,boylankolchin:dry.mergers,
jesseit:kinematics}
and may distinguish merger histories 
with preferences for certain orbital configurations (possible if e.g.\ massive 
galaxies accrete minor companions preferentially along filaments).
These more detailed, 
``second-order'' structural parameters may therefore 
represent a more sensitive probe of the re-merger history (i.e.\ the 
history of subsequent mergers after the original, spheroid 
forming gas-rich merger), while the ``first-order'' 
parameter (the light profile) retains a memory of 
the degree of dissipation imprinted in the {\em original} 
gas-rich merger.

\acknowledgments We thank 
Marijn Franx, 
Josh Younger, and
Barry Rothberg 
for helpful discussions and contributed data sets used in this paper. We also 
thank the anonymous referee for helpful advice on the 
content herein. This work
was supported in part by NSF grants ACI 96-19019, AST 00-71019, AST 02-06299, 
and AST 03-07690, and NASA ATP grants NAG5-12140, NAG5-13292, and NAG5-13381. 
Support for TJC was provided by the W.~M.\ Keck Foundation.
JK's work was supported in part by NSF grant AST 06-07490.

\bibliography{/Users/phopkins/Documents/lars_galaxies/papers/ms}

\clearpage
\longtabler
\begin{\tableset}{lccccccccccccc}
\tablecolumns{14}
\tabletypesize{\scriptsize}
\tablecaption{Fits to Core Ellipticals\label{tbl:core.fits}}
\tablewidth{0pt}
\tablehead{
\colhead{Name} &
\colhead{Ref.} &
\colhead{${\rm N}_{\rm phot}$} &
\colhead{$M_{\ast}$} &
\colhead{$M_{V}$} &
\colhead{$\sigma$} &
\colhead{$R_{e}$} &
\colhead{$\epsilon$} &
\colhead{$100\,a_{4}/a$} &
\colhead{$(v/\sigma)^{\ast}$} &
\colhead{$n_{s}$ (fit)} &
\colhead{$n_{s}$ (sim)} &
\colhead{$f_{e}$ (fit)} & 
\colhead{$f_{sb}$ (sim)} \\
\colhead{(1)} &
\colhead{(2)} &
\colhead{(3)} &
\colhead{(4)} &
\colhead{(5)} &
\colhead{(6)} &
\colhead{(7)} &
\colhead{(8)} &
\colhead{(9)} &
\colhead{(10)} &
\colhead{(11)} &
\colhead{(12)} &
\colhead{(13)} &
\colhead{(14)} 
}
\startdata
NGC 0584 & 2,3 &  5 & $11.27$ & $-21.91$ & $217$ & $3.39$ & $0.30$ & $ 1.50$ & $1.55$ & $2.92^{+0.05}_{-0.19}$ & $2.77^{+0.45}_{-0.20}$ & $0.052^{+0.004}_{-0.019}$ & $0.103^{+0.191}_{-0.066}$ \\
NGC 0720 & 2 &  4 & $11.64$ & $-22.20$ & $247$ & $6.31$ & $0.39$ & $ 0.35$ & $0.23$ & $1.82^{+0.81}_{-0.20}$ & $2.39^{+1.07}_{-0.34}$ & $0.177^{+0.043}_{-0.010}$ & $0.096^{+0.131}_{-0.061}$ \\
NGC 0741 & 2 &  1 & $12.25$ & $-23.27$ & $293$ & $13.18$ & $0.15$ & -- & -- & $3.10$ & $2.41^{+0.69}_{-0.08}$ & $0.026$ & $0.076^{+0.030}_{-0.040}$ \\
NGC 0777 & 3 &  3 & $11.65$ & $-22.92$ & $348$ & $7.17$ & -- & -$ 0.20$ & $0.28$ & $2.33^{+0.16}_{-0.16}$ & $2.98^{+1.94}_{-0.82}$ & $0.118^{+0.006}_{-0.006}$ & $0.076^{+0.105}_{-0.047}$ \\
NGC 1316 & 2 &  1 & $12.27$ & $-23.32$ & $250$ & $7.24$ & $0.37$ & $ 1.00$ & $0.91$ & $2.11$ & $2.48^{+0.79}_{-0.76}$ & $0.075$ & $0.105^{+0.114}_{-0.035}$ \\
NGC 1374 & 2 &  1 & $10.71$ & $-20.57$ & $207$ & $2.57$ & $0.11$ & -- & -- & $4.37$ & $1.95^{+1.32}_{-0.35}$ & $0.051$ & $0.194^{+0.115}_{-0.092}$ \\
NGC 1399 & 2 &  4 & $11.56$ & $-22.07$ & $359$ & $4.17$ & $0.12$ & $ 0.10$ & $0.25$ & $2.61^{+0.08}_{-0.11}$ & $3.07^{+0.44}_{-0.26}$ & $0.142^{+0.004}_{-0.002}$ & $0.108^{+0.127}_{-0.048}$ \\
NGC 1407 & 3 &  2 & $11.31$ & $-22.20$ & $285$ & $3.32$ & $0.05$ & -$ 0.20$ & $0.84$ & $1.22^{+0.26}_{-0.26}$ & $2.53^{+0.71}_{-0.36}$ & $0.189^{+0.033}_{-0.033}$ & $0.103^{+0.123}_{-0.071}$ \\
NGC 1600 & 2,3 &  4 & $11.93$ & $-23.21$ & $321$ & $12.96$ & $0.33$ & -$ 1.20$ & $0.05$ & $1.32^{+0.26}_{-0.31}$ & $3.50^{+0.49}_{-1.06}$ & $0.224^{+0.023}_{-0.020}$ & $0.087^{+0.048}_{-0.040}$ \\
NGC 1700 & 2 &  4 & $11.49$ & $-21.95$ & $234$ & $4.37$ & $0.29$ & $ 0.90$ & $0.80$ & $4.44^{+1.04}_{-0.57}$ & $2.95^{+1.78}_{-0.79}$ & $0.011^{+0.005}_{-0.004}$ & $0.144^{+0.163}_{-0.055}$ \\
NGC 2832 & 2 &  2 & $12.52$ & $-23.76$ & $330$ & $31.62$ & $0.30$ & -$ 0.30$ & $0.12$ & $2.93^{+0.04}_{-0.04}$ & $3.19^{+1.48}_{-0.55}$ & $0.081^{+0.004}_{-0.004}$ & $0.070^{+0.048}_{-0.025}$ \\
NGC 3379 & 2,3 &  7 & $11.03$ & $-21.14$ & $221$ & $2.45$ & $0.10$ & $ 0.20$ & $0.82$ & $2.54^{+0.32}_{-0.58}$ & $3.06^{+0.49}_{-1.81}$ & $0.124^{+0.052}_{-0.018}$ & $0.190^{+0.137}_{-0.092}$ \\
NGC 3607 & 3 &  3 & $10.88$ & $-21.22$ & $248$ & $1.68$ & $0.16$ & -- & $0.92$ & $1.80^{+0.07}_{-0.32}$ & $3.11^{+1.48}_{-0.87}$ & $0.126^{+0.016}_{-0.000}$ & $0.229^{+0.180}_{-0.123}$ \\
NGC 3608 & 2 &  3 & $11.02$ & $-21.12$ & $195$ & $3.31$ & $0.18$ & -$ 0.20$ & $0.44$ & $2.68^{+0.20}_{-0.92}$ & $3.78^{+1.15}_{-1.14}$ & $0.108^{+0.074}_{-0.004}$ & $0.167^{+0.152}_{-0.077}$ \\
NGC 3640 & 2,3 &  6 & $11.17$ & $-21.53$ & $180$ & $3.75$ & $0.21$ & -$ 0.20$ & $1.48$ & $3.21^{+1.14}_{-0.17}$ & $2.92^{+1.05}_{-0.53}$ & $0.098^{+0.116}_{-0.060}$ & $0.154^{+0.192}_{-0.080}$ \\
NGC 3842 & 2 &  1 & $12.19$ & $-23.18$ & $316$ & $17.78$ & $0.17$ & -- & -- & $3.09$ & $3.69^{+0.37}_{-0.65}$ & $0.045$ & $0.085^{+0.048}_{-0.038}$ \\
NGC 4168 & 3 &  3 & $10.91$ & $-21.45$ & $182$ & $3.94$ & $0.11$ & -- & $0.26$ & $2.55^{+0.22}_{-0.56}$ & $2.90^{+1.28}_{-0.66}$ & $0.081^{+0.000}_{-0.006}$ & $0.077^{+0.144}_{-0.047}$ \\
NGC 4261 & 1,3 &  4 & $11.41$ & $-22.33$ & $294$ & $11.09$ & $0.21$ & -$ 1.30$ & $0.10$ & $2.40^{+0.64}_{-0.64}$ & $2.78^{+1.70}_{-0.38}$ & $0.105^{+0.026}_{-0.012}$ & $0.105^{+0.124}_{-0.068}$ \\
NGC 4278 & 2,3 &  8 & $10.70$ & $-20.52$ & $259$ & $2.02$ & $0.16$ & -- & $0.71$ & $3.91^{+0.45}_{-0.56}$ & $3.32^{+1.47}_{-1.85}$ & $0.119^{+0.016}_{-0.017}$ & $0.234^{+0.112}_{-0.121}$ \\
NGC 4291 & 2,3 &  4 & $10.71$ & $-20.63$ & $259$ & $1.60$ & $0.25$ & -$ 0.40$ & $0.52$ & $1.87^{+1.84}_{-0.23}$ & $2.92^{+0.59}_{-0.80}$ & $0.182^{+0.055}_{-0.008}$ & $0.230^{+0.116}_{-0.106}$ \\
NGC 4365 & 1,2,3 &  7 & $11.59$ & $-22.44$ & $269$ & $13.83$ & $0.24$ & -$ 1.10$ & $0.08$ & $2.61^{+1.12}_{-0.44}$ & $2.87^{+1.02}_{-0.53}$ & $0.067^{+0.037}_{-0.002}$ & $0.105^{+0.147}_{-0.053}$ \\
NGC 4374 & 1,3 &  4 & $11.10$ & $-22.43$ & $287$ & $8.29$ & $0.14$ & -$ 0.40$ & $0.09$ & $2.12^{+1.13}_{-0.34}$ & $2.57^{+2.11}_{-0.35}$ & $0.126^{+0.011}_{-0.010}$ & $0.215^{+0.170}_{-0.119}$ \\
NGC 4382 & 1,2,3 &  4 & $11.26$ & $-22.43$ & $196$ & $8.43$ & $0.19$ & $ 0.59$ & $0.33$ & $3.89^{+1.11}_{-1.35}$ & $3.11^{+0.63}_{-0.80}$ & $0.034^{+0.008}_{-0.010}$ & $0.090^{+0.086}_{-0.054}$ \\
NGC 4406 & 1,2,3 &  7 & $11.36$ & $-22.66$ & $250$ & $19.72$ & $0.29$ & -$ 0.70$ & $0.18$ & $2.81^{+0.90}_{-0.47}$ & $3.09^{+1.11}_{-0.91}$ & $0.033^{+0.011}_{-0.003}$ & $0.098^{+0.040}_{-0.068}$ \\
NGC 4472 & 1,2,3 &  7 & $11.70$ & $-23.12$ & $287$ & $16.34$ & $0.16$ & -$ 0.30$ & $0.47$ & $2.45^{+0.65}_{-0.95}$ & $3.32^{+0.91}_{-0.93}$ & $0.091^{+0.010}_{-0.047}$ & $0.045^{+0.088}_{-0.016}$ \\
NGC 4473 & 2,3 &  6 & $10.91$ & $-20.82$ & $178$ & $2.56$ & $0.43$ & $ 0.90$ & $0.40$ & $2.59^{+0.32}_{-0.75}$ & $3.49^{+1.27}_{-0.91}$ & $0.162^{+0.075}_{-0.036}$ & $0.216^{+0.095}_{-0.111}$ \\
NGC 4486 & 1,2 &  4 & $11.93$ & $-22.95$ & $360$ & $31.44$ & $0.12$ & $ 0.01$ & $0.11$ & $2.78^{+1.93}_{-0.22}$ & $2.71^{+0.56}_{-0.52}$ & $0.054^{+0.031}_{-0.001}$ & $0.083^{+0.103}_{-0.048}$ \\
NGC 4489 & 1,3 &  3 & $9.82$ & $-18.56$ & $49$ & $1.10$ & $0.12$ & -$ 0.20$ & $1.48$ & $1.06^{+0.13}_{-0.13}$ & $3.06^{+0.43}_{-0.76}$ & $0.214^{+0.021}_{-0.022}$ & $0.223^{+0.123}_{-0.117}$ \\
NGC 4552 & 1,2,3 &  7 & $11.25$ & $-21.46$ & $261$ & $7.74$ & $0.07$ & $ 0.01$ & $0.28$ & $2.58^{+1.18}_{-0.33}$ & $2.79^{+1.46}_{-0.83}$ & $0.136^{+0.025}_{-0.021}$ & $0.190^{+0.214}_{-0.092}$ \\
NGC 4589 & 2,3 &  6 & $11.00$ & $-21.45$ & $215$ & $4.38$ & $0.20$ & -- & $0.57$ & $3.40^{+0.81}_{-0.34}$ & $2.72^{+1.74}_{-0.74}$ & $0.045^{+0.004}_{-0.004}$ & $0.103^{+0.123}_{-0.066}$ \\
NGC 4636 & 1,2,3 &  8 & $11.38$ & $-21.96$ & $191$ & $16.28$ & $0.18$ & -$ 0.10$ & $0.25$ & $2.71^{+0.76}_{-0.89}$ & $2.91^{+0.82}_{-1.31}$ & $0.056^{+0.007}_{-0.017}$ & $0.076^{+0.064}_{-0.043}$ \\
NGC 4649 & 1,2,3 &  7 & $11.73$ & $-22.62$ & $341$ & $10.52$ & $0.17$ & -$ 0.50$ & $0.46$ & $2.77^{+0.66}_{-1.00}$ & $3.32^{+1.05}_{-0.86}$ & $0.083^{+0.037}_{-0.012}$ & $0.103^{+0.116}_{-0.070}$ \\
NGC 4874 & 2 &  2 & $12.37$ & $-23.49$ & $290$ & $61.66$ & $0.09$ & -$ 0.30$ & $0.22$ & $2.83^{+0.01}_{-0.01}$ & $4.22^{+0.46}_{-1.27}$ & $0.036^{+0.000}_{-0.000}$ & $0.074^{+0.093}_{-0.027}$ \\
NGC 4889 & 2,3 &  4 & $12.20$ & $-23.62$ & $381$ & $17.67$ & $0.35$ & -$ 0.25$ & $0.05$ & $1.58^{+0.56}_{-0.70}$ & $3.10^{+0.80}_{-0.67}$ & $0.145^{+0.033}_{-0.026}$ & $0.073^{+0.055}_{-0.028}$ \\
NGC 5061 & 2 &  1 & $11.53$ & $-22.01$ & $194$ & $3.55$ & $0.04$ & -- & -- & $3.55$ & $3.07^{+0.12}_{-0.90}$ & $0.075$ & $0.194^{+0.143}_{-0.103}$ \\
NGC 5322 & 3 &  3 & $11.34$ & $-22.21$ & $224$ & $4.68$ & -- & -$ 0.90$ & $0.26$ & $2.48^{+0.26}_{-0.26}$ & $1.88^{+2.92}_{-1.21}$ & $0.086^{+0.011}_{-0.011}$ & $0.103^{+0.080}_{-0.066}$ \\
NGC 5419 & 2 &  1 & $12.30$ & $-23.37$ & $315$ & $15.14$ & $0.21$ & -- & -- & $3.78$ & $2.89^{+0.33}_{-0.32}$ & $0.049$ & $0.087^{+0.045}_{-0.039}$ \\
NGC 5490 & 3 &  3 & $11.35$ & $-22.13$ & $301$ & $6.31$ & $0.19$ & -- & $0.27$ & $3.55^{+0.39}_{-0.39}$ & $2.75^{+1.92}_{-0.79}$ & $0.108^{+0.005}_{-0.006}$ & $0.106^{+0.121}_{-0.062}$ \\
NGC 5557 & 2,3 &  4 & $11.31$ & $-22.28$ & $260$ & $5.94$ & $0.21$ & -- & $0.12$ & $2.78^{+0.61}_{-1.41}$ & $3.01^{+4.33}_{-0.71}$ & $0.088^{+0.052}_{-0.006}$ & $0.105^{+0.101}_{-0.068}$ \\
NGC 5576 & 2 &  3 & $11.13$ & $-21.31$ & $188$ & $3.80$ & $0.30$ & -$ 0.50$ & $0.22$ & $4.87^{+2.67}_{-0.69}$ & $2.96^{+0.38}_{-0.43}$ & $0.240^{+0.055}_{-0.157}$ & $0.192^{+0.116}_{-0.094}$ \\
NGC 5813 & 2,3 &  6 & $11.20$ & $-21.68$ & $238$ & $9.82$ & $0.16$ & $ 0.01$ & $0.51$ & $3.16^{+0.04}_{-0.11}$ & $2.93^{+1.32}_{-0.84}$ & $0.044^{+0.013}_{-0.001}$ & $0.102^{+0.081}_{-0.041}$ \\
NGC 5982 & 2 &  1 & $11.50$ & $-21.97$ & $250$ & $5.01$ & $0.27$ & -- & -- & $3.02$ & $2.77^{+0.45}_{-0.24}$ & $0.217$ & $0.118^{+0.111}_{-0.034}$ \\
NGC 6166 & 2 &  1 & $12.55$ & $-23.80$ & $300$ & $67.61$ & $0.28$ & -- & $0.08$ & $2.73$ & $3.09^{+0.15}_{-0.56}$ & $0.074$ & $0.089^{+0.137}_{-0.042}$ \\
NGC 6702 & 3 &  3 & $11.45$ & $-22.26$ & $182$ & $6.39$ & $0.23$ & -$ 0.40$ & $0.18$ & $3.11^{+0.57}_{-0.57}$ & $1.85^{+1.95}_{-1.85}$ & $0.082^{+0.027}_{-0.002}$ & $0.089^{+0.131}_{-0.060}$ \\
NGC 7052 & 3 &  3 & $11.48$ & $-22.46$ & $275$ & $9.77$ & $0.45$ & $ 0.01$ & $0.34$ & $1.74^{+0.16}_{-0.20}$ & $3.51^{+2.55}_{-2.83}$ & $0.087^{+0.015}_{-0.009}$ & $0.085^{+0.023}_{-0.039}$ \\
NGC 7385 & 3 &  3 & $11.95$ & $-23.53$ & $259$ & $10.27$ & $0.12$ & $ 0.01$ & $0.16$ & $1.83^{+1.04}_{-0.37}$ & $1.55^{+2.25}_{-1.55}$ & $0.107^{+0.017}_{-0.017}$ & $0.087^{+0.056}_{-0.040}$ \\
NGC 7619 & 2,3 &  4 & $11.60$ & $-22.60$ & $337$ & $6.85$ & $0.24$ & $ 0.20$ & $0.53$ & $2.88^{+0.39}_{-0.34}$ & $2.67^{+0.64}_{-1.03}$ & $0.105^{+0.007}_{-0.006}$ & $0.103^{+0.119}_{-0.066}$ \\
NGC 7785 & 2,3 &  4 & $11.48$ & $-22.46$ & $291$ & $4.56$ & $0.42$ & -$ 1.50$ & $0.47$ & $1.75^{+1.42}_{-0.25}$ & $2.53^{+3.06}_{-2.02}$ & $0.092^{+0.006}_{-0.026}$ & $0.046^{+0.109}_{-0.016}$ \\
IC 1459 & 2 &  3 & $11.81$ & $-22.51$ & $311$ & $4.47$ & $0.26$ & -- & $0.22$ & $2.08^{+2.39}_{-0.56}$ & $4.67^{+0.00}_{-2.18}$ & $0.185^{+0.073}_{-0.131}$ & $0.126^{+0.150}_{-0.041}$ \\
IC 4296 & 3 &  3 & $11.75$ & $-23.18$ & $323$ & $5.86$ & $0.10$ & $ 0.00$ & $0.64$ & $1.81^{+0.01}_{-0.05}$ & $2.71^{+0.53}_{-0.38}$ & $0.139^{+0.005}_{-0.005}$ & $0.085^{+0.098}_{-0.051}$ \\
IC 4329 & 2 &  2 & $12.63$ & $-23.95$ & $270$ & $41.69$ & $0.15$ & -- & -- & $2.64^{+0.65}_{-0.65}$ & $3.19^{+1.48}_{-0.25}$ & $0.045^{+0.024}_{-0.024}$ & $0.079^{+0.058}_{-0.034}$ \\
A0076-M1 & 2 &  1 & $12.07$ & $-22.96$ & -- & $5.95$ & -- & -- & -- & $3.15$ & $3.07^{+0.12}_{-0.13}$ & $0.086$ & $0.117^{+0.136}_{-0.031}$ \\
A0119-M1 & 2 &  2 & $12.67$ & $-24.01$ & -- & $45.71$ & -- & -- & -- & $2.30^{+0.03}_{-0.03}$ & $2.91^{+0.98}_{-0.95}$ & $0.025^{+0.001}_{-0.001}$ & $0.047^{+0.058}_{-0.001}$ \\
A0168-M1 & 2 &  2 & $12.16$ & $-23.12$ & -- & $31.62$ & -- & -- & -- & $3.67^{+0.22}_{-0.22}$ & $3.18^{+0.51}_{-1.01}$ & $0.005^{+0.001}_{-0.001}$ & $0.047^{+0.058}_{-0.001}$ \\
A0193-M1 & 2 &  1 & $12.61$ & $-23.91$ & -- & $63.91$ & -- & -- & -- & $3.74$ & $3.76^{+0.91}_{-1.12}$ & $0.037$ & $0.076^{+0.082}_{-0.031}$ \\
A0194-M1 & 2 &  1 & $12.11$ & $-23.03$ & -- & $23.84$ & -- & -- & -- & $2.94$ & $2.76^{+2.16}_{-0.12}$ & $0.056$ & $0.074^{+0.058}_{-0.031}$ \\
A0195-M1 & 2 &  2 & $11.89$ & $-22.65$ & -- & $14.23$ & -- & -- & -- & $3.43^{+0.06}_{-0.06}$ & $2.91^{+2.01}_{-0.27}$ & $0.060^{+0.002}_{-0.002}$ & $0.081^{+0.057}_{-0.045}$ \\
A0260-M1 & 2 &  1 & $12.36$ & $-23.47$ & -- & $31.67$ & -- & -- & -- & $2.64$ & $2.89^{+0.33}_{-0.36}$ & $0.067$ & $0.060^{+0.047}_{-0.014}$ \\
A0295-M1 & 2 &  2 & $12.15$ & $-23.11$ & -- & $39.81$ & -- & -- & -- & $3.05^{+0.01}_{-0.01}$ & $3.10^{+0.72}_{-0.94}$ & $0.035^{+0.000}_{-0.000}$ & $0.047^{+0.058}_{-0.002}$ \\
A0347-M1 & 2 &  1 & $12.01$ & $-22.85$ & -- & $26.47$ & -- & -- & -- & $2.75$ & $2.77^{+0.45}_{-0.20}$ & $0.127$ & $0.087^{+0.054}_{-0.043}$ \\
A0376-M1 & 2 &  1 & $12.43$ & $-23.60$ & -- & $47.86$ & -- & -- & -- & $3.50$ & $2.77^{+0.45}_{-0.20}$ & $0.032$ & $0.047^{+0.058}_{-0.002}$ \\
A0397-M1 & 2 &  1 & $12.33$ & $-23.42$ & -- & $33.88$ & -- & -- & -- & $2.72$ & $2.77^{+0.45}_{-0.20}$ & $0.079$ & $0.084^{+0.051}_{-0.039}$ \\
A0496-M1 & 2 &  4 & $12.82$ & $-24.28$ & -- & $58.88$ & -- & -- & -- & $3.45^{+0.14}_{-0.10}$ & $2.80^{+0.53}_{-0.36}$ & $0.030^{+0.005}_{-0.006}$ & $0.065^{+0.085}_{-0.018}$ \\
A0533-M1 & 2 &  4 & $11.91$ & $-22.68$ & -- & $21.38$ & -- & -- & -- & $4.69^{+1.22}_{-1.87}$ & $2.89^{+0.45}_{-0.29}$ & $0.034^{+0.002}_{-0.003}$ & $0.085^{+0.082}_{-0.042}$ \\
A0548-M1 & 2 &  4 & $11.95$ & $-22.75$ & -- & $20.89$ & -- & -- & -- & $3.80^{+2.80}_{-0.42}$ & $2.84^{+0.39}_{-0.40}$ & $0.009^{+0.025}_{-0.002}$ & $0.077^{+0.096}_{-0.032}$ \\
A0634-M1 & 2 &  1 & $11.92$ & $-22.70$ & -- & $16.60$ & -- & -- & -- & $3.24$ & $2.89^{+0.33}_{-0.32}$ & $0.319$ & $0.086^{+0.049}_{-0.040}$ \\
A0779-M1 & 2 &  1 & $12.54$ & $-23.78$ & -- & $36.23$ & -- & -- & -- & $3.01$ & $2.89^{+0.33}_{-0.32}$ & $0.076$ & $0.074^{+0.056}_{-0.028}$ \\
A0999-M1 & 2 &  1 & $11.78$ & $-22.45$ & -- & $13.80$ & -- & -- & -- & $2.74$ & $2.77^{+0.45}_{-0.20}$ & $0.179$ & $0.087^{+0.086}_{-0.050}$ \\
A1016-M1 & 2 &  1 & $11.67$ & $-22.26$ & -- & $22.41$ & -- & -- & -- & $4.82$ & $3.13^{+0.06}_{-0.96}$ & $0.013$ & $0.089^{+0.072}_{-0.042}$ \\
A1142-M1 & 2 &  1 & $12.01$ & $-22.85$ & -- & $24.84$ & -- & -- & -- & $4.99$ & $2.64^{+0.60}_{-0.11}$ & $0.022$ & $0.089^{+0.080}_{-0.030}$ \\
A1177-M1 & 2 &  1 & $12.41$ & $-23.56$ & -- & $66.74$ & -- & -- & -- & $4.31$ & $2.64^{+0.60}_{-0.11}$ & $0.028$ & $0.074^{+0.080}_{-0.027}$ \\
A1314-M1 & 2 &  1 & $12.27$ & $-23.31$ & -- & $96.02$ & -- & -- & -- & $4.34$ & $2.53^{+0.71}_{-0.36}$ & $0.034$ & $0.085^{+0.083}_{-0.038}$ \\
A1367-M1 & 2 &  1 & $12.12$ & $-23.05$ & -- & $14.86$ & -- & -- & -- & $2.88$ & $2.91^{+0.30}_{-0.31}$ & $0.046$ & $0.078^{+0.054}_{-0.032}$ \\
A1631-M1 & 2 &  1 & $12.29$ & $-23.34$ & -- & $32.36$ & -- & -- & -- & $4.28$ & $2.64^{+0.30}_{-0.11}$ & $0.025$ & $0.074^{+0.059}_{-0.028}$ \\
A1656-M1 & 2 &  1 & $12.55$ & $-23.81$ & -- & $22.55$ & -- & -- & -- & $2.20$ & $2.64^{+0.57}_{-0.11}$ & $0.116$ & $0.059^{+0.048}_{-0.013}$ \\
A2040-M1 & 2 &  1 & $12.35$ & $-23.46$ & -- & $51.29$ & -- & -- & -- & $1.64$ & $2.91^{+0.18}_{-0.75}$ & $0.093$ & $0.076^{+0.064}_{-0.031}$ \\
A2052-M1 & 2 &  1 & $12.11$ & $-23.04$ & $250$ & $189.2$ & $0.24$ & -- & -- & $3.48$ & $2.75^{+0.52}_{-0.38}$ & $0.002$ & $0.089^{+0.127}_{-0.042}$ \\
A2147-M1 & 2 &  1 & $12.18$ & $-23.16$ & -- & $36.31$ & -- & -- & -- & $4.44$ & $2.64^{+0.57}_{-0.11}$ & $0.049$ & $0.059^{+0.059}_{-0.013}$ \\
A2162-M1 & 2 &  1 & $12.15$ & $-23.11$ & -- & $26.01$ & -- & -- & -- & $3.04$ & $3.54^{+0.15}_{-1.38}$ & $0.129$ & $0.088^{+0.077}_{-0.041}$ \\
A2197-M1 & 2 &  1 & $12.45$ & $-23.63$ & -- & $33.91$ & -- & -- & -- & $3.38$ & $2.53^{+0.71}_{-0.36}$ & $0.125$ & $0.074^{+0.057}_{-0.028}$ \\
A2572-M1 & 2 &  1 & $12.31$ & $-23.39$ & -- & $49.16$ & -- & -- & -- & $4.26$ & $3.13^{+0.11}_{-0.60}$ & $0.013$ & $0.089^{+0.095}_{-0.043}$ \\
A2589-M1 & 2 &  1 & $12.65$ & $-23.98$ & -- & $74.18$ & -- & -- & -- & $4.53$ & $2.89^{+0.33}_{-0.36}$ & $0.008$ & $0.084^{+0.097}_{-0.038}$ \\
A2593-M1 & 2 &  1 & $12.52$ & $-23.75$ & -- & $21.00$ & -- & -- & -- & $1.76$ & $2.64^{+0.57}_{-0.11}$ & $0.029$ & $0.070^{+0.083}_{-0.024}$ \\
A2877-M1 & 2 &  1 & $12.63$ & $-23.94$ & -- & $24.28$ & -- & -- & -- & $2.79$ & $2.91^{+0.33}_{-0.75}$ & $0.098$ & $0.087^{+0.048}_{-0.041}$ \\
A3144-M1 & 2 &  1 & $11.68$ & $-22.28$ & -- & $10.96$ & -- & -- & -- & $3.19$ & $2.64^{+0.36}_{-0.48}$ & $0.155$ & $0.120^{+0.110}_{-0.046}$ \\
A3193-M1 & 2 &  2 & $11.97$ & $-22.78$ & -- & $15.17$ & -- & -- & -- & $4.04^{+0.00}_{-0.00}$ & $2.64^{+0.60}_{-0.11}$ & $0.036^{+0.003}_{-0.003}$ & $0.090^{+0.082}_{-0.043}$ \\
A3367-M1 & 2 &  2 & $11.69$ & $-22.30$ & -- & $12.52$ & -- & -- & -- & $3.84^{+0.31}_{-0.31}$ & $2.77^{+0.45}_{-0.24}$ & $0.011^{+0.002}_{-0.002}$ & $0.087^{+0.054}_{-0.051}$ \\
A3376-M1 & 2 &  2 & $12.26$ & $-23.29$ & -- & $35.48$ & -- & -- & -- & $1.90^{+0.36}_{-0.36}$ & $3.17^{+0.32}_{-0.64}$ & $0.133^{+0.033}_{-0.033}$ & $0.084^{+0.084}_{-0.037}$ \\
A3395-M1 & 2 &  1 & $12.79$ & $-24.23$ & -- & $81.28$ & -- & -- & -- & $2.75$ & $2.64^{+0.60}_{-0.11}$ & $0.013$ & $0.074^{+0.094}_{-0.027}$ \\
A3526-M1 & 2 &  2 & $12.88$ & $-24.38$ & -- & $20.63$ & -- & -- & -- & $2.82^{+0.01}_{-0.01}$ & $2.83^{+0.39}_{-0.26}$ & $0.005^{+0.000}_{-0.000}$ & $0.047^{+0.058}_{-0.001}$ \\
A3528-M1 & 2 &  1 & $12.83$ & $-24.30$ & -- & $184.2$ & -- & -- & -- & $5.25$ & $3.09^{+0.15}_{-0.56}$ & $0.018$ & $0.086^{+0.104}_{-0.039}$ \\
A3532-M1 & 2 &  1 & $12.99$ & $-24.58$ & -- & $200.7$ & -- & -- & -- & $4.36$ & $3.20^{+1.73}_{-0.71}$ & $0.018$ & $0.085^{+0.098}_{-0.038}$ \\
A3554-M1 & 2 &  1 & $12.66$ & $-23.99$ & -- & $52.48$ & -- & -- & -- & $1.55$ & $2.53^{+0.71}_{-0.36}$ & $0.039$ & $0.061^{+0.074}_{-0.015}$ \\
A3556-M1 & 2 &  1 & $12.46$ & $-23.65$ & -- & $19.95$ & -- & -- & -- & $3.01$ & $2.53^{+0.66}_{-0.36}$ & $0.051$ & $0.078^{+0.055}_{-0.031}$ \\
A3558-M1 & 2 &  1 & $13.19$ & $-24.92$ & -- & $81.28$ & -- & -- & -- & $1.95$ & $2.53^{+0.66}_{-0.36}$ & $0.070$ & $0.090^{+0.135}_{-0.042}$ \\
A3559-M1 & 2 &  1 & $12.58$ & $-23.86$ & -- & $45.19$ & -- & -- & -- & $3.94$ & $2.53^{+0.66}_{-0.36}$ & $0.043$ & $0.074^{+0.061}_{-0.028}$ \\
A3562-M1 & 2 &  1 & $12.84$ & $-24.32$ & -- & $110.8$ & -- & -- & -- & $1.63$ & $2.77^{+0.17}_{-0.24}$ & $0.083$ & $0.089^{+0.128}_{-0.042}$ \\
A3564-M1 & 2 &  1 & $11.91$ & $-22.68$ & -- & $5.62$ & -- & -- & -- & $1.79$ & $2.91^{+0.18}_{-0.38}$ & $0.123$ & $0.102^{+0.096}_{-0.065}$ \\
A3570-M1 & 2 &  1 & $11.83$ & $-22.54$ & -- & $23.44$ & -- & -- & -- & $8.62$ & $2.77^{+0.45}_{-0.20}$ & $0.022$ & $0.090^{+0.093}_{-0.042}$ \\
A3571-M1 & 2 &  1 & $13.40$ & $-25.30$ & -- & $117.9$ & -- & -- & -- & $2.51$ & $2.89^{+0.33}_{-0.32}$ & $0.004$ & $0.105^{+0.139}_{-0.051}$ \\
A3574-M1 & 2 &  1 & $12.53$ & $-23.77$ & -- & $34.85$ & -- & -- & -- & $3.25$ & $2.89^{+0.33}_{-0.32}$ & $0.020$ & $0.074^{+0.064}_{-0.028}$ \\
A3656-M1 & 2 &  2 & $12.42$ & $-23.58$ & -- & $69.98$ & -- & -- & -- & $4.58^{+0.60}_{-0.60}$ & $2.77^{+0.45}_{-0.20}$ & $0.046^{+0.022}_{-0.022}$ & $0.084^{+0.083}_{-0.038}$ \\
A3677-M1 & 2 &  1 & $11.64$ & $-22.21$ & -- & $26.30$ & -- & -- & -- & $5.99$ & $2.77^{+0.45}_{-0.20}$ & $0.083$ & $0.103^{+0.103}_{-0.044}$ \\
A3716-M1 & 2 &  2 & $12.52$ & $-23.75$ & -- & $53.70$ & -- & -- & -- & $2.95^{+0.44}_{-0.44}$ & $2.49^{+0.75}_{-0.52}$ & $0.029^{+0.003}_{-0.003}$ & $0.060^{+0.081}_{-0.014}$ \\
A3736-M1 & 2 &  1 & $12.65$ & $-23.98$ & -- & $41.69$ & -- & -- & -- & $2.87$ & $2.64^{+0.30}_{-0.11}$ & $0.074$ & $0.047^{+0.058}_{-0.001}$ \\
A3742-M1 & 2 &  1 & $11.64$ & $-22.20$ & -- & $6.35$ & -- & -- & -- & $1.95$ & $2.72^{+0.17}_{-0.75}$ & $0.349$ & $0.117^{+0.115}_{-0.044}$ \\
A3744-M1 & 2 &  1 & $12.01$ & $-22.86$ & -- & $32.73$ & -- & -- & -- & $6.75$ & $2.59^{+0.07}_{-0.35}$ & $0.129$ & $0.087^{+0.080}_{-0.041}$ \\
A3747-M1 & 2 &  2 & $11.89$ & $-22.65$ & -- & $14.45$ & -- & -- & -- & $3.50^{+0.46}_{-0.46}$ & $2.81^{+0.43}_{-0.65}$ & $0.035^{+0.003}_{-0.003}$ & $0.090^{+0.051}_{-0.053}$ \\
A4038-M1 & 2 &  1 & $11.91$ & $-22.68$ & -- & $13.24$ & -- & -- & -- & $2.95$ & $2.89^{+0.33}_{-0.36}$ & $0.077$ & $0.090^{+0.054}_{-0.044}$ \\
\enddata
\tablenotetext{ \, }{{\footnotesize Compiled and fitted parameters for the confirmed core ellipticals 
in our observed samples. Columns show: (1) Object name. (2) Source for surface brightness 
profiles, where $1=$\citet{jk:profiles}, $2=$\citet{lauer:bimodal.profiles}, $3=$\citet{bender:data}.
(3) Total number of different surface brightness profiles in our combined samples for 
the given object. (4) Stellar mass [$\log{M_{\ast}/M_{\sun}}$]. (5) $V$-band absolute magnitude. 
(6) Velocity dispersion [km/s]. (7) Effective (half-light) radius of the 
{\em total} light profile [kpc]. (8) Ellipticity. (9) Boxy/diskyness. (10) Rotation. 
(11) Outer Sersic index $n_{s}$ of the two-component best-fit profile. Where multiple 
profiles are available for the same object, we show the median and minimum/maximum 
range of fitted $n_{s}$ values. 
(12) Range of outer Sersic indices fit in the same manner to the best-fit simulations, 
at $t\approx1-3$\,Gyr after the merger when the system has relaxed.
(13) Fraction of light in the inner or extra light component of the fits. 
Where multiple 
profiles are available for the same object, we show the median and minimum/maximum 
range of fitted values. 
(14) Fraction of light from stars produced in the central, merger-induced starburst 
in the best-fit simulations ($\pm$ the approximate interquartile range allowed). \\
}}
\end{\tableset}

\clearpage

\begin{appendix}

\section{Fits to the Sample of Kormendy et.\ al.\ 2008}
\label{sec:appendix:jk}

\begin{figure*}
    \centering
    \plotter{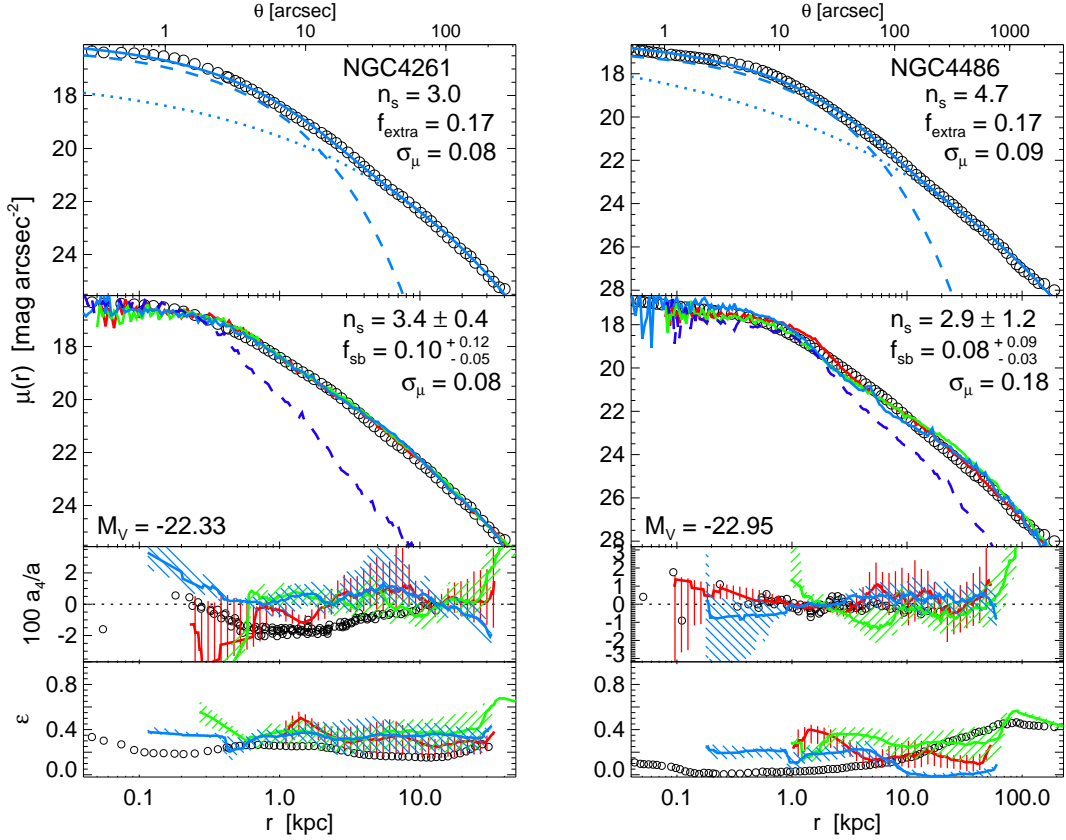}
    \caption{As Figure~\ref{fig:jk1}, but in log-log space. 
    Surface brightness profiles -- decomposed into 
    dissipationless and dissipational/extra light components -- 
    are shown for core ellipticals in and around the 
    Virgo cluster$^{\ref{foot:4261}}$. Open circles show the observations, from \citet{jk:profiles}. 
    These are the highest-mass core ellipticals in Virgo
    ($\sim6-8\,\mstar$).
    {\em Upper:} Observed V-band surface brightness profile with our 
    two component best-fit model (solid, dashed, and dotted lines show the 
    total, inner/extra light component, and outer/pre-starburst component). 
    The best-fit outer Sersic index, extra light fraction, and rms residuals about the 
    fit are shown.
    {\em  Lower:} Colored lines show the corresponding surface brightness 
    profiles from the three simulations in our library which correspond 
    most closely to the observed system. Dashed line shows the 
    profile of the starburst light in the best-matching simulation. 
    The range of outer Sersic indices in the simulations (i.e.\ across sightlines for 
    these objects) and range of starburst mass fractions which match the 
    observed profile are shown, with the rms residuals of the observations about the 
    best-fit simulation$^{\ref{foot:explainfits}}$. 
    {\em Bottom:} Observed disky/boxy-ness ($a_{4}$) and ellipticity profiles, 
    with the median (solid) and $25-75\%$ range (shaded) corresponding profile 
    from the best-fitting simulations above. Note that these are not fitted for in any sense. 
    Figures~\ref{fig:jk2.log}-\ref{fig:jk5.log}
    show the other core ellipticals in the sample, ranked from most to least massive.
    \label{fig:jk1.log}}
\end{figure*}

\begin{figure*}
    \centering
    \plotter{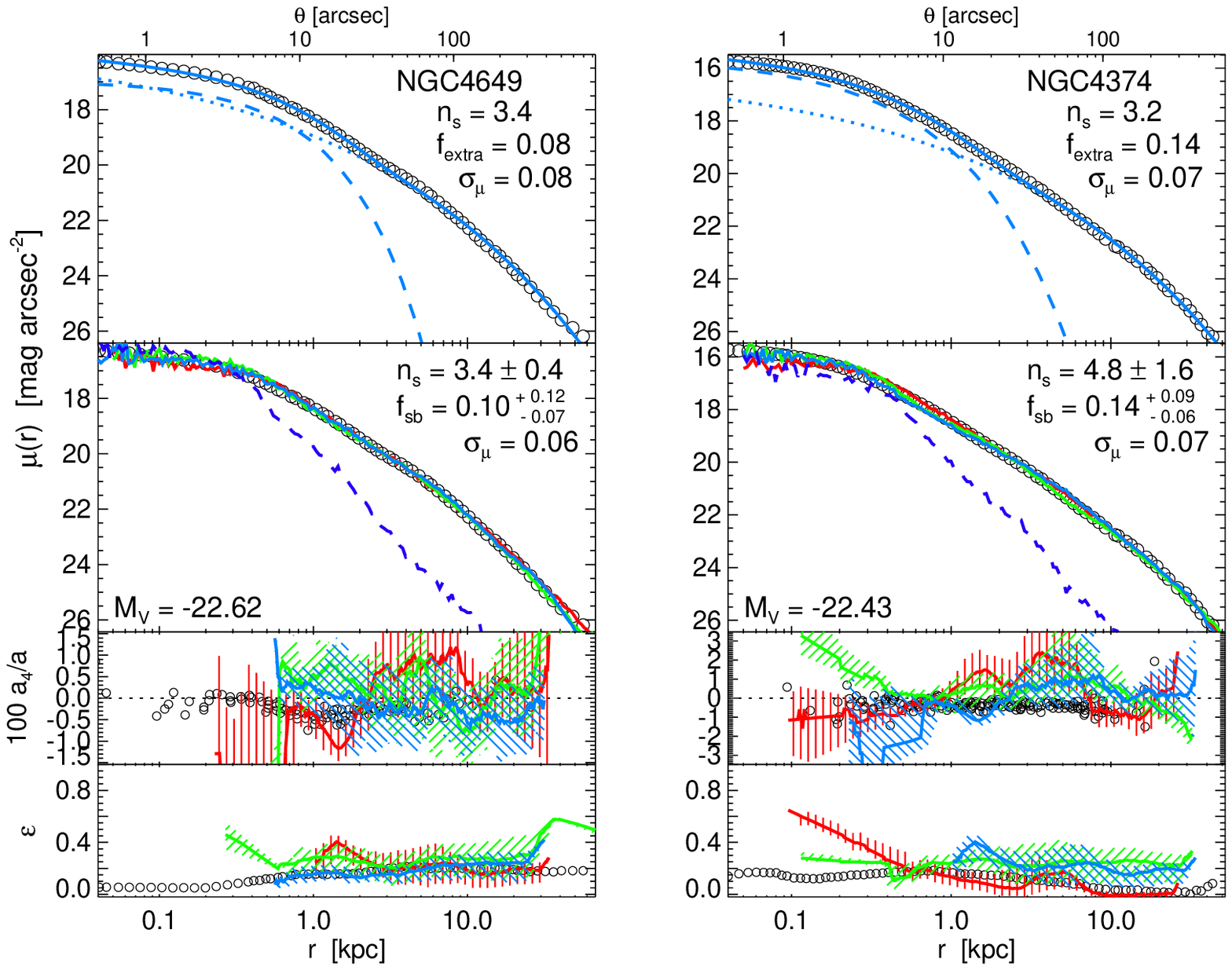}
    \caption{The next most massive core ellipticals ($\sim5-6\,\mstar$). 
    (As Figure~\ref{fig:jk2}, but in log-log space.)
    \label{fig:jk2.log}}
\end{figure*}

\begin{figure*}
    \centering
    \plotter{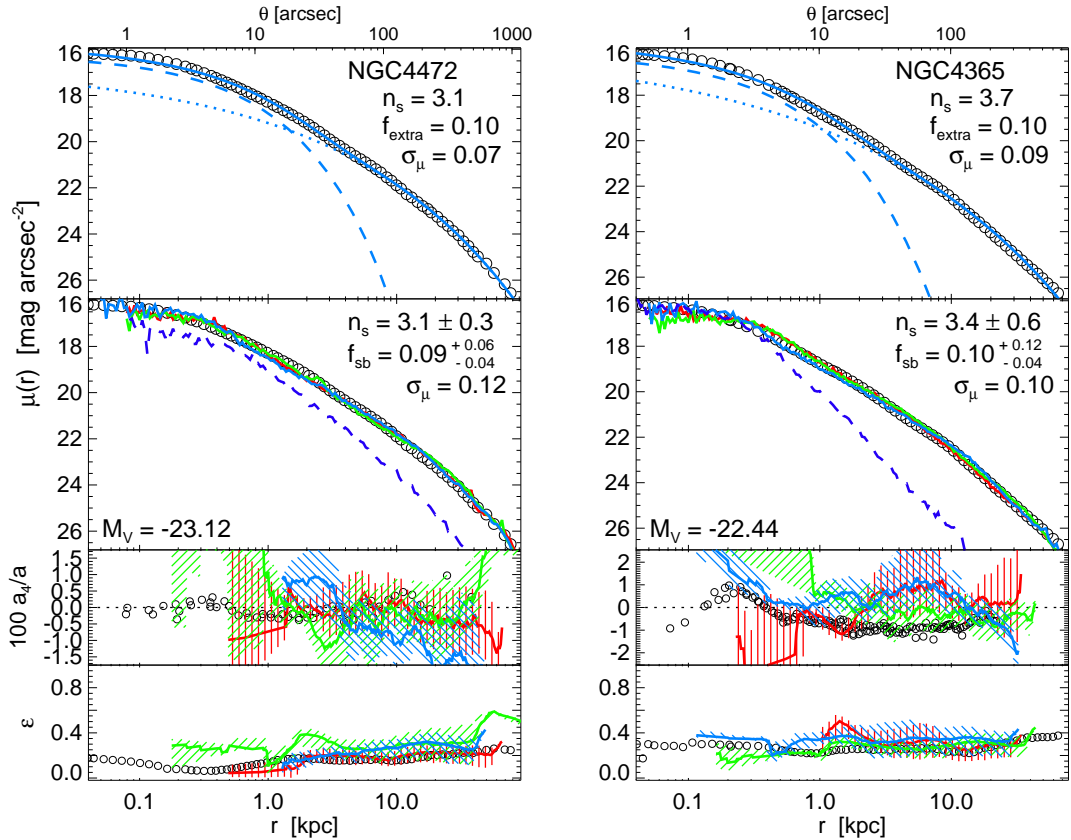}
    \caption{Similar massive core ellipticals ($\sim4-5\,\mstar$).
    (As Figure~\ref{fig:jk3}, but in log-log space.) 
    \label{fig:jk3.log}}
\end{figure*}

\begin{figure*}
    \centering
    \plotter{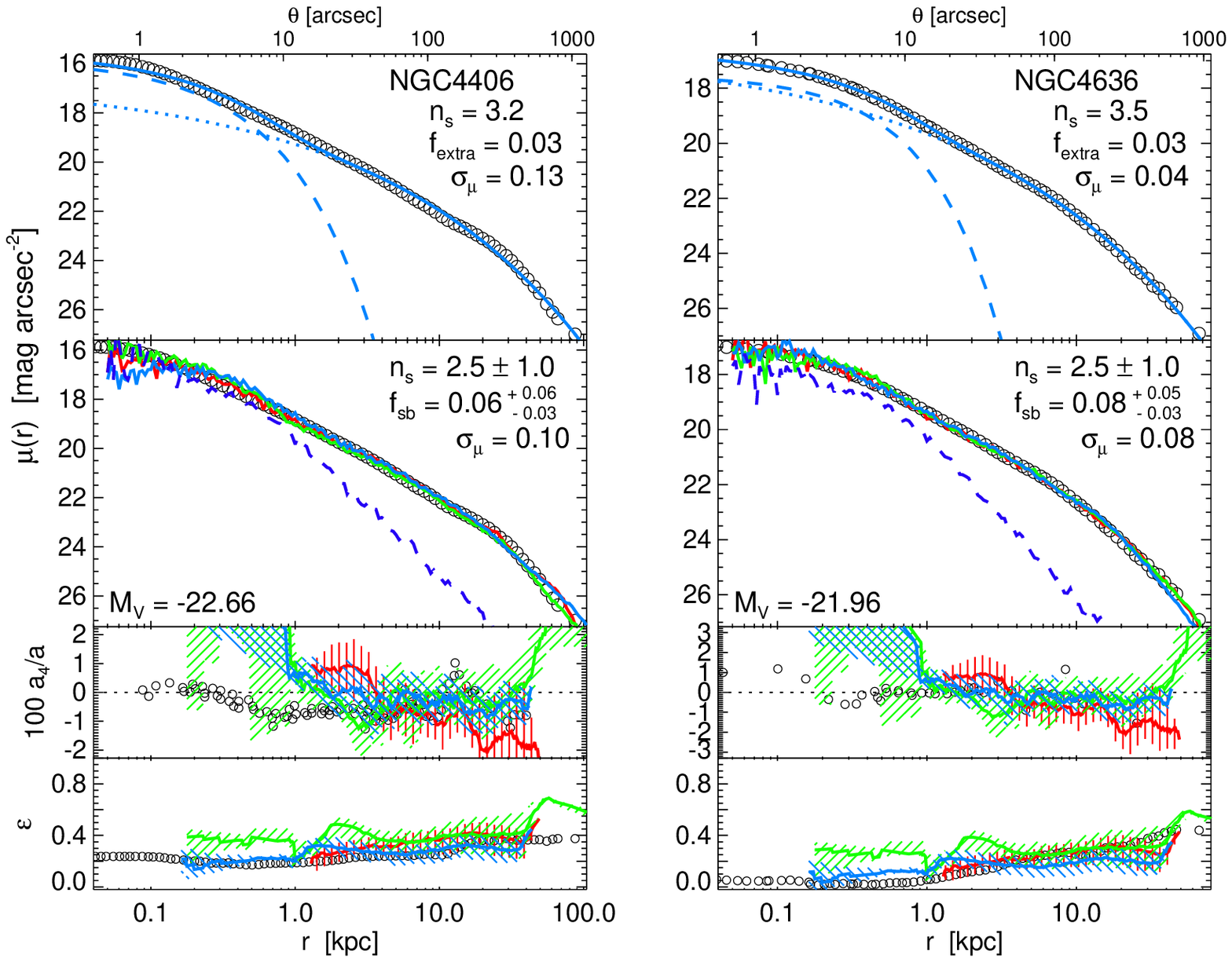}
    \caption{The next most massive core ellipticals ($\sim2-3\,\mstar$).   
    (As Figure~\ref{fig:jk4}, but in log-log space.)
    \label{fig:jk4.log}}
\end{figure*}

\begin{figure*}
    \centering
    \plotter{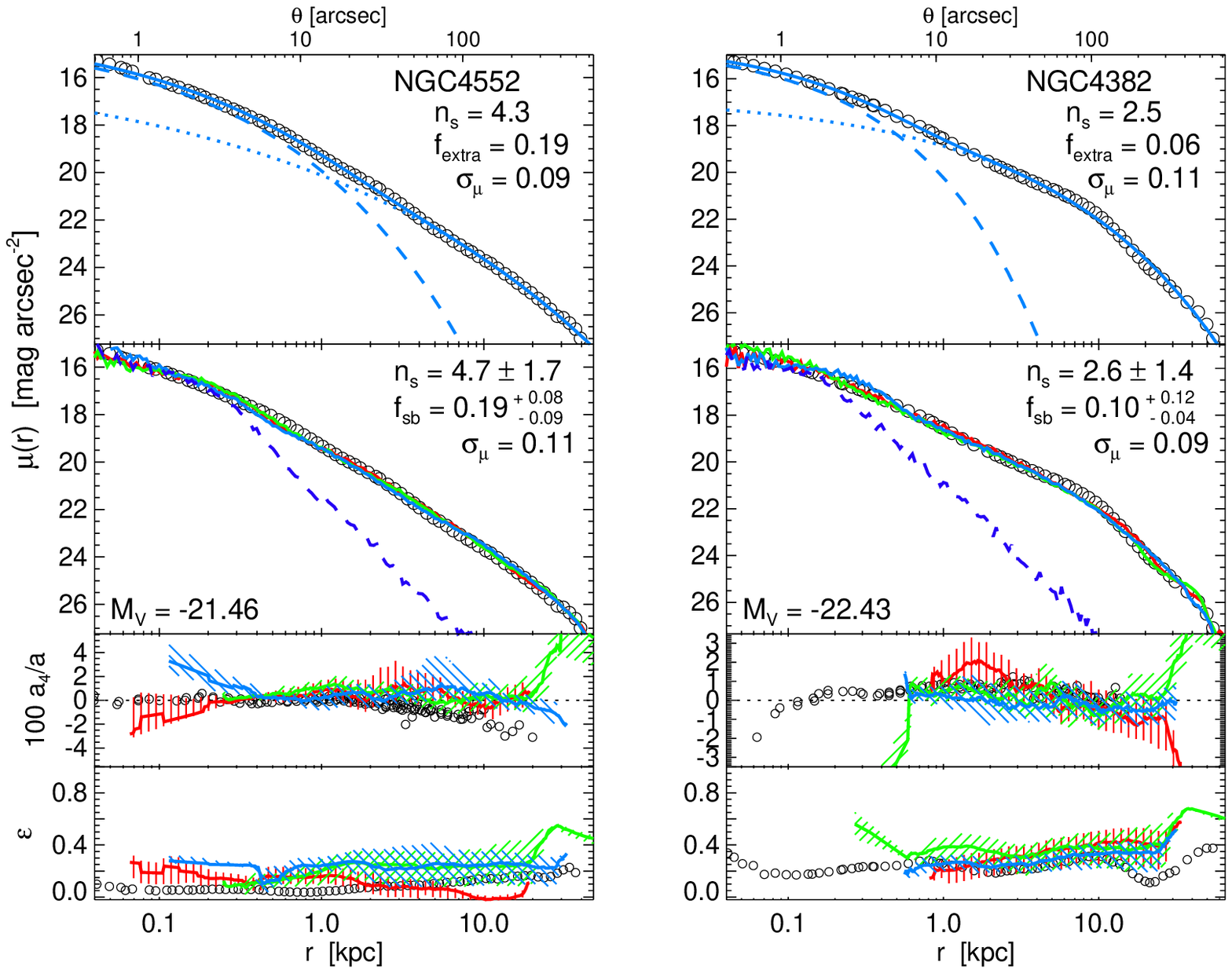}
    \caption{The least massive core ellipticals in the \citet{jk:profiles} 
    Virgo sample ($\sim1-2\,\mstar$). (As Figure~\ref{fig:jk5}, but in log-log space.)
    \label{fig:jk5.log}}
\end{figure*}

In Figures~\ref{fig:jk1.log}-\ref{fig:jk5.log} we reproduce Figures~\ref{fig:jk1}-\ref{fig:jk5}, 
but with profiles shown in log-log projection as opposed to $r^{1/4}$ projection. 


\end{appendix}

\end{document}